\documentclass[twocolumn,trackchanges]{aastex63}

\usepackage{mathrsfs}
\usepackage{xcolor}
\newcommand{\gr}{{$\gamma$-ray}}

\shorttitle{multi-frequency neutrino sources}
\shortauthors{Chang et al.}

\begin{document}

\title{Hunting for neutrino emission from multi-frequency variable sources}

\correspondingauthor{Yu-Ling Chang and Donglian Xu}\email{ylchang@sjtu.edu.cn,donglianxu@sjtu.edu.cn}

\author[0000-0002-0196-3496]{Yu-Ling Chang}
\affiliation{Tsung-Dao Lee Institute, Shanghai Jiao Tong University, Shanghai 201210, P.R.China}

\author[0000-0001-8166-6602]{Bruno Arsioli}
\affiliation{University of Trieste and INFN, via Valerio 2, I-34127 Trieste, Italy.}

\affiliation{ICRANet, P.zza della Repubblica 10, I-65122, Pescara, Italy}

\author[0000-0002-4300-5130]{Wenlian Li}
\affiliation{Tsung-Dao Lee Institute, Shanghai Jiao Tong University, Shanghai 201210, P.R.China}

\author[0000-0003-1639-8829]{Donglian Xu}
\affiliation{Tsung-Dao Lee Institute, Shanghai Jiao Tong University, Shanghai 201210, P.R.China}
\affiliation{INPAC and School of Physics and Astronomy, Shanghai Jiao Tong University, Shanghai 200240, P.R.China}
\affiliation{Shanghai Laboratory for Particle Physics and Cosmology, Key Laboratory for Particle Physics and Cosmology(MOE), Shanghai 200240, P.R.China}

\author[0000-0002-1908-0536]{Liang Chen}
\affiliation{Key Laboratory for Research in Galaxies and Cosmology, Shanghai Astronomical Observatory, Chinese Academy of Sciences, 80 Nandan Road, Shanghai 200030, P.R.China}



\begin{abstract}
Pinpointing the neutrino sources is crucial to unveil the mystery of high-energy cosmic rays. 
The search for neutrino-source candidates from coincident neutrino-photon signatures and objects with particular electromagnetic flaring behaviors can increase our chances of finding neutrino emitters. 
In this paper, we first study the temporal correlations of astrophysical flares with neutrinos, considering a few hundreds of multi-frequency sources from ALMA, WISE, {\it Swift}, and {\it Fermi} in the containment regions of IceCube high-energy alerts. 
Furthermore, the spatial correlations between blazars and neutrinos are investigated using the subset of 10-year IceCube track-like neutrinos with around 250 thousand events. 
In a second test, we account for 2700 blazars with different types of flaring stages in addition to their position. 
No significant neutrino emissions were found from our analyses. 
Our results indicate an interesting trend showing that the infrared flaring phases of WISE blazars might be correlated with the arrival times of the neutrino alerts. 
A possible overflow of neutrinos associated with two of our selected blazar samples is discussed in details. 
One is characterized by a significant flaring lag in infrared with respect to \gr s, like seen for TXS\,0506+056, and the other is characterized by highly simultaneous infrared and \gr\ flares. 
Our investigation suggests the need to improve current multi-frequency light-curve catalogs to pair with the advent of more sensitive neutrino observatories. 


\end{abstract}



\section{Introduction} \label{intro}
The origin of high-energy cosmic rays is one of the most important open questions for more than a century. 
Neutrinos are ideal messengers for tracking the origin of cosmic rays as they are undeflected when traveling through space.
Since IceCube reported the detection of high-energy astrophysical neutrinos 
\citep{IceCube2013a,IceCube2015,IceCube2016,IceCube2021b}, identifying the sources of those neutrino events is one of the most pressing challenges in modern astrophysics. 
With no significant anisotropy found from the diffuse flux of astrophysical neutrinos, a substantial fraction of the observed neutrino flux is expected to be of extragalactic origin. 
Blazars, a special type of Active Galactic Nuclei (AGNs) - and the most energetic sources of continuous non-thermal radiation in the Universe \citep{Padovani2017} - have long been suggested to be one of the most promising astrophysical counterparts for neutrinos (for a recent review, see \citet{Giommi2021a}). 
Neutrinos are expected to be produced via the decays of charged pions generated from the interaction of photons with protons accelerated within blazar jets. 
The charged pions are accompanied by neutral pions that decay into very-high-energy (VHE) \gr s; therefore, \gr\ photons are direct companions for neutrinos. A relationship between the two messengers is expected if they are produced from the hadronic process at the same site. 

The first hint of a statistical connection between high synchrotron peaked blazars (HSPs/HBLs) and neutrinos was reported by \citet{Padovani2016}. They suggested a chance probability of association between 2FHL\footnote{2FHL: Second {\it Fermi}-LAT Catalog of High-Energy Sources, \citet{Fermi2FHL} } HBLs with IceCube events to be $\sim 0.4-1.3\%$, depending on \gr\ fluxes. 
Applying the ``energetic test" presented in \citet{Padovani2014}, \citet{Padovani2016} further reported $\sim 5$ probable HBLs counterparts for IceCube neutrinos. 
Other searches for neutrino counterparts with samples of \gr\ blazars found no evidence of \gr\ emission associated with IceCube neutrino events \citep{Brown2015,Palladino2017,kraub2018}. 

The importance of multi-frequency data to single out the most likely candidates for neutrino events has been highlighted in \citet{Padovani2016}. 
\citet{Righi2019a} and \citet{Franckowiak2020} analyzed the {\it Fermi}-detected blazars located within neutrino containment regions with a multi-frequency approach. 
Some of those potential neutrino blazars show temporal coincidence between \gr\ flares and neutrino events, but yet there is no compelling evidence to conclude on the association between the \gr\ photons and the IceCube neutrinos. 
Besides, \citet{Luo2020} suggested that the multi-frequency selected blazar sample, 5BZCAT\footnote{The 5th edition of the Roma-BZCAT \citet{Massaro2015}}, showed no significant correlation with the IceCube alert list. 
Later, \citet{Giommi2020} reported a $\sim 3.2 \sigma$ correlation excess with \gr\ HBLs and IBLs\footnote{According to the peak frequency of synchrotron radiation ($\nu^{S}_{\rm peak}$), blazars are divided into high-(HSP/HBL: $\nu^{S}_{\rm peak} \geq 10^{15}~{\rm Hz}$), intermediate-(ISP/IBL: $10^{14}~{\rm Hz} \leq \nu^{S}_{\rm peak} < 10^{15}~{\rm Hz}$), and low-(LSP/LBL: $\nu^{S}_{\rm peak} < 10^{14}~{\rm Hz}$) peaked sources respectively\citep{Abdo2010a}} in the vincinity of IceCube high-energy track-like events. 
They identified probable \gr\ blazar counterparts for IceCube neutrinos using the VOU-Blazars tool \citep{Chang2020}, designed to find blazar/AGN candidates with multi-frequency data from the Virtual Observatory\footnote{\url{http://www.ivoa.net}}. 

The VHE \gr s produced inevitably during the photo-hadronic process might cascade down to lower energy due to the absorption within the source or in further interactions with the extragalactic background light via photon-photon annihilation \citep{Franckowiak2020}. 
The source environment of astrophysical neutrino counterparts might be optically thick to GeV \gr s. 
Indeed, most of the neutrino activity has no \gr\ flare companion.  
Given that the pionic \gr\ photons may cascade down to X-ray band in blazar jets with strong photons fields, a stacking analysis of {\it Swift} BASS\footnote{The BAT AGN Spectroscopic Survey, \citet{Baumgartner2013}} objects \citep{Goswami2021} and a time-dependent search using the position of X-ray selected blazars from 5BZCAT \citep{Sharma2021} were proposed. 

Recently, \citet{Plavin2020} and \citet{Plavin2021} found that the positions of radio-bright blazars are statistically coincident with arrival directions of neutrino events at $4\sigma$ level, when considering a complete flux-density-limited sample of radio-loud (jetted) AGNs selected from VLBI Radio Fundamental Catalog\footnote{\url{http://astrogeo.org/rfc/}}. 
Radio emissions above $10$~GHz\footnote{RATAN-600 \citep{Korolkov1979} monitoring data} were found to increase around neutrino arrival times for those potential neutrino VLBI-selected blazars. 
This was later confirmed by \citet{Hovatta2021} with OVRO 15 GHz light curves\footnote{\url{https://sites.astro.caltech.edu/ovroblazars/}}. 
While \citet{Zhou2021} found no significant correlation between the same population of radio-bright blazars and IceCube 10-year track-like events in their stacking analyses. 
\citet{Illuminati2021} further performed a time-dependent search for neutrino flares from the direction of those radio-bright blazars but found no significant ANTARES flares. 

The contribution of blazars to the observed astrophysical neutrino flux has been constrained by cross-correlation joint stacking analyses between IceCube datasets and blazar samples. 
The stacking limits depend on the assumption that all the stacked sources have similar neutrino spectral shapes. 
Indexes of 2 \citep{Huber2019} and 2.5 \citep{Aartsen2017a} are generally applied in the stacking analyses \citep{Smith2021}, motivated by Fermi acceleration and the spectra of diffuse neutrino flux. 
From the stacking, blazars' contributions are found to be $\lesssim 15 -27\%$. 
Constraints obtained from other methods, such as multiplets and auto-correlation \citep{Murase2014,Yuan2020,Bartos2021}, or prediction according blazar hadronic models or observed data are all consistent with the stacking limits \citep{Padovani2015,Padovani2016,Murase2014}. 
All these results suggest that blazars may not be the dominant sources of the IceCube diffuse neutrino flux. 
Since these limits only allow constraining the average neutrino emission for sources on a long time scale, it is possible that individual sources can outshine these limits over a shorter period, such as their flaring phases \citep{Huber2019}. 

As the correlation of \gr\ radiation with neutrino emissions may not be straightforward due to the cascades, looking for associations in other wavebands might give us interesting hints on neutrino sources. 
Additionally, blazar flares are suggested as promising transients for neutrino production \citep{Oikonomou2019,Murase2018}, and studies based on blazars' flaring properties might bring about new insight. 

Here we propose a series of analyses to search for neutrino emissions from flaring sources with multi-frequency data.
Our works consist of two parts. 
First, we study the multi-frequency sources inside the containment regions of IceCube alerts, investigating the correlation between flaring phases in various wavebands and the arrival time of the alerts. 
The second part of our analyses will focus on blazars, analyzing their light curves in two different bands: low frequency (infrared) and high frequency (\gr). 
We aim to study the correlation with neutrinos among blazars with different types of multi-frequency activities, taking into account the flaring stages from promising neutrino blazar TXS\,0506+056. 
No previous studies have looked into the neutrino emission from blazars considering their multi-frequency light curves and accounting for correlations of multi-frequency flaring phases. 

In section~\ref{txsactivity}, we introduce the multi-frequency behaviors of TXS\,0506+056, which will be investigated throughout this paper. 
The neutrino data samples, the multi-frequency catalogs, and the blazar samples used in this work are described in section~\ref{data}, and the selection of our source lists is shown in section~\ref{sample}. 
In section~\ref{flares} and \ref{blazarlc}, we study the correlations of neutrinos with the electromagnetic flares of multi-frequency sources and blazars with different multi-frequency flaring stages. 
We discuss and summarize our results in section~\ref{discuss} and \ref{conclude}. 

\section{Multi-frequency flares of TXS\,0506+056} \label{txsactivity}
In Sep. 2017, the flaring state of a bright \gr\ blazar, TXS\,0506+056, was found in spatial and temporal coincidence with a 290 TeV neutrino alert, IceCube-170922A, at $3\sigma$ significance \citep{IceCube2018a}. 
This neutrino event was accompanied by strong flares of TXS\,0506+056 across the electromagnetic spectrum. 
A time-dependent search for archival neutrino flares with 9.5 years of IceCube data found a $3.5\sigma$ excess of $\sim 13.5$ neutrinos in 2014-2015 from the same direction \citep{IceCube2018b}. 
The multi-messenger association of 2017 alerts and 2014-2015 flares with TXS\,0506+056 revealed this blazar as the first likely extragalactic high-energy neutrino source and triggered a considerable interest in the nature of counterparts of astrophysical neutrinos. 

The case of TXS\,0506+056 demonstrates that the observed coincident activities of neutrinos and photons would greatly increase the probability of identifying the counterparts of IceCube events \citep{IceCube2018a}. 
The highly variable characteristics of blazars, with flux that could increase at least a factor of two in a day, makes them play an indispensable role in finding astrophysical counterparts of high energy neutrinos. 
\citet{Murase2018} argue that neutrinos from blazars can be dominated by the flares in the standard leptonic scenario for their gamma-ray emission. 
\citet{Oikonomou2019} and \citet{Stathopoulos2021} estimated the neutrino emissions associated to \gr\ and X-ray flaring periods of 12 Fermi bright blazars (for which simultaneous observations exist) and another 66 blazars \citep[observed more than 50 times with the Swift X-ray Telescope - XRT,][]{Giommi2021b}, respectively. 
Those works predicted the highest rates of muon neutrinos to be $\sim 1.2-3.0 \ \rm{yr}^{-1}$, concerning the X-ray flares of Mrk\,421 and the \gr\ flares of AO\,0235+164 and OJ\,287. 

\begin{figure} [h!]
\begin{center}
\includegraphics[width=1.\linewidth]{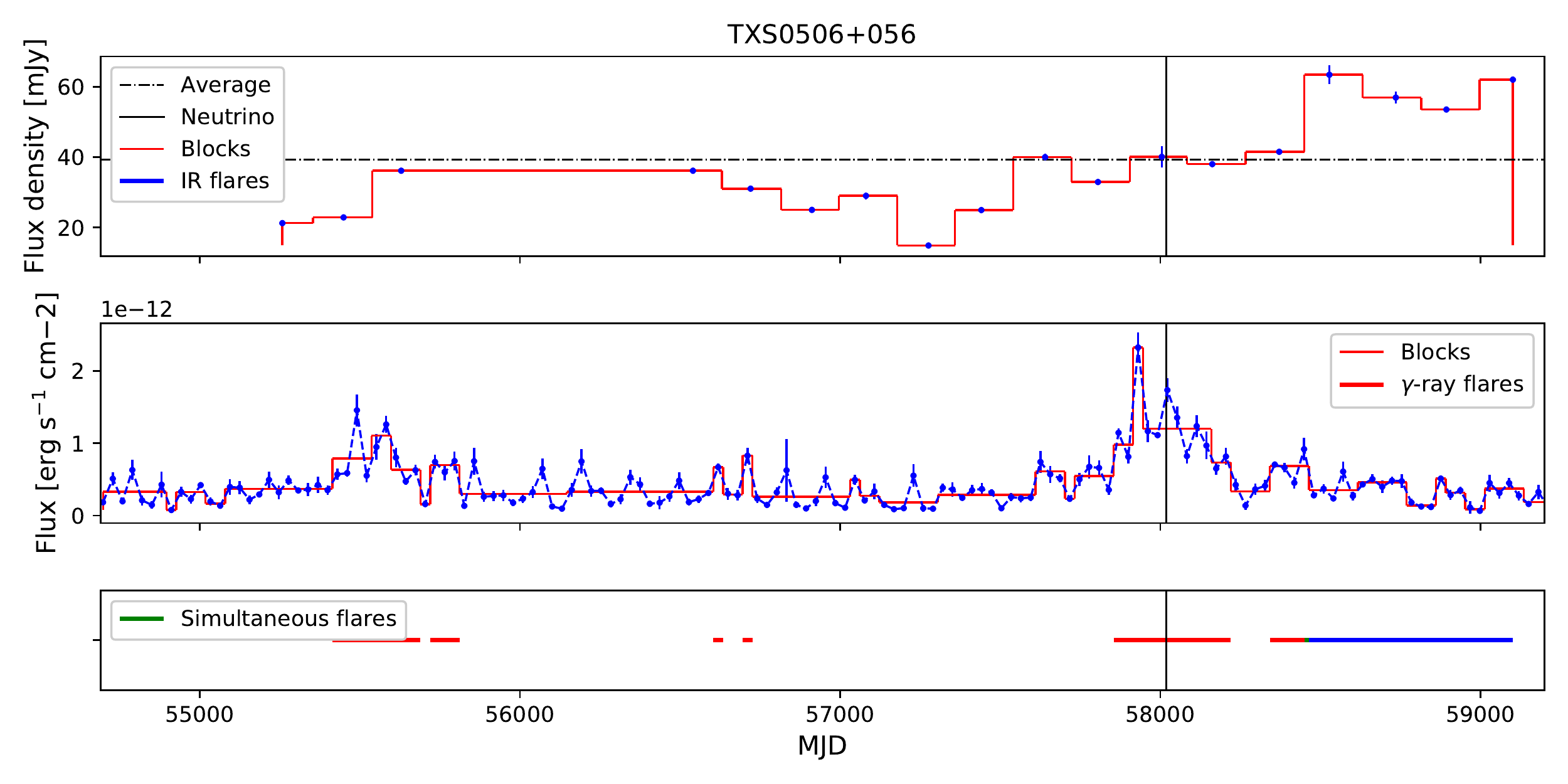}
\end{center}
\caption{Light curves of TXS\,0506+056 showing non-simultaneous infrared and \gr\ flares. The upper panel is the infrared light curve from WISE multi-epoch data, and the middle panel is the {\it Fermi} \gr\ light curve analyzed with Fermi Science Tool. The lower panel represents the flaring periods in infrared (blue thick lines) and \gr\ (red thick lines) as well as the simultaneous flaring stages (green lines). Black lines represent the arrival time of IceCube-170922A.}
\label{txs0506}
\end{figure}

Considering the multi-frequency activity of TXS\,0506+056 around the arrival time of IceCube-170922A, the strongest flaring period in the low-energy band (radio and infrared) and in the high-energy band (\gr s) are not simultaneous. 
There is a significant lag of $\sim 300$ days between the \gr\ and radio/infrared flares. 
Figure~\ref{txs0506} shows the non-simultaneous infrared and \gr\ light curves of TXS\,0506+056 with the time lag between flares are estimated with the Bayesian Block Algorithm \citep{Scargle2013}. 
The infrared data correspond to the WISE (Wide-field Infrared Survey Explorer, \citet{Wright2010}) mission, and the \gr\ data are obtained from {\it Fermi}-LAT observation (following the analysis described in section~\ref{fermiana}). 
In addition, PKS\,1502+106 and J0242+1101 also show long-lasting 15GHz radio flares in coincidence with IceCube-190730A \citep{Franckowiak2020} and Antares flares \citep{Illuminati2021}, respectively. 
Both blazars show a significant lag of radio to \gr\, flaring stages. 

In general, lags in different energy bands might arise from changes in the source's environment, causing different emission zones to shine in different energies. 
Taking the most plausible neutrino blazar, TXS\,0506+056, for example, many studies have shown difficulties of reconciling both neutrino (IceCube170922A and 2014-2015 flares) and multi-frequency activities through a single emission model \citep{Murase2018,Reimer2019,Rodrigues2019,Petropoulou2020}. 
A single-zone lepto-hadronic model might be able to describe the contribution from the high-energy alert \citep{Ansoldi2018,Keivani2018,Cerruti2019,Gao2019,Righi2019b}, but the modeling requires a subdominant hadronic component and assumes the presence of sufficient photons with right energy from the external field. 
Moreover, no single-zone scenario can explain the high neutrino flux from 2014-2015 flares and -at the same time- satisfy the constraints from the simultaneous spectral energy distributions (SED). 
The neutrino flares, surprisingly, were not accompanied by any photon flare (given all the observed information, \citet{IceCube2018b}), indicating that not all the neutrino emission is necessarily correlated to electromagnetic activity, especially in \gr\ \citep{Halzen2019,Kun2021}. 

Inspired by the time lag and the remarkable characters of TXS\,0506+056, we try to recognize and compare the high stage in both low-frequency (infrared/radio) and high-frequency (\gr) light curves for selected blazar samples, and identify sources with multi-frequency activity similar to that of TXS\,0506+056. 
By investigating and selecting blazars like TXS\,0506+056 according to their multi-frequency flaring stages, we might be able to identify promising neutrino counterparts effectively. 

A number of analyses are performed throughout the paper, and there are combinations of source samples that are built with multiple selection criteria. In Table~\ref{anatable}, we summarize all the analyses and source lists as well as their corresponding selection and motivation.

\section{Multi-messenger Samples and Data} \label{data}

\subsection{IceCube data samples} \label{neutrino}
IceCube has been announcing high-energy neutrino alerts since the spring of 2016, bringing about a total of 67 real-time alerts up to the end of May 2021, with a relatively high probabilities of those events being of astrophysical origin. 
Those alerts passed the selection criteria of the real-time alert system \citep{Aartsen2017b} and generally have energy $\geq 100$~TeV.
Another 35 archival events from 2010 to 2016 fulfill the same criteria before the operation of the real-time alert system, summing up a total of 102 events considered in our analyses. 
The lists of real-time and archival alerts/events are taken from the Gamma-ray Coordinates Network (GCN) / Astrophysical Multimessenger Observatory Network (AMON) Notices and IceCube website\footnote{\url{https://gcn.gsfc.nasa.gov/amon_hese_events.html}, \url{https://gcn.gsfc.nasa.gov/amon_ehe_events.html}, \url{https://gcn.gsfc.nasa.gov/amon_icecube_gold_bronze_events.html}, \url{https://icecube.wisc.edu/science/data/TXS0506_alerts}}.

Apart from high-energy neutrino alerts and archival events which would have qualified as real-time alerts, IceCube published a sample of track-like neutrino events collecting 10 years data  \citep[from 2008 to 2018,][]{IceCube2021a} that was assembled for neutrino point-source searches and used in the IceCube's 10-year time-integrated point-source analysis \citep{IceCube2020}. 
This sample covers a broader energy range than the high-energy alert list, containing events with E $<$ 100 TeV. 
To study only events with a higher probability of being of astrophysical origin and that are well reconstructed, we required events with reconstructed energy $\gtrsim 60$~TeV \citep[which is the same cut applied in][]{Padovani2016} and with angular uncertainty $\leq 5$ degrees. 
The cuts result in 250,821 well-reconstructed high-energy track events studied in this paper. 

\subsection{Blazar samples} \label{blasamp}
Three blazar samples are used in our studies. 

\textbf{The 3HSP:} The 3HSP catalog is a multi-frequency selected sample of 2013 HSP and HSP candidates \citep{Chang2019}. 
The 3HSP catalog is currently the most extensive and complete HSP catalog and an ideal sample to study the statistical properties (such as completeness, evolution, etc.) of blazars. 
In this study, we further consider 78 extra HSPs which should be included in the 3HSP catalog in the future. The updated version of the 3HSP catalog, which contains 2081 sources, is currently available through the Virtual Observatory\footnote{http://bsdc.icranet.org/threehsp/q/cone/form}. 
A second and more complete version of the 3HSP catalog will be published soon.

\textbf{The 5BZCAT:} We also selected blazars from the 5BZCAT catalog \citep{Massaro2015}, which consists of 3561 robust blazars, all confirmed via optical spectroscopy. 
Even though the 5BZCAT is a compilation of blazars found by many different methods and thus not a complete sample, it is the largest catalog of confirmed blazars with optical spectral observations. 

\textbf{The WIBRaLS:} The last samples we use are the WISE Blazar-like radio-loud sources, named WIBRaLS and WIBRaLS2 catalogs\footnote{In this paper, we use WIBRaLS to represent both WIBRaLS and WIBRaLS2 catalogs} \citep{DABrusco2014, DABrusco2019}. Those catalogs contain a total of $\sim 12415$ blazar candidates and are the largest samples of their kind to date.
The WIBRaLS catalogs are samples of infrared selected radio-loud blazar candidates with WISE\citep{AllWISEcat} mid-infrared colors similar to that of confirmed \gr\ blazars. 

In addition, there are 18 ${\it Fermi}$ 4LAC-associated blazars \citep{Fermi4LAC} within the containment regions of the IceCube alerts that are not cataloged in 3HSP, 5BZCAT, or WIBRaLS. 
Among those 18 sources, 11 are related to alerts which have relatively large angular uncertainty and were not considered in our analyses. 
In total, we have collected a meta-blazar sample with 15424 blazars and blazar candidates from three extensive blazar catalogs and 4LAC.

\subsection{Multi-frequency data} \label{mwcatalog}
Here we describe the multi-frequency data used in this paper, from millimeter radio up to \gr.

\textbf{ Millimeter Radio:} The millimeter multi-epochs data were obtained from ALMA Calibration Catalog \citep[ACC][]{Bonato2019}, which is an astronomical-measurement database of calibration sources that are mostly bright blazars observed in seven different bands (ranging from 84GHz to 950GHz). 
We used the band 3 data (84-116 GHz) to describe most of our sources. 
For 16 sources with {\it Fermi} counterparts in our meta blazar sample, we consider millimeter data other than band 3, with preference to band 4 (125-163GHz), band 6 (211-275GHz), and band 7 (275-373GHz). 
The ACC light curves have different time intervals varying from days to years between May 2011 and July 2018. 
For those observations with time separation smaller than 15 days, we combined the bins and took their average value. 

\textbf{Infrared:} This paper used $4.6 \mu m$ WISE W2 infrared light curves from AllWISE Multiepoch photometry dataset \citep[AllWISEMEP,][]{AllWISEMEP} and NEOWISE \citep[Near-Earth Object WISE,][]{Mainzer2014,NEOWISE} data release, with observing time ranging from January 2010 to December 2020. 
The WISE light curves have a large time interval (of several hundred days), and the observations are usually centered in 1-2 days along several months. 
Thus, we combined the infrared data with time separation smaller than 15 days, averaging the signal and removing the outliers with flux values that lie outside $3-4\sigma$ of the mean. 
Those with signal-to-noise ratio $\leq$ 2 are also removed. 
After the combination, the mean interval is roughly 180 days. 
Moreover, there is a break between 55600 to 56500 MJD which does affect the identification of relevant flaring periods, resulting from the gap between the AllWISEMEP and NEOWISE surveys. 
We manually added artificial points 180 days after the beginning and before the end of the break to remedy the data gap.
The flux and error of the artificial points are based in the average infrared flux for each source.

\textbf{X-ray:} We consider the 3 keV multi-epoch observation data from {\it Swift} XRT, covering December 2005 to October 2020. The data is based on blazars frequently observed by {\it Swift} and was made available in \citet{Giommi2019} and \citet{Giommi2021b}. 
Similar to the data pre-processing in millimeter and infrared, the X-ray observations with time separation smaller than 15 days were combined and averaged. 

\textbf{\gr :} The \gr\ data is retrieved from the aperture photometric light curves\footnote{\url{ https://fermi.gsfc.nasa.gov/ssc/data/access/lat/10yr_catalog/ap_lcs.php}} of {\it Fermi}-LAT 4FGL-DR2 catalog \citep{Fermi4LAT}. 
Those aperture light curves are binned evenly with 30 days intervals since June of 2008. 
By using 10-year-average photon indexes in the 4FGL-DR2 catalog, we converted the photon fluxes of the aperture light curves from 0.1-200 GeV to 0.8-200 GeV energy fluxes in units of ${\rm erg}~{\rm cm^{-2}}~{\rm s^{-1}}$, focusing on the high energy band to avoid the contamination from nearby sources due to the large point spread function at lower energy.

\section{Source lists} \label{sample}

\subsection{Multi-frequency sources in IceCube alert regions} \label{mwsource}
The first samples we will study in this paper are multi-frequency sources in the vicinity of IceCube alert regions. 
Starting from the four multi-frequency catalogs described in section~\ref{mwcatalog}, we selected the sources located in $\delta < |40^{\circ}|$ and $b > |10^{\circ}|$, with similar location distribution concerning most IceCube alerts and to avoid the complicated Galactic plane region. 

Given that we want to study the correlations between the sources' flaring activity and IceCube alerts, selecting sources with a higher probability of being variable is more effective. 
That is, for infrared sources, we took only those with a counterpart in the infrared-selected WIBRaLS catalog and those with $1.4$~GHz flux $\geq 100$~mJy. 
By applying these cuts, we obtained only bright and blazar-like infrared sources, as blazars are the most variable sources in the infrared sky. 
For \gr\ sources, only those with variability index $> 18.48$ are selected. 
The {\it Fermi}-LAT team applies the variability index obtained with the likelihood ratio test to evaluate the variability of a \gr\ source, considering the fluxes in several time intervals. 
In the 4FGL-DR2 catalog, a value greater than 18.48 indicates $<1\%$ chance of being a steady source. 
We note that the X-ray and millimeter catalogs we used are mainly for blazars, and the selected soures are already highly variable. 

Further criteria are applied to the uneven-binned X-ray and millimeter samples to remove the ones with insufficient data. 
Specifically, only millimeter and X-ray sources with more than five detections at different epochs are considered. 
All the above criteria combined lead to a selection of 445 ACC, 3179 WIBRaLS, 876 XRT, and 992 4FGL-DR2 sources , with 21, 214, 44, and 93 of them in the containment region of IceCube alerts, respectively. 
Here we applied a factor of 1.1, 1.3, and 1.5 to the $90\%$ containment region of the alerts, taking into account that $10\%$ of the candidates expected to be outside of the $90\%$ containment area and the possible systematic errors \citep{Giommi2020}. 
The optimal factor was obtained for each multi-frequency catalog via a simulation, which is described in section~\ref{timewindow}. 

\subsection{Blazars with non-simultaneous multi-frequency flares}  \label{txsblazar}

The second part of our source lists is made of blazars with multi-frequency activity similar to that of TXS\,0506+056 (see section~\ref{txsactivity} and Figure~\ref{txs0506}). This section describes our selection steps to obtain a sample of such blazars. Our steps are summarized as follows:
\begin{enumerate}
    \item Cut the meta-blazar sample with WISE and {\it Fermi} 4FGL-DR2 data, defining the Fermi-Infrared Blazar Sample (FIBS) with 2700 sources. 
    \item Identify the optimal segmentation and flares with Bayesian Blocks Algorithm. 
    \item Remove those with weak or ambiguous variability.
    \item Select non-simultaneous multi-frequency flaring sources like TXS\,0506+056. 
    \item Refine the selection by computing dedicated light curves with a {\it Fermi} likelihood analysis.
\end{enumerate}

\subsubsection{Fermi-Infrared Blazar Sample} \label{fibs}

We begin by cutting our meta-blazar sample considering two different wavebands of light-curve data. 
Given the completeness and accessibility of multi-wavelength light curves, here we focus on objects with both {\it Fermi} 4FGL-DR2 and WISE multi-epoch data, to study the flaring behavior in $\gamma$-rays and infrared. 
Among the meta-blazar sample obtained in section~\ref{blasamp}, only 2700 sources have both \gr\ and infrared light curves available. 
We call these 2700 blazars the ``Fermi-Infrared Blazar Sample (FIBS)" with 1011, 1622, and 1547 sources in the 3HSP, 5BZCAT, and WIBRaLs catalogs, respectively.
There are 1335 FIBS sources cataloged in at least two blazar catalogs, and 152 of them are in all three catalogs. 

The IceCube Collaboration has been searching for neutrino excess from several lists of objects \citep{IceCube2011,IceCube2013b,IceCube2014,IceCube2017,IceCube2019,IceCube2020}. 
Among them, there are eight blazars in FIBS for which the correlation with astrophysical neutrinos that has reached the significance of the order of p-value $\leq 0.05$ for at least one of the IceCube all-sky point-source searches. 
Moreover, 90 out of 2700 FIBS sources are located within the 90\% containment region of 102 IceCube high-energy alerts (section~\ref{neutrino}). 
TXS\,0506+056 blazar is known for having a relatively high significance in IceCube point-source analysis and by its direct association with a track event IceCube-170922A and the neutrino excess in $2014-2015$ \citep{IceCube2018a,IceCube2018b,Padovani2018}. 
The 97 sources either in the alert containment regions or associated with a weak neutrino excess signal (at $95\%$ significance level) are called {\it potential neutrino sources} throughout the paper. 

In search for potential neutrino blazars, the correlation between radio and \gr\ flares is also of interest. 
Especially, the promising neutrino blazar TXS\,0506+056 shows a significant lag between the radio and \gr\ flaring phases. 
However, there is currently no other public and long-term monitored radio data available. 
The data from the millimeter ACC catalog is the best option for us, even though the data do not cover the full time range of {\it Fermi} light curves and sometimes are triggered by high state in other wavebands. 
In our meta-blazar sample, there are 504 sources with both {\it Fermi} and ACC multi-epoch data available. 

\subsubsection{Multi-frequency flaring stages} \label{mwstage}

Our next step is to identify flares from the light curves of FIBS and {\it Fermi}-ACC sub-sample. 
A flaring period can be objectively identified through the Bayesian Blocks Algorithm, aiming to find the optimal segmentation of the data in the observation interval \citep{Scargle2013}. 
In this work, we use Bayesian Blocks astropy implementation\footnote{\url{ https://docs.astropy.org/en/stable/api/astropy.stats.bayesian_blocks.html}} 


to detect statistically significant variations in multi-frequency light curves with more than two observations at different time epochs. 
We chose the prior which makes the algorithm sensitive to variations that are significant at~$99\%$ confidence level (a false alarm probability 0.01), identifying only strong and clear flaring episodes and avoiding tentative flares. 
If the algorithm recognizes no flare for a given light curve, we lower the confidence level to $95\%$ to search for potential weaker flaring periods of the source. 
Our purpose is to select as many sources like TXS\,0506+056 as we can. 

In \citet{Abdo2010b}, the ``bright state'' for observations in a light curve are defined as $F_{i}-\sigma_{i} > { \langle F_{i} \rangle } + 1.5\times S$ 
, where $F_{i}$ and $\sigma_{i}$ represent the flux (density) and error of each detection of the light curve, while ${ \langle F_{i} \rangle }$ and S are the mean and standard deviation of the light curve. 
Following and slightly adapting the above criterion, we keep those blocks with $ F_{\rm blocks} > F_{\rm quiescent} + 1.3\times {\rm S}$ as in flaring state. 
$F_{\rm blocks}$ is the average flux (density) in a block, and $F_{\rm quiescent}$  is the quiescent flux (density). 
The quiescent flux of a source is the mean flux value of the faintest $\sim 30\%$ points detected in the quiescent phase. 
We excluded the detections with the lowest flux values when estimating the quiescent flux as they might be outliers with low statistics. 
The quiescent flux is used in this paper since that the average flux ${ \langle F_{i} \rangle}$ sometimes could be twice as bright as the quiescent state, especially for a highly variable source flaring in most of the observing period. 
Besides, considering that we are comparing the average flux for multiple detections in a block, the flaring block are selected with $1.3\times S$ instead of the factor of 1.5 in \citet{Abdo2010b}. 
We note that sometimes for the infrared light curves, the $F_{\rm quiescent}$ is much lower than the mean value due to the variability and quality of the WISE data, and we further require $ F_{\rm blocks} > F_{\rm average} + {\rm S}$ to be considered as in flaring state. 

\begin{figure} [h!]
\begin{center}
\includegraphics[width=1.\linewidth]{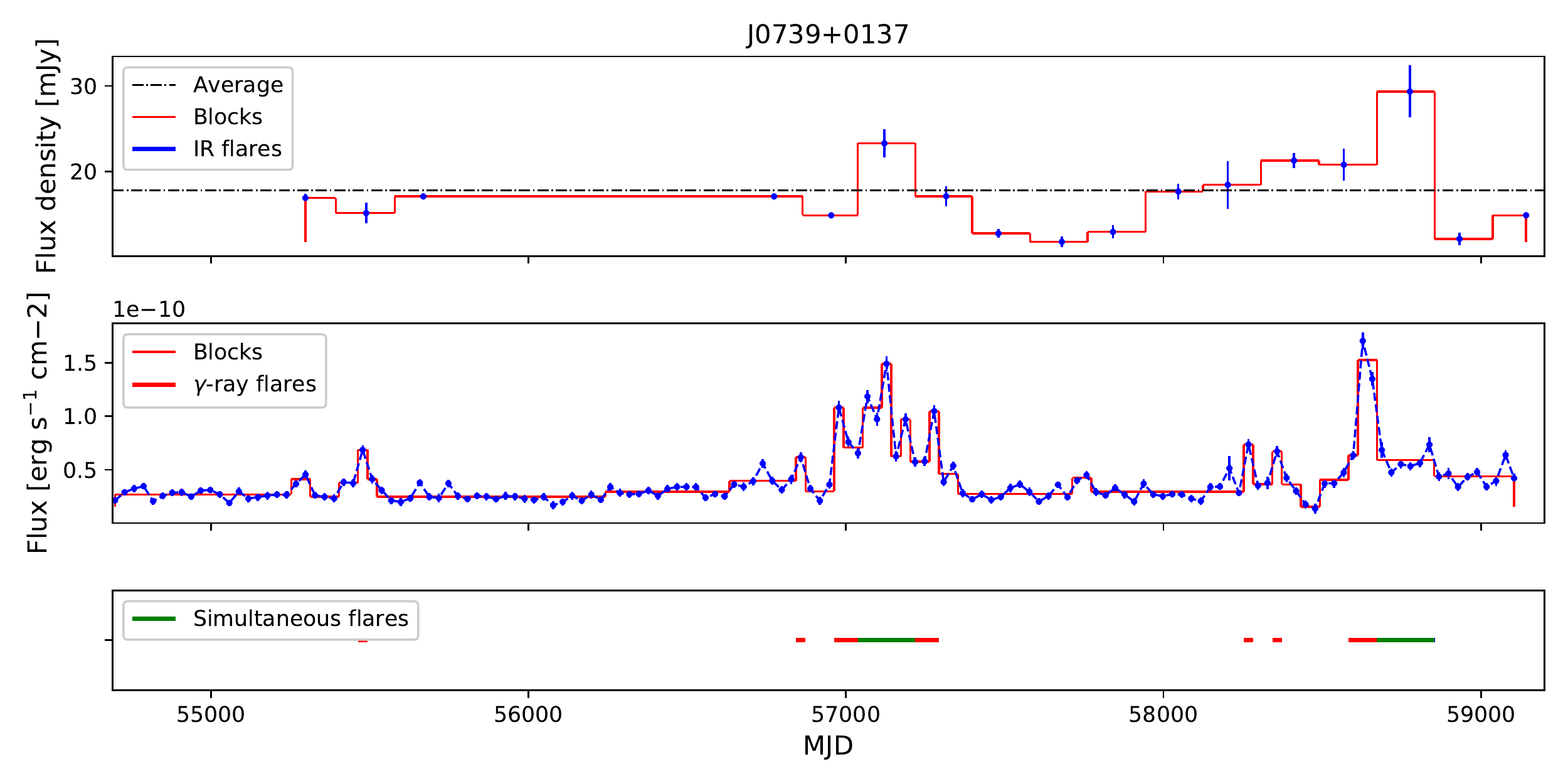}
\end{center}
\caption{Example of CFS light curves with J0739+0137. The upper panel is the infrared light curve from WISE multi-epoch data, and the middle panel is the \gr\ aperture light curve from {\it Fermi}-LAT. The lower panel represents the flaring periods in infrared (blue thick lines) and \gr\ (red thick lines) as well as the simultaneous flaring stages (green lines).}
\label{cfslc}
\end{figure}

Once the bright phases in both the high-energy and low-energy light curves of each blazar are identified, the ``multi-frequency epoch" of a blazar could be divided into three flaring stages: high energy flaring, low energy flaring, and simultaneous flaring stages. 
Sources with no flare identified in both light curves are not considered. 
The high energy flaring stage is defined as the time-interval with brighter flux in the high energy light curve but relatively low flux (quiet state) in low energy; and the low energy flaring stage is defined vice versa. 
The simultaneous flaring stage refers to time intervals with both high-energy and low-energy flux in the high state. 
We define sources with no simultaneous flaring stage as ``Non-simultaneous Flaring Sources" (NFS).
On the contrary, sources with simultaneous flaring stages longer than half of the whole infrared and \gr\ flaring time are considered as ``Correlated Flaring Sources" (CFS). 
An example of the infrared and \gr\ light curves for sources in CFS control group are illustrated in Figure~\ref{cfslc}. 


\subsubsection{Non-variable sources and uncertain sources} \label{uncertain}

We removed cases with no significant variability or cases where the light curves resemble small fluctuations close to background levels. 
Those {\it potential neteutrino sources} are not dropped out. 

{\bf Non-variable sources:} For sources with {\it Fermi} variability index $\leq 18$, which are treated as non-variable, we exclude the ones with Normalized Excess Variance\footnote{$\sigma_{\rm NXS}=\frac{{\rm S}^2-<{\sigma_{i}}^2>}{<F_{i}>^2}$, where $F_{i}$ and $\sigma_{i}$ represent the flux (density) and its associated error in time bin $i$, and S is the flux standard deviation concerning the entire light curve.} $\sigma_{\rm NXS} < 0.0001$ or with $F_{\rm blocks}^{\rm max}$/$F_{\rm blocks}^{\rm min} <$ 1.7. 
The ratio 1.7 is roughly the median of {\it Fermi non-variable sources}. 

{\bf Uncertain sources:} If the ratio of standard deviation between the non-flaring phases and the whole observing time of a source is less than 0.9, it is regarded as ambiguous source with extremely fluctuating light curve. 
The standard deviation ratio value of 0.9 is approximately the third quartile of whole FIBS sources. 

We removed a total of 886 FIBS sources in this step. 
In the end, only 1353 FIBS sources remained to be selected in the next step. 

\subsubsection{Selection of non-simultaneous flaring sources} \label{select}
As we have removed non-variable and uncertain sources, we begin to select those blazars with low-energy flares that are not simultaneous with high-energy flares, like TXS\,0506+056 (Figure~\ref{txs0506}). 
Once there is no simultaneous flaring stage between low and high frequency flares, we directly selected those with long radio/infrared flares that happen after rapid high-energy {\it Fermi} flares and low-energy flaring lag smaller than 500 days. 
If the simultaneous flaring stages in a source only consists of a small fraction (usually not occupied by more than one-fourth of the successive flaring period in two light curves), the source is still considered as good candidates and are picked when relatively long low-frequency flares follow short high-frequency flares. 
On the contrary, we remove those cases where the majority of the multi-frequency flaring periods are simultaneous and complex (i.e., when there are multiple \gr\ flares contained inside a single infrared flare) because these multi-frequency behaviors are not the ones we see for the TXS\,0506+056. 

The above criteria led to a selection of 171 and 21 TXS-like sources from the FIBS sources and the ACC-{\it Fermi} sub-sample, with a significant lag of infrared and millimeter flares with respect to the \gr\ flares. 
We note that these two lists are not the final samples as they are selected based on the {\it Fermi} aperture light curves. 
Further refining of these two preliminary source lists will be discussed in the next sections. 



\subsubsection{Likelihood \gr\ analysis} \label{fermiana}
In previous sections, we have obtained two preliminary lists of sources with multi-frequency flaring activity like TXS\,0506+056 using {\it Fermi} aperture photometric light curves. 
However, aperture photometry is not suitable for detailed scientific analysis. 
The aperture light curves were only used to screen a large amount of FIBS sources in the first place and reduce the extensive computation tasks for thousands of \gr\ light curves via likelihood analysis. 
Additional confirmation and filtering of those pre-selected sources are required to make sure our selection was robust. 
We hence performed a binned likelihood light-curve analysis with the Fermi Science Tool\footnote{\url{https://fermi.gsfc.nasa.gov/ssc/data/analysis/software/}} for all pre-selected sources obtained in last section. 

Taking the same time interval as the aperture \gr\ light curves, we applied a 30-day bin in the {\it Fermi} likelihood analysis from MJD 54682 to 59257. 
The energy range covers from 900 MeV to 500 GeV, focusing on a slightly higher energy range when compared to the aperture light curves. 
A higher energy band was chosen mainly to reduce the computation load and avoid contamination at lower energy. 
We scanned for three different pivot energies (1.0, 2.5, and 5.0 GeV) for every pre-selected source and time bin and kept the pivot energy that leads to the lowest uncertainty associated with the normalization (N$_0$) and photon index ($\Gamma$). 
For time bins where the test statistics (TS) are very similar for all pivot energies, we kept the ones with a higher signal-to-noise ratio. 
For some particular cases where the likelihood analysis did not converge, an extra run with pivot energy of 2 GeV was tried. 
Detections with ${\rm TS} \leq 5$, ${\rm Flux_{err}} \geq {\rm Flux}$, or ${{\rm Photon\ Index} (\Gamma)  > 5.5}$ are considered as upper limits, and the limits of energy fluxes are set to $3 \times 10^{-14}~{\rm erg}~{\rm s}^{-1}~{\rm cm}^{-2}$. 
This value was applied to represent an average upper-limit level for our sources since most of the the detections deemed as limits have energy fluxes smaller than $3 \times 10^{-14}~{\rm erg}~{\rm s}^{-1}~{\rm cm}^{-2}$. 

We reevaluated the flaring phases of our pre-selected TXS-like sources from FIBS and ACC-{\it Fermi} sub-sample with these new light curves one by one. 
This procedure results in 32 (11) sources that remained in the infrared- (millimeter-) \gr\ sub-sample. 
To avoid missing possible neutrino emitters, we adopted a slightly different criteria to help identify sources with relatively weak gamma-ray flares. 
We rerun the Bayesian Blocks Algorithm assuming the gamma-ray flux upper limits to be of $10^{-14}~{\rm erg}~{\rm s}^{-1}~{\rm cm}^{-2}$. 
In this way, the source 3HSPJ064850.5-694522 was identified at the border of our main selection criteria, and therefore added to our TXS-like sample. 

A large number of sources removed might be a consequence of the large flux fluctuations associated with {\it Fermi} aperture light curves (especially for faint sources, where low photometric fluxes are usually associated with non-detection in the likelihood analysis, ending up flagged as flux upper limits). 
Contamination of the photons from nearby sources in the aperture light curves at low energy and the gap between MJD 55600 to 56500 in WISE multi-epoch data, additionally, would cause bogus selections. 
We note that some of the \gr\ flares from the dedicated light-curve analysis sometimes are associated with sources that produced only 1-2 detection(s) in the entire light curve. 
However, given that we have scanned the pivot energy with several values, carefully removed the problematic detections, and objectively and systematically identified the flaring phases with Bayesian Blocks Algorithm, the detected variation/flares could be considered as robust. 

\subsubsection{Lists of selected blazars and control groups} \label{groups}

Finally, we have a total of 32 and 11 sources with TXS-like multi-frequency activity selected from FIBS and from ACC-{\it Fermi} sub-sample, respectively. 
To check whether those selected sources are more probable neutrino emitters concerning the remained ones, we took NFS and CFS samples (section~\ref{mwstage}) selected from FIBS as control groups. 
We note that some sources without simultaneous flares in two different energy bands have already been selected as TXS-like sources, and they are not included in the control groups. 
Sources with extremely ambiguous flares are removed from the control groups as well. 
Our control groups, in the end, contain 409 NFS and 62 CFS among FIBS sources. 
The four groups of sources are totally independent. 

Given that the ACC multi-epoch data is much more sparse than the WISE and {\it Fermi} light curves, it is meaningless to search for highly overlapped millimeter and \gr\ flares, as well as orphan flares. 
Therefore, there are no control groups for the millimeter-selected sample. 
We note that the selection concerning the ACC data suffers from a heavy incompleteness, and the 11 sources selected here are just meant to compare the results from millimeter with those from infrared. 
The final number of sources in each selected list and control group are shown in Table~\ref{numtable}, and the full list and all the light curves (as well as the Bayesian Blocks results) of selected TXS-like sources from the FIBS are illustrated in the Appendix~\ref{tabletxscfs} and \ref{mwlcselect}.

\section{Time correlations between multi-frequency flares and the neutrino alerts} \label{flares}
Our first analysis is to explore the correlations between the flaring period of multi-frequency sources and the arrival time of the neutrino alerts.


\subsection{Time-dependent analysis} \label{analysisa}

To study if the bright state epochs are consistent with the arrival time of the IceCube alerts, we follow the methods in \citet{Plavin2020} and perform a time-dependent analysis for multi-frequency sources (ACC, WIBRaLS, XRT, and 4FGL-DR2, selected in section~\ref{mwsource}) inside the IceCube alert regions. 
Given that most ICeCube alerts are close to the Celestial equator and to focus on extragalactic objects, here we consider only 80 IceCube alerts with $b > |10^{\circ}|$ and $\delta < |40^{\circ}|$. 

In the time-dependent analysis, an observable ${\rm R(t=0)}$ was defined as the ratio of the average flux (density) within the neutrino time window ($\Delta T$) and outside this time window. 
\begin{equation}
{\rm R(t=0)}=\frac{{\rm average~flux~(density)~within}~\Delta T}{{\rm average~flux~(density)~outside}~\Delta T}
\end{equation}
where $t=0$ means that the observing times of the multi-frequency data were not shifted manually. 
$\Delta T$ is the time window around the detected time of the neutrino alert that contains the source which is being considered. 
Higher ${\rm R(t)}$ values imply that flares tend to occur within the $\Delta T$ of arrival of IceCube alerts. 
We normalized the multi-frequency fluxes by the quiescent flux (average flux of faintest 30\% detection for a source, see section~\ref{mwstage}) of each source to avoid bias from different sources. 
The normalized fluxes are defined as ``flare levels", which are the ratio between the flux of each detection and the quiescent flux. 

\subsubsection{Defining the time window $\Delta T$ values} \label{timewindow}
The $\Delta T$ is determined by Monte Carlo simulation with the smallest p-value from several arbitrary trial values for 4FGL and WIBRaLS sources. 
For each $\Delta T$ trial value, we scrambled the R.A. of the IceCube alerts 10,000 times using ${\rm R(t=0)}$ as the reference test statistic, and the p-value is the probability that shifted alert positions yield a higher test statistic. 
As described in section~\ref{mwsource}, we enlarged the 90\% containment region of the alerts with several trial factors, 1.1, 1.3, and 1.5, to account for the systematic error of the alerts in the simulation. 
In case the enlarge factor improves the significance of correlations, it will be further used for the time-dependent analysis. 

\begin{figure} [h!]
\begin{center}
\includegraphics[width=0.9\linewidth]{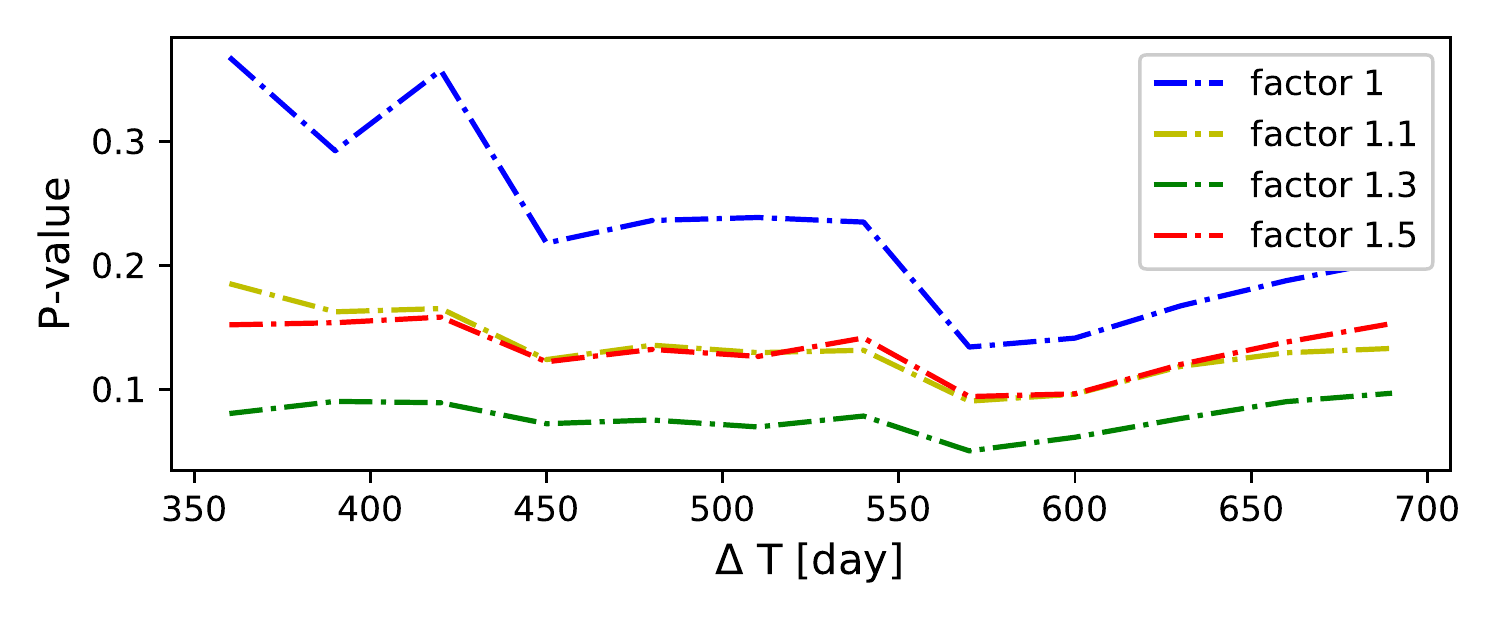}
\includegraphics[width=0.9\linewidth]{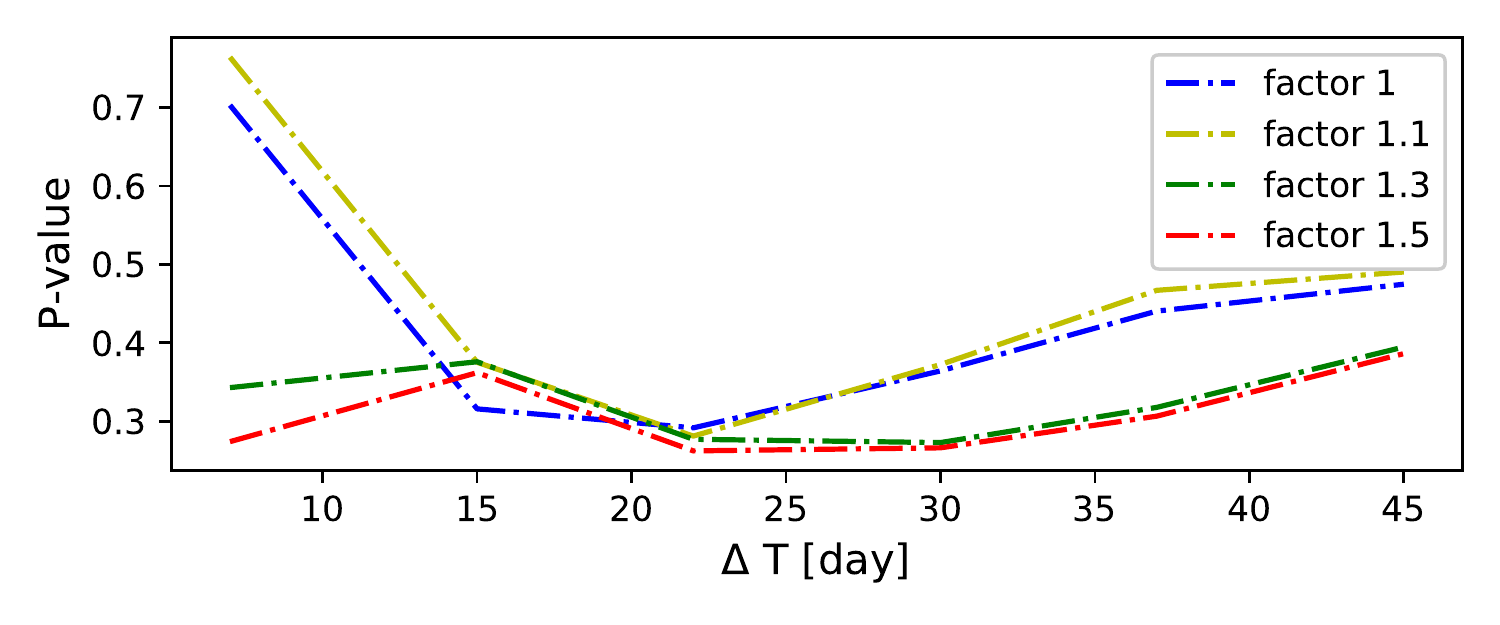}
\end{center}
\caption{P-values with respect to different $\Delta T$ (time window) trials. The upper panel is for WIBRaLS sources, and the lower panel is for {\it Fermi} 4FGL-DR2 sources.}
\label{simtwindow}
\end{figure}

According to simulation results in Figure~\ref{simtwindow}, we selected $\Delta T=570$ days for WIBRaLS sources and $\Delta T=22$ days for {\it Fermi} 4FGL-DR2 sources. 
The most significant signal was obtained when extending the containment region with a factor of 1.3 for the WIBRaLS sample and a factor of 1.5 for the 4FGL sample. 
We note that the minimum and second minimum values in the curve of factor 1.5 for the 4FGL sources are very close to each other. 
This minimum is also close to the one in the curve with factor 1.1, which seems to converge more reasonably. 
Therefore, we also perform the analysis considering an extension of factor 1.1. 
For ACC and XRT sources, we used the medium time interval of roughly 30 days and did not enlarge the 90\% containment regions. 
The simulations do not converge well for them since their light curves are not equally binned, and those factors for the containment region did not lead to more significant results. 
Finally, we consider 21, 165, 44, and 93 ACC, WIBRaLS, XRT, and 4FGL-DR2 sources within the contaminant regions of IceCube alerts, though with different extended factors of the regions.

\subsubsection{Sliding the observing time}
To test the correlation between electromagnetic flaring times and neutrino alert times, the observing times of multi-frequency sources are shifted with a time parameter $t$, while the neutrino arrival time and time window remain the same. 
The correlation could be demonstrated once the highest flux ratio ${\rm R(t)}$ is centered at the original observing time $t=0$. 
The observing times were shifted between $-3600 < t < 3600$ days, roughly spanning 10 years before and after the real observation, with shifted bin roughly equal to half of the time window $\Delta T$. 
We note that the shifted bin of infrared data is 360 days, not half of the $\Delta T$, given that the simulated time window for WIBRaLS sources is too large. 
IceCube has observed neutrino data and published alerts for around 10 years, including ``alert-like" archival neutrino events before the real-time alert system was established. 
By choosing to shift the light curves from 10 years before the actual observing date to 10 years after that, the latest multi-frequency observation would be shifted through the time window of the first IceCube alert. 
 
The maximum ${\rm R(t)}$ value might be at the center ($t=0$, without shifting the observing time) by chance.  
We estimated the chance probability of flares coinciding with the alert arrival time with control samples via Monte Carlo simulation. 
The control samples were built by randomly selecting the same number of sources as the experimental sample from those outside the alert region and assigning the ``alert time'' for each randomly selected source. 
We repeat all the steps above with the control samples and obtained randomized ${\rm R(t)}$, simulating the ratio of average flux (density) inside and outside the time window $\Delta T$ with random sources and alert times for 10,000 times. 
The randomized ${\rm R(t)}$ will be used to estimate the significance of the correlation between electromagnetic flares and IceCube alerts.

\subsection{Correlations with IceCube alerts} \label{alertflare}
\begin{figure} [h!]
\begin{center}
\includegraphics[width=1.\linewidth]{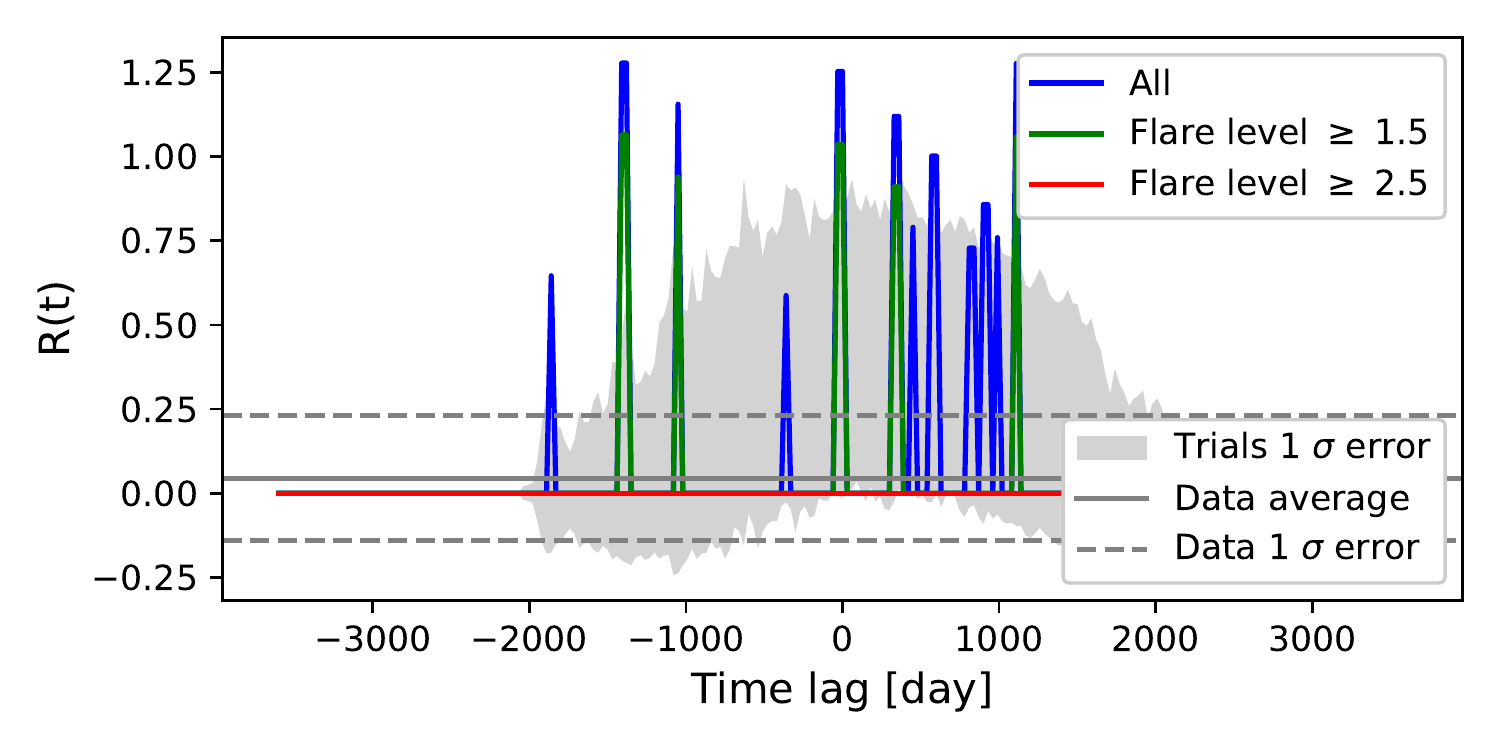}
\end{center}
\caption{The ratio of ACC flux density averaged over a 30-day window to the average flux density outside it. The data average and $1\sigma$ statistical error are calculated with the observed experimental data, while the trials $1\sigma$ statistical error are obtained from random trials of the control samples. }
\label{timedelayacc}
\end{figure}

\begin{figure} [h!]
\begin{center}
\includegraphics[width=1.\linewidth]{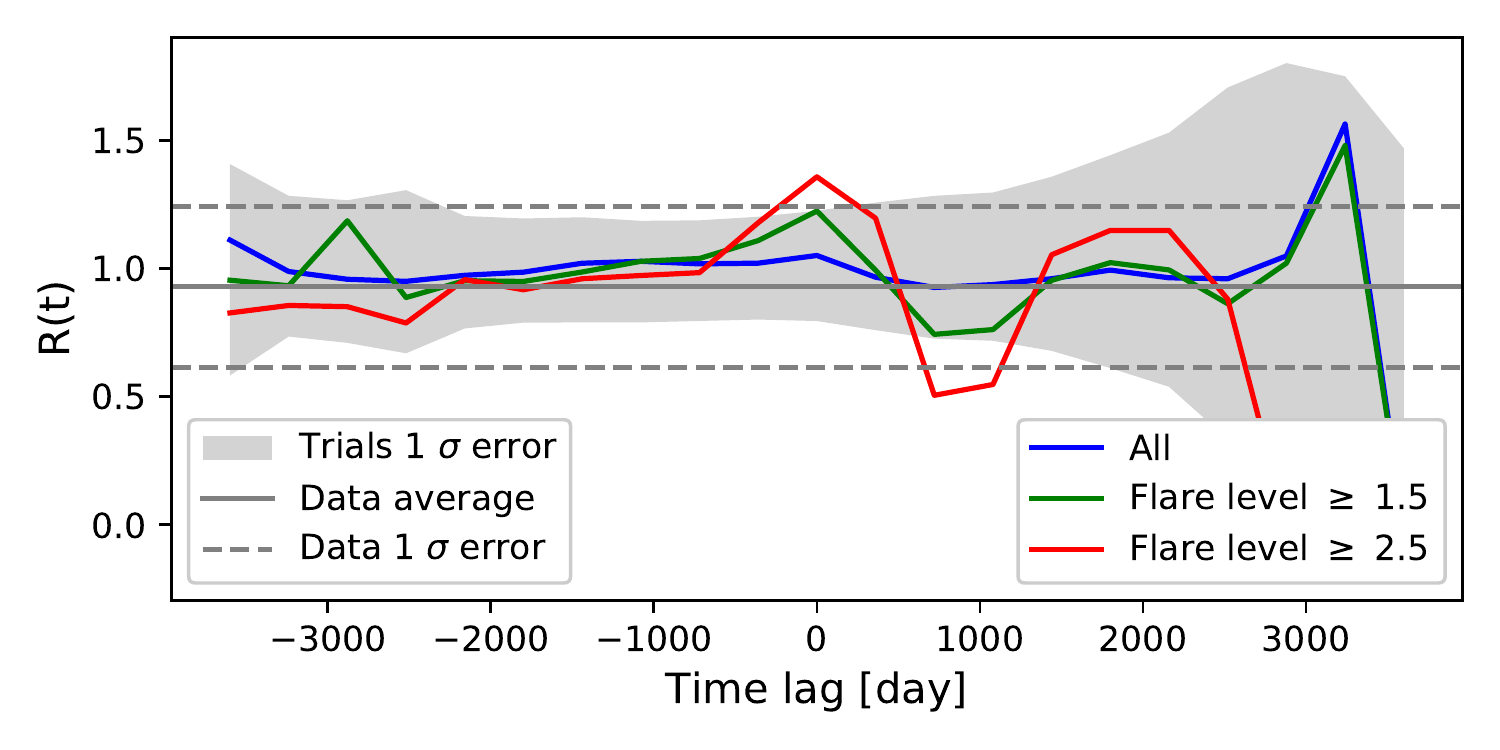}
\includegraphics[width=1.\linewidth]{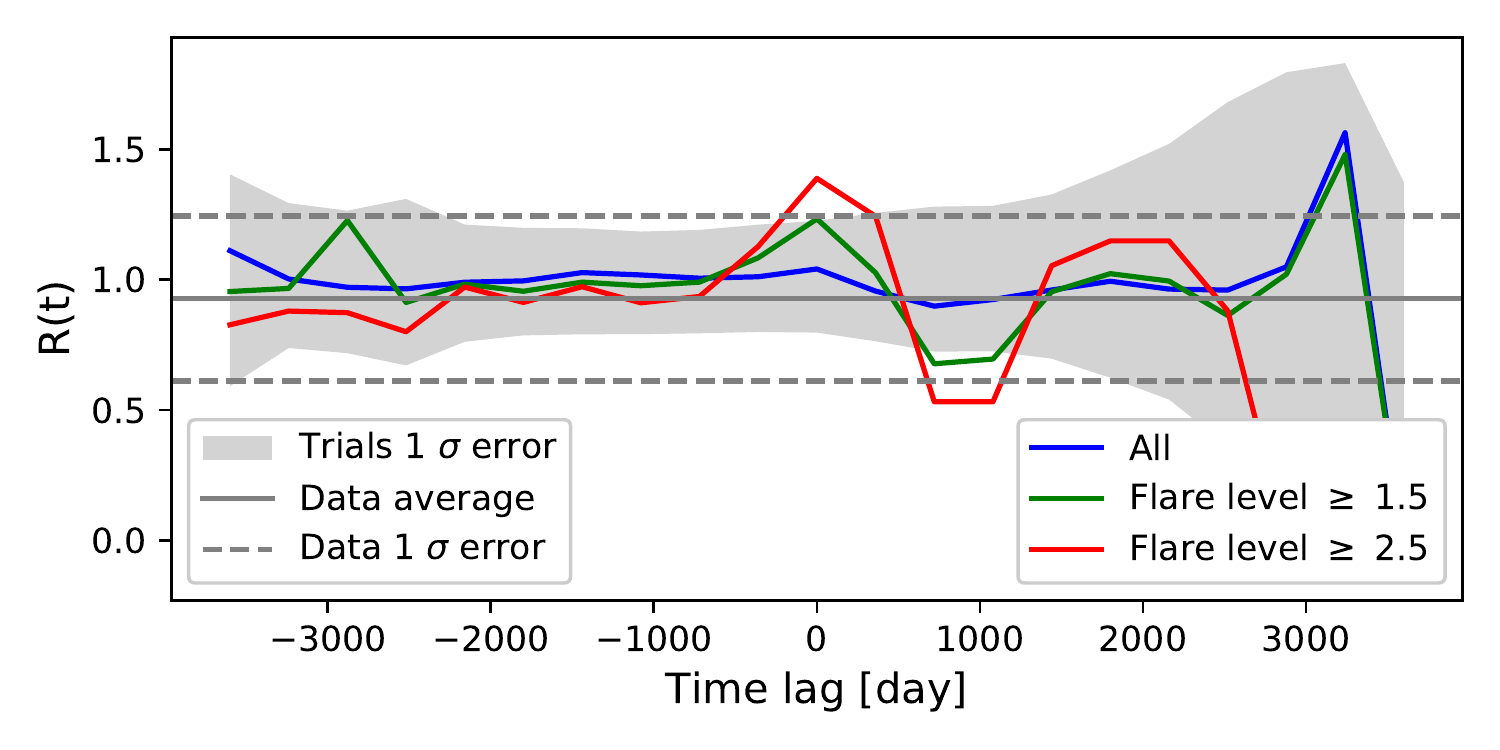}
\end{center}
\caption{The ratio of WIBRaLS flux density averaged over a 630-day window to the average flux density outside it. Up: all WIBRaLS sources. Down: TXS\,0506+056 and PKS 1502+106 are removed from the sample. Definition of legends are the same as Figure~\ref{timedelayacc}.}
\label{timedelaywibrl}
\end{figure}

\begin{figure} [h!]
\begin{center}
\includegraphics[width=1.\linewidth]{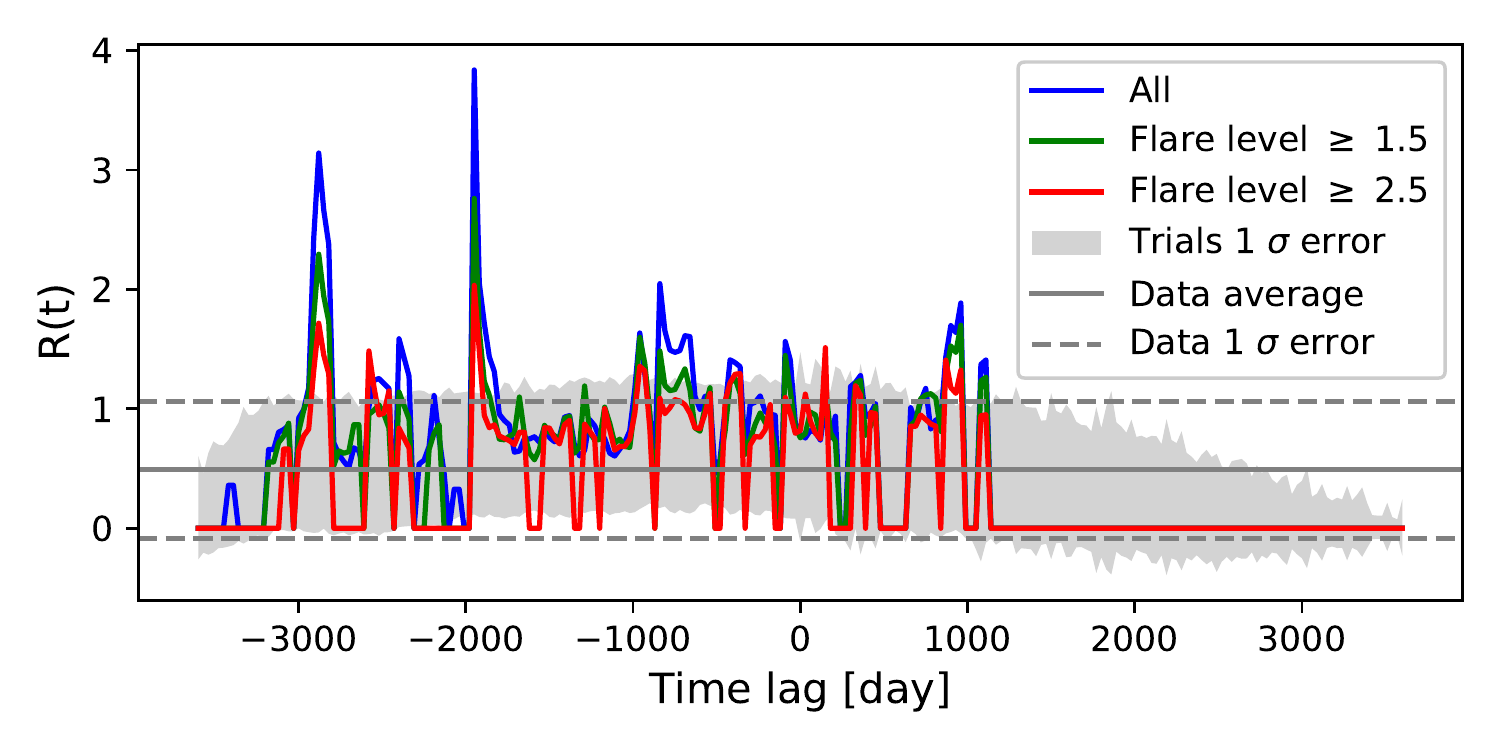}
\end{center}
\caption{The ratio of XRT flux averaged over a 30-day window to the average flux outside it. Definition of legends are the same as Figure~\ref{timedelayacc}.}
\label{timedelayouxrt}
\end{figure}

\begin{figure} [h!]
\begin{center}
\includegraphics[width=1.\linewidth]{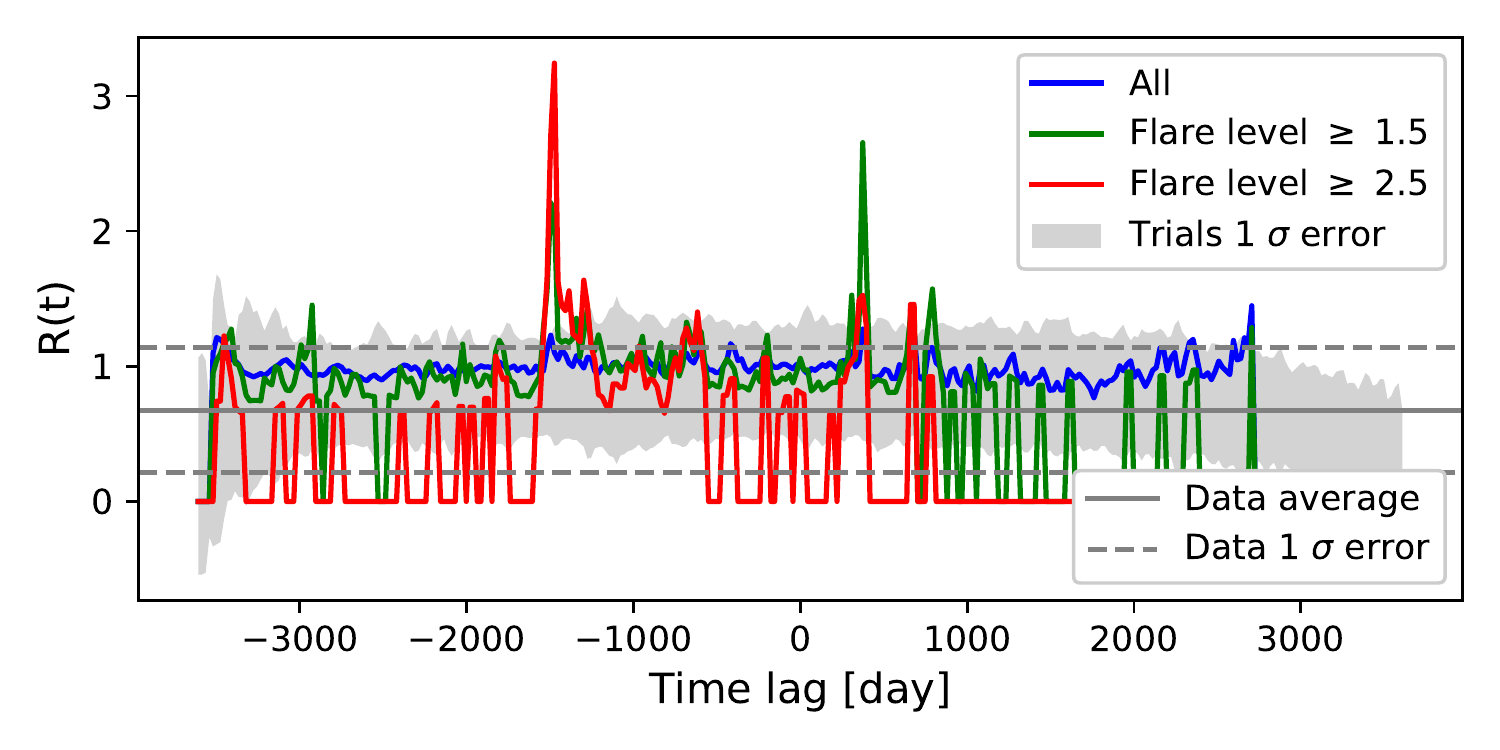}
\includegraphics[width=1.\linewidth]{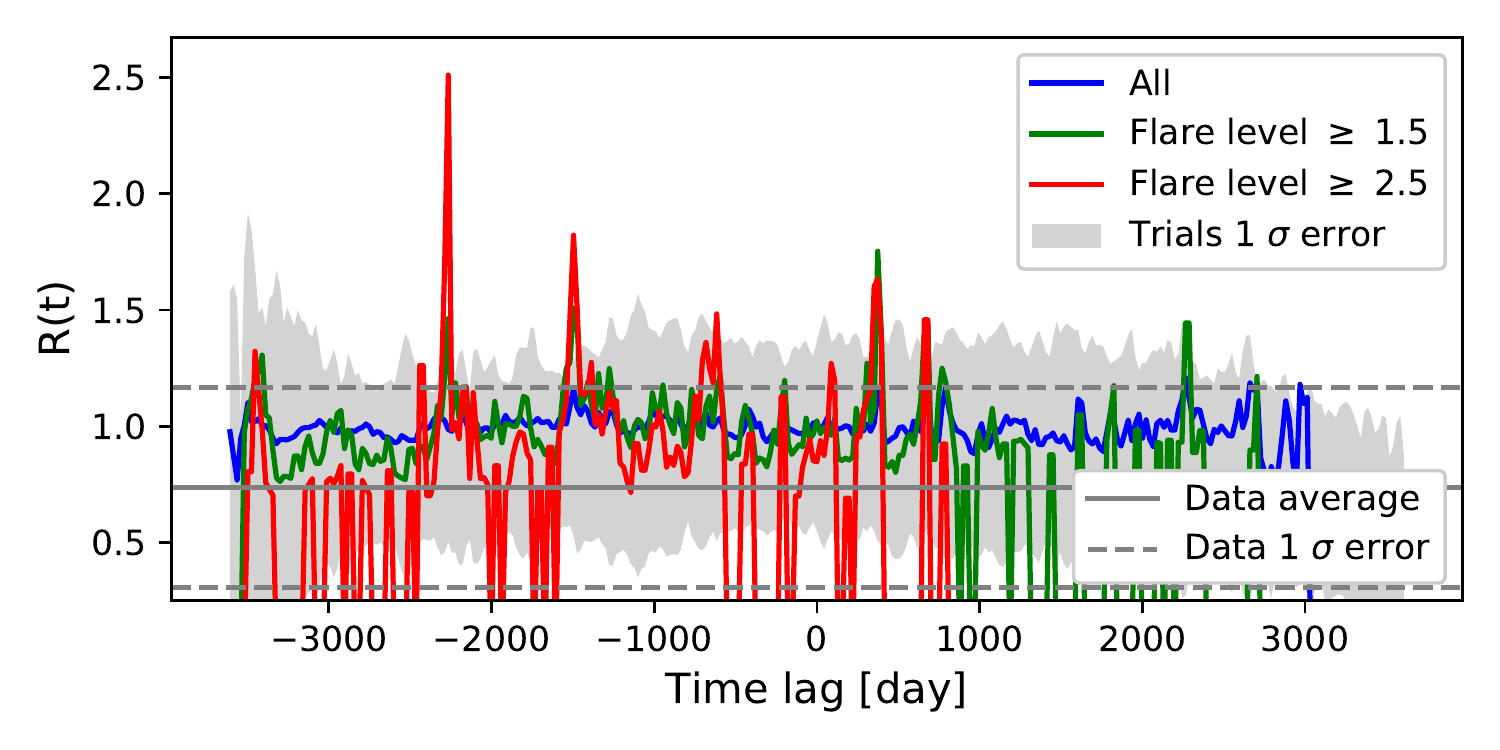}
\end{center}
\caption{The ratio of 4FGL flux averaged over a 22-day window to the average flux outside it. Up: with 90\% containment region extended by a factor of 1.1. Down: factor of 1.5. Definition of legends are the same as Figure~\ref{timedelayacc}.}
\label{timedelayfermi}
\end{figure}

The temporal correlation between neutrino alerts and multi-frequency flares are shown from Figure~\ref{timedelayacc} to \ref{timedelayfermi}. 
The X-axes are the delays of the multi-frequency detections, and the Y-axes represent the gathering of flares around the neutrino time window $\Delta T$. 
To investigate the variation of the correlation concerning the strength of the flares, we tested the relationships for several flare levels cuts. 
We selected the flare level values of 1.5 and 2.5 as our flare-criteria, simply because they are roughly the mean and $1\sigma$ upper limit of the multi-frequency light curves' variability. 

The ${\rm R(t)}$ does not center at any time lag value in Figure~\ref{timedelayacc}. 
We do not find any evidence suggesting a correlation between the millimeter flares and neutrino alerts, but it should be noted that the millimeter light curves are not equally binned concerning some of the detections taken from the target of opportunity observations.

According to Figure~\ref{timedelaywibrl}, the maximum ${\rm R(t)}$ value (apart from the peak at around 3000 days) is at the time lag $t=0$ for strong flares with flare level $\geq 1.5$ or $\geq 2.5$, implying that the strong infrared flares might be correlated with the neutrino arrival time. 
However, the correlation is not significant as the highest ${\rm R(t=0)}$ is just above the $1\sigma$ statistical trial error. 
We then removed the two promising neutrino blazars, TXS\,0506+056 and PKS 1502+106, to evaluate if the possible association is driven by these two blazars. 
Without the two neutrino blazars, we could still tell the ${\rm R(t=0)}$ peak from the figure, suggesting that the possible trend is not driven by them.  
Other blazars might also play a role in the association between infrared flares and neutrino alerts. 
We note that the ${\rm R(t)}$ peak at $t \sim 3000$~days is too far away from the neutrino arrival time and thus we do not consider it to be associated with the neutrino events. 
This peak is caused by the flares from only one source and the signature has low statistics as there are only a few sources/flares considered in the situation of large shift time. 

There are ${\rm R(t)}$ peaks around 2000 and 3000 days before the neutrino arrival time in Figure~\ref{timedelayouxrt}. 
However, no peak stands out around the central region, which might be more interesting when considering the association with the neutrino alert. 
Our results suggest that the correlation between X-ray flares and neutrino events is not obvious, and probably the results are affected by the poor time coverage and uneven time interval of the Swift XRT data. 
A further investigation with dedicated X-ray time coverage might be fruitful in the future. 

Figure~\ref{timedelayfermi} suggests that the \gr\ flaring periods are not apparently correlated with the arrival time of the neutrino alerts. 
In the upper panel, several sources with extremely bright flares dominate the results and cause the peaks several hundred days delayed and roughly 1500 days ahead of the neutrino alerts. 
We have no evidence that the peak at $t=-1500$ days is related to neutrino emissions, but the possible correlation at $t \sim 300$ days might be more likely connected to a hadronic process. 
However, no $R(t)$ peak found after removing the few brightest sources in the figure. 
The peaks are also rendered when we consider more sources with a larger area of the neutrino containment region, and there are several weak peaks in the lower panel. 
This indicates that the systematical association between \gr\ flares and neutrinos might not exist, and the hadronic emission in \gr\ is complicated and might depend on sources. 

\begin{figure} [h!]
\begin{center}
\includegraphics[width=1.\linewidth]{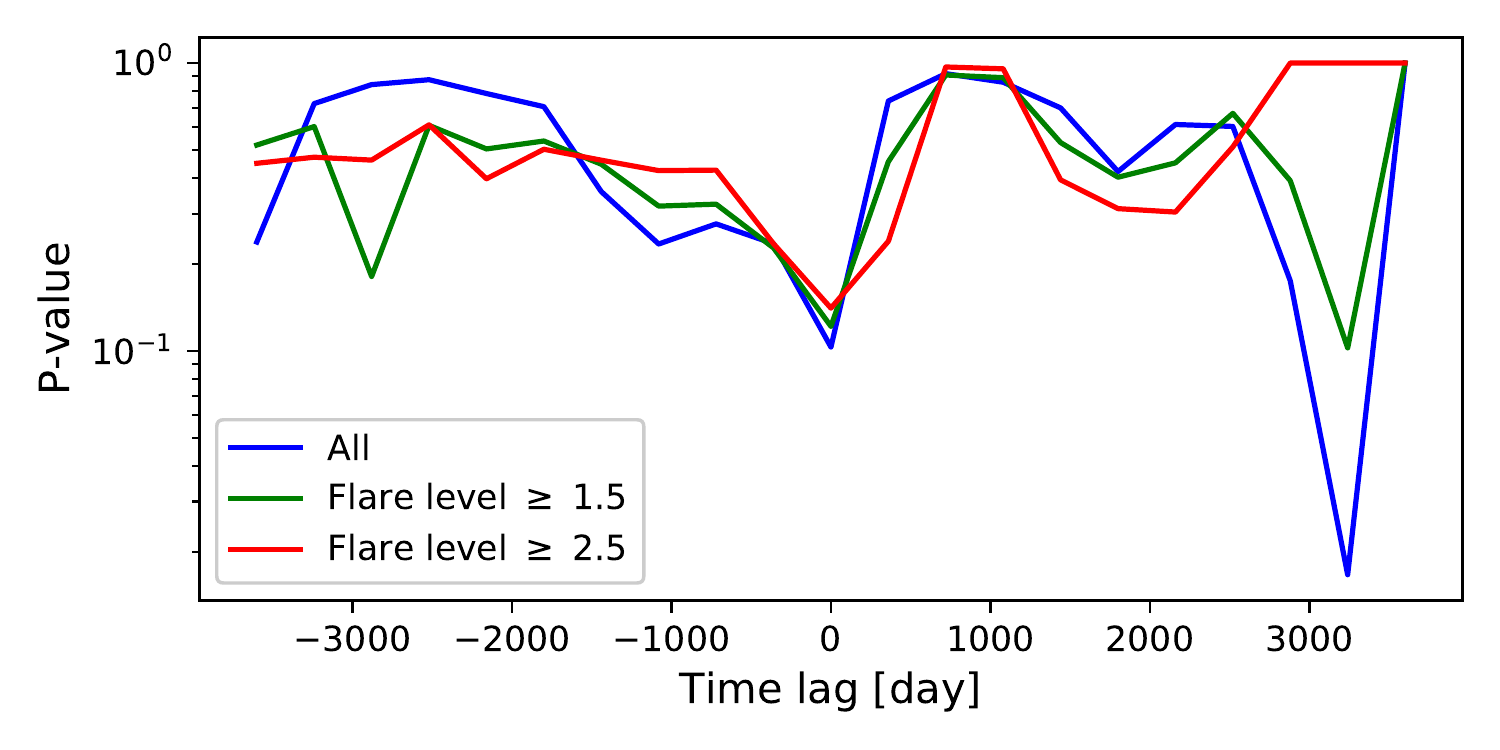}
\end{center}
\caption{Pre-trial p-values of the ratio of averaged flux over a 630-day window to that outside it for WIBRaLS sample.}
\label{timedelaypwib}
\end{figure}

Among our time-dependent analyses for different multi-frequency samples, only the infrared one show some interesting sign of possible correlation with neutrinos. 
We estimated the significance level of the correlations by calculating the probability that the randomly selected sources and assigned alert times (from the control samples) could lead to higher ${\rm R(t)}$ values only  by chance.
The chance probability is illustrated as pre-trial p-value in Figure~\ref{timedelaypwib}. 
We did not consider the trial factors for the pre-trial values as the correlation between infrared flares and IceCube alerts is far from significant. 
As shown in Figure~\ref{timedelaypwib}, the smallest pre-trial p-value close to the neutrino arrival time ($t=0$) is $\lnsim 0.1$. 

\section{Spatial correlations between blazars and the neutrino-track events} \label{blazarlc}
Another part of our analysis aims to study the correlations of neutrinos with different subgroups of blazars divided by their multi-frequency activity. Three tests are carried out in this section. The first two tests are performed for the infrared-\gr-cut FIBS sources to investigate the overall multi-frequency flaring correlations and simultaneity of the whole population of blazars. The purpose is to compare the multi-frequency properties of typical blazars and those that might associate with IceCube neutrinos and understand the potential bias in our analyses. In the last test, we applied our selected lists of TXS-like blazars (see section~\ref{txsblazar}) to study the neutrino-blazar correlations w.r.t. electromagnetic flaring stages.

\subsection{Multi-frequency correlations of blazars} \label{dcfcheck}

We study the average time lag of the infrared flares with respect to the \gr\ ones for all the FIBS sources to check if the similar time lag seen in the TXS\,0506+056 is a common phenomenon among blazars. 
Cross-correlation between two multi-frequency light curves for a source could be estimated with Discrete Correlation Function (DCF). 
The time lag for each source is retrieved from the time bin with maximum cross-correlation value. 
Here, the time lag is also evaluated with the fitting of DCF cross-correlated matrix, and we took the Gaussian mean as a representation of the time lag of the two light curves. 

\begin{figure} [h!]
\begin{center}
\includegraphics[width=1.\linewidth]{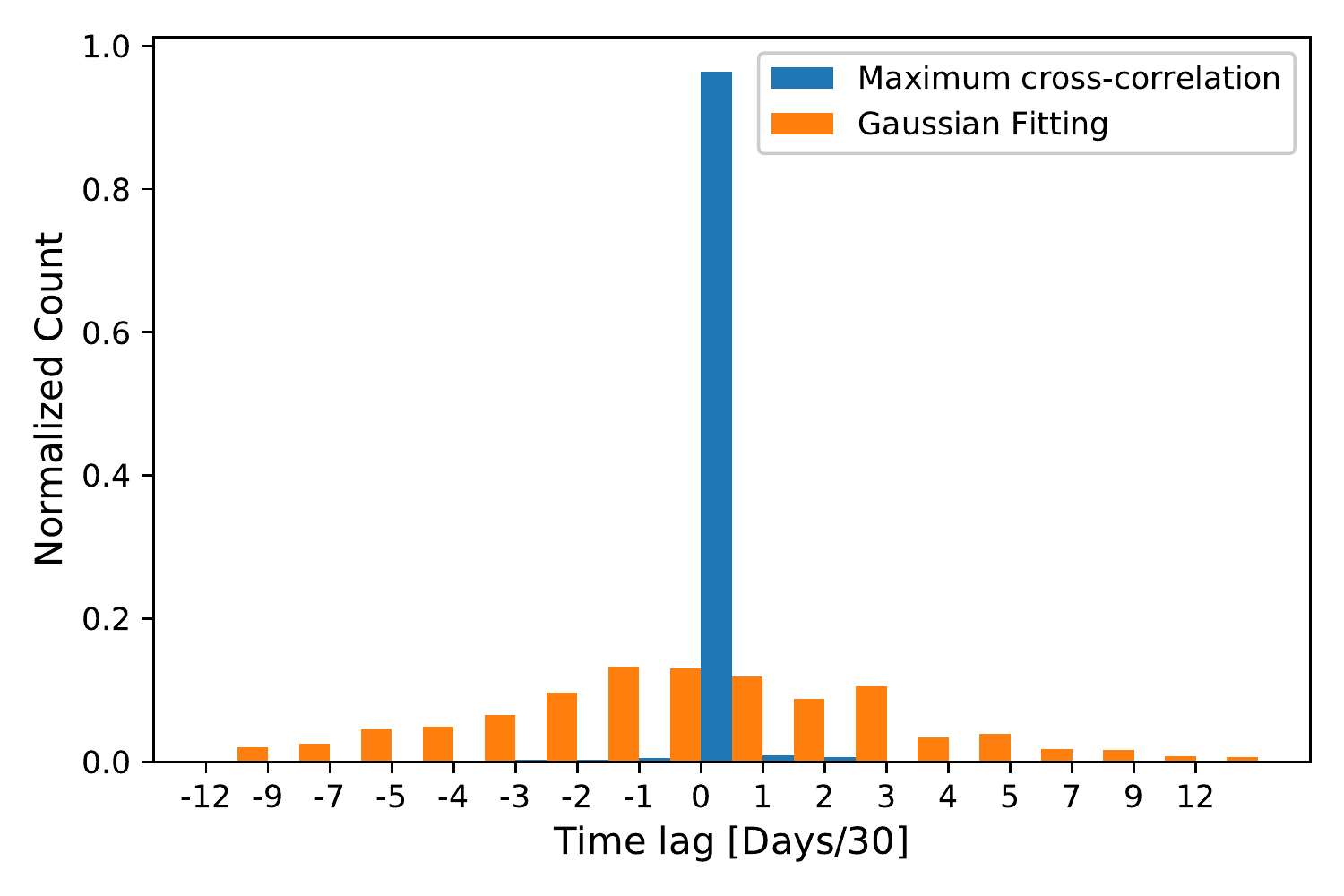}
\end{center}
\caption{Correlation between infrared and {\it Fermi} light curves. The time lags of flaring in infrared are estimated with the Discrete Correlation Function. The blue and orange bars represent the time lag corresponding to the maximum cross-correlation and corresponding to the Gaussian fitted mean of the cross-correlated matrix between two light curves, respectively.}
\label{irdcf}
\end{figure}

Figure~\ref{irdcf} illustrates the distribution of time lag of flaring in infrared compared with the activity in \gr\ for FIBS sources. 
According to the figure, the time lag obtained directly with the time bin with maximum cross-correlation value substantially centers at bin 0-30 days, while that obtained with the Gaussian fitting is more scattered but still gathers around the zero. 
On average, there is no significant delay of infrared to \gr\ flares for most blazars. 
Blazars like TXS\,0506+056, with an infrared time lag of $\sim 300$ days, represent only a tiny fraction of the total FIBS sources. 

\subsection{Simultaneity of multi-frequency flares of blazars} \label{nusource}

In this section, we present a comparison between the multi-frequency flaring phase for blazars that are probabably related to IceCube neutrinos and that for the entire blazar population. 
Blazars within $90\%$ IceCube alert containment regions or associated with a weak neutrino excess signal (p-value $\leq 0.05$) in IceCube point-source analyses are considered as {\it potential neutrino sources} (section~\ref{fibs}). 
We would like to test whether the high-energy and low-energy flares of those potential neutrino blazars are more likely to be simultaneous or ``orphan".  

\begin{figure} 
\begin{center}
\includegraphics[width=1.\linewidth]{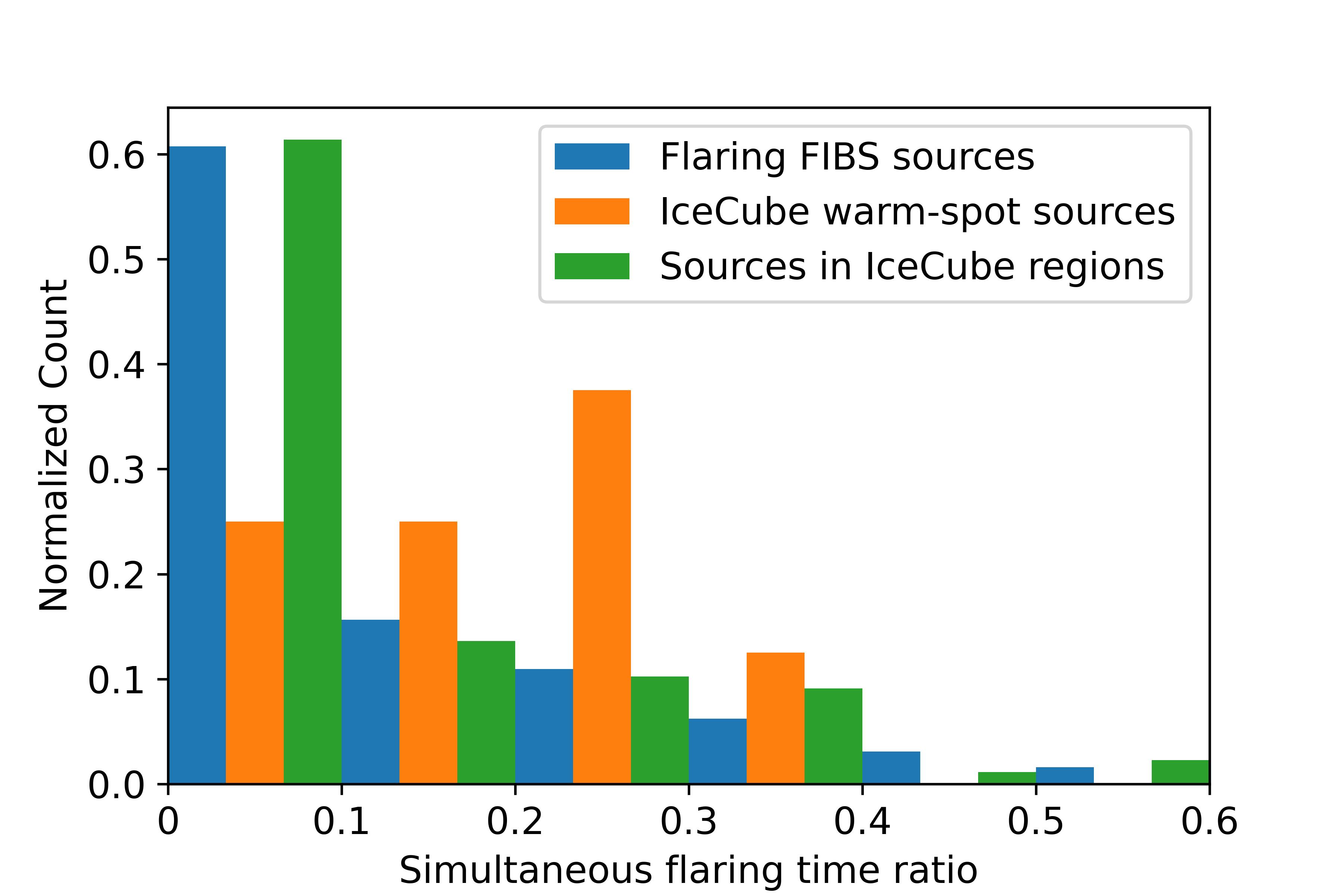}
\end{center}
\caption{Distribution of the simultaneous flaring period (overlap flaring time) ratio in infrared and \gr. The blue bars indicate the FIBS sources with both flaring in infrared and \gr, and the orange and green bars are the ``IceCube Warm Spots" and ``IceCube Region" sources, respectively. See text for more details.}
\label{irichist}
\end{figure}

Figure~\ref{irichist} shows the distribution of the ratio of the simultaneous flaring stages to the whole flaring period in infrared and \gr\ (simultaneous flaring time ratio) for FIBS (sub-)samples. 
To simplify, here we called those with $\geq 95\%$ significance level in IceCube's point-source searches ``IceCube Warm Spots." 
On the other hand, sources within the containment regions of IceCube alerts are defined as ``IceCube Region" sources. 
The flaring times are defined with the Bayesian Blocks Algorithm, and details of how we defined the simultaneous flaring stages in a light curve are written in section~\ref{mwstage}. 
We note that here we assumed the $90\%$ IceCube containment region is a rectangle instead of an approximate ellipse to speed up the simulation. 
Besides, only sources with flares in both WISE and {\it Fermi} light curves are plotted in the histogram.
The number of FIBS sources and {\it potential neutrino sources} considered in this test are a bit different, with 2239 infrared and \gr\ flaring blazars among FIBS, 8 ``IceCube Warm Spots", and 88 ``IceCube Region" sources applied. 

From Figure~\ref{irichist}, it is clear that the distribution of the simultaneous flaring time ratio for ``IceCube Warm Spots" is substantially different from that of typical FIBS sources and ``IceCube Region" sources, suggesting that those with a weak neutrino signal in IceCube's previous point-source analyses tend to have simultaneous flares in infrared and \gr. 

\begin{figure} [h!]
\begin{center}
\includegraphics[width=1.\linewidth]{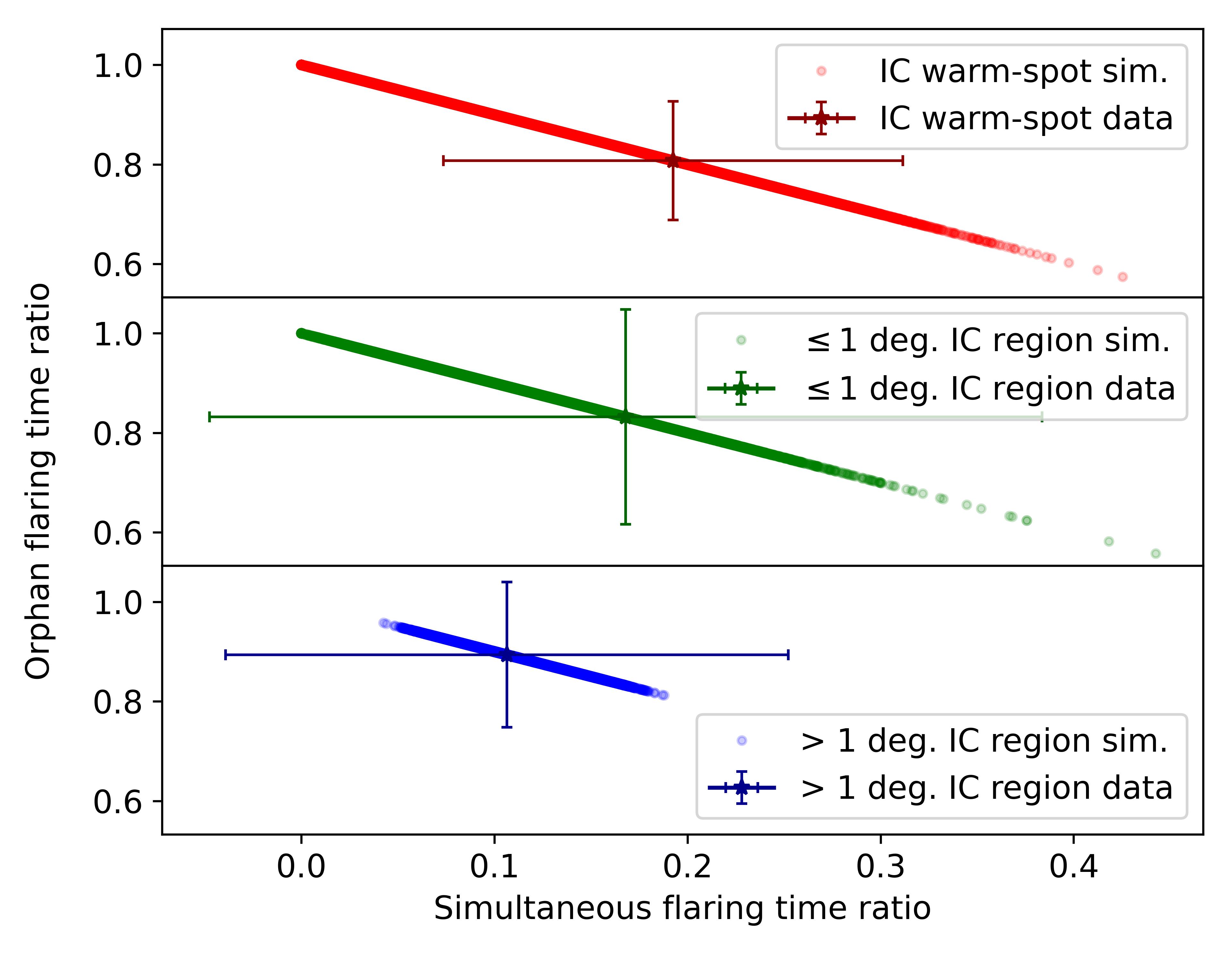}
\end{center}
\caption{The ratio of orphan flares versus simultaneous multi-frequency flares with observed and simulated data. The stars represent the averaged values of the observed data, while the circles mean those from the simulations. The red, green, and blue points are ``IceCube Warm Spots", ``IceCube Region" sources, with a distance smaller than 1$^\circ$, and ``IceCube Region" sources with a distance larger than 1$^\circ$.}
\label{iroverlap}
\end{figure}

The significance levels are estimated using Monte Carlo simulation. 
For ``IceCube Region" sources, we randomly scrambled the RA of IceCube alerts $N=10^5$ times and counted the number of iterations with a simultaneous flaring period longer than the original RA, and the number is denoted M. 
The p-value is then defined as $(M+1)/(N+1)$. 
While for ``IceCube Warm Spots", in each iteration, we randomly selected the same number of insignificant sources as the ones with $95\%$ significance level instead of shifting the RA. 
Simulation results are illustrated in Figure~\ref{iroverlap} and Table~\ref{simtable}.
In the simulation, we further divided the ``IceCube Region" sources by their distance to the reconstructed center of the alerts, trying to study the influences of the distance to the alert center on the significance level. 
Among 88 sources with both infrared and \gr\ flares in their light curves and within the IceCube $90\%$ containment region, 14 of them are closer than one degree to the alerts' center. 

\begin{table} [h!]
\begin{center}
    \begin{tabular}{cc}
    \hline
     Sources & p-value \\
    \hline
    \hline
    IceCube warm-spot sources  & 0.038 \\
    \hline
    Sources in IceCube region with a distance $ \leq 1^\circ$ & 0.218 \\
    \hline
    Sources in IceCube region with a distance $> 1^\circ$ & 0.466 \\
    \hline
    \end{tabular}
\caption{Pre-trial p-values for three sub-groups of {\it potential neutrino soruces.}}
\label{simtable}
\end{center}
\end{table}

According to Figure~\ref{iroverlap} and Table~\ref{simtable}, ``IceCube Warm Spots" tend to have overlap and simultaneous flaring period in low and high frequency, at a $96.2\%$ significance level, consistent with the histogram in Figure~\ref{irichist}. 
Besides, the pre-trial p-value is smaller for those sources closer to the IceCube alerts, but still with a low significance level and the value larger than 0.2. 
Given that the significance level is not high, only pre-trail p-values are shown in this test. 

\subsection{Time-integrated analysis-Correlations with IceCube 10-year track-like neutrinos} \label{icecubeana} 

The possible neutrino emissions from blazars based on their multi-frequency activity, which is one of the main purposes of this paper, will then be thoroughly investigated. 
Since the most probable neutrino blazar TXS\,0506+056 has a significant flaring lag in infrared with respect to \gr\ (Figure~\ref{txs0506}), we speculate that the neutrino blazars' low-frequency and high-frequency flaring stage might not be simultaneous. 
In section~\ref{groups}, we have built three groups of sources selected from the FIBS sources, that is, one trial group with 32 selected sources with TXS-like multi-frequency flaring behavior and two control groups according to their multi-frequency flaring stages: sources without simultaneous infrared and \gr\ flares and sources with highly simultaneous flares (NFS and CFS, see also Figure~\ref{cfslc}). 
Eleven sources with ACC and {\it Fermi} flaring activity similar to that of TXS\,0506+056 were also selected for comparison.

We first check on the difference in the potential association of neutrinos among selected groups of blazars by counting the number of {\it potential neutrino sources} in these groups. 
This provides an alternative way to investigate the correlation between neutrinos and blazars. 
Table~\ref{numtable} shows the number and fraction of blazars potentially associated with IceCube neutrinos in each group.  
The total number of sources in each list is represented with a bracket. 
Compared with the control groups, the fraction of {\it potential neutrino sources} for TXS-like sources selected from both FIBS and ACC-{\it Fermi} sub-sample are higher. 
The relatively high fraction for selected TXS-like sources from the infrared-\gr\ sub-sample indicates that sources with multi-frequency light curve behaviors similar to TXS\,0506+056 are more likely to be related to neutrinos, though the statistic is low given the small number of sources selected. 
It should be noted that the high fraction for sources selected from the ACC-{\it Fermi} sub-sample, on the contrary, is tentative and biased given that the multi-epoch data from the ACC catalog is far from complete. 

\begin{table} [h!]
\begin{center}
    \begin{tabular}{lccc}
    \hline
      & Selected TXS-like & CFS & NFS \\
    \hline
    \hline
    Infrared-\gr\ & 4(32) & 2(62) & 32(409) \\
    (FIBS) & 12.50\% & 3.23\% & 7.82\% \\
    \hline
    Millimeter-\gr\ & 2(12) & --- & --- \\
    (ACC-{\it Fermi}) & 16.67\% & --- &  --- \\
    \hline
    \end{tabular}
\caption{Number and fraction of {\it potential neutrino sources} in our selected lists and control groups. The values inside the brackets represent the total number of sources in each selected list and control group. Example light curves of TXS-like sources and CFS samples are illustrated in Figure~\ref{txs0506} and \ref{cfslc}, and detailed selections of those groups of sources are described in Section~\ref{txsblazar}. }
\label{numtable}
\end{center}
\end{table}

To test our hypothesis, we made use of IceCube 10 years of public data for point-source searches \citep{IceCube2021a} and conducted a time-independent analysis to study the spatial correlation between the track-like neutrino events and the four groups of blazars. 
We calculated the number of neutrino events that have at least one blazar located inside their uncertainty regions ($N_{observed}$) then used Monte Carlo simulation to estimate how many neutrino events are around our groups of blazars by chance. 
The simulation was iterated $10^{5}$ times with all blazars' positions randomly scrambled in R.A., and the expected number of neutrinos close to our blazars by chance ($N_{expected}$) is the average of those simulated number of events nearby. 
By subtracting $N_{expected}$ from $N_{observed}$, we got the excess number of neutrinos associated with our groups of blazars. 
The excesses of neutrino counts represent the fact that we observed more neutrino around our test blazars' positions than expected. 
The p-values are estimated based on whether the number of neutrinos around our blazar samples is higher than what is expected to arise by chance.


\begin{figure} [h!]
\begin{center}
\includegraphics[width=1.\linewidth]{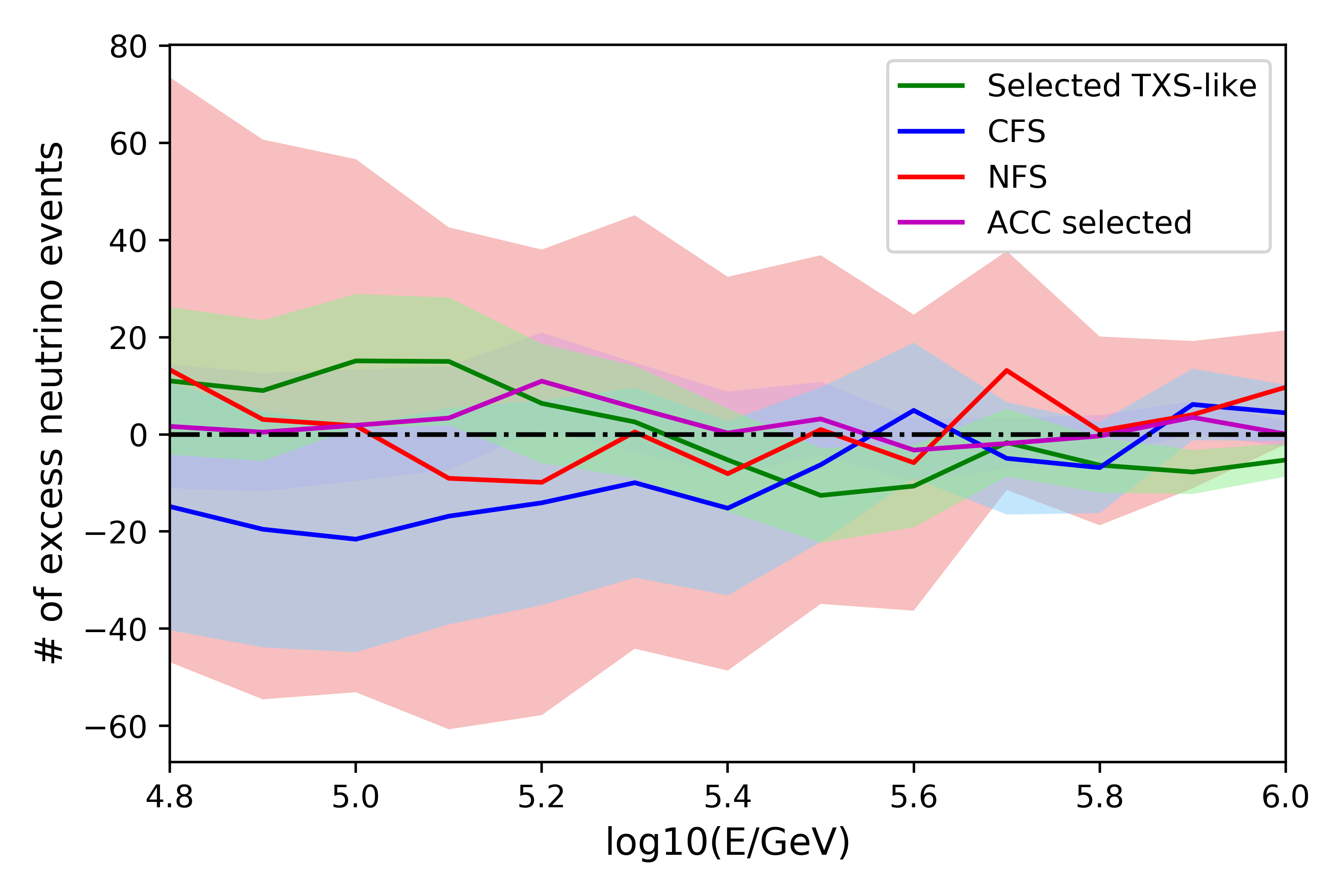}
\end{center}
\caption{Excess number of neutrinos around the position of four groups of blazars w.r.t. the reconstructed energy. Filled regions show the $1\sigma$ statistical errors from random trials of simulation. }
\label{figsim}
\end{figure}

\begin{figure} [h!]
\begin{center}
\includegraphics[width=1.\linewidth]{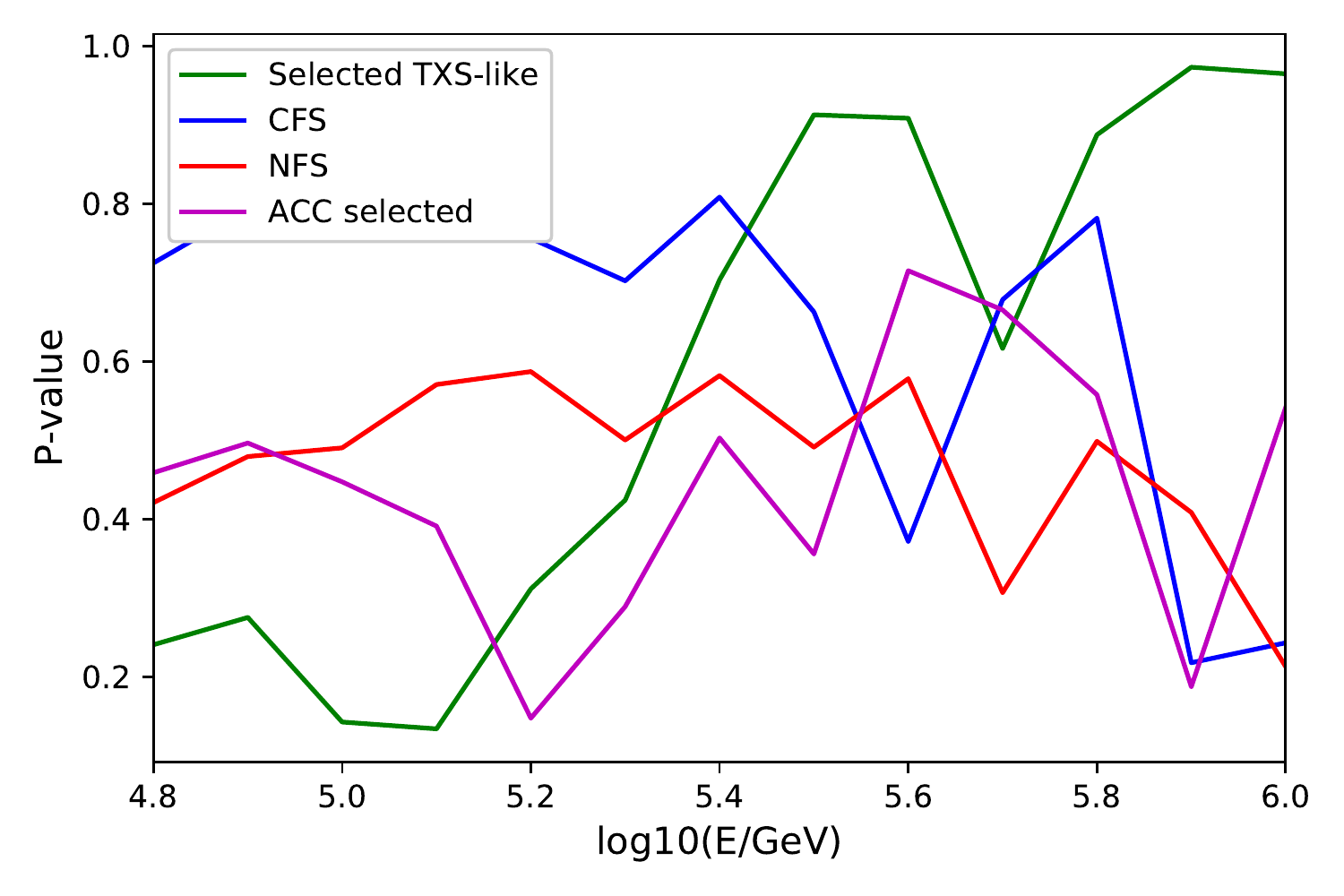}
\end{center}
\caption{Pre-trial p-values of the excess number of neutrino around the position of four groups of blazars w.r.t. the reconstructed energy. }
\label{figpval}
\end{figure}

The excess number of neutrinos around the position of four groups of blazars as well as the corresponding p-values with respect to the reconstructed energy of 10-year IceCube events are illustrated in Figure~\ref{figsim} and \ref{figpval}. 
Table~\ref{simresult} shows the measured and expected number of neutrino ($N_{observed}$ and $N_{expected}$) around four groups of blazars. 
The figures suggest that, at energy bins between $\sim 60$ to $\sim 150$~TeV, generally there is a small neutrino overflow in the vicinity of blazars with non-simultaneous infrared and \gr\ flares (those similar to TXS\,0506+056), with an average overflow of $\sim 10-15$ neutrinos. 
This overflow is not significant, with a pre-trial p-value around $0.13-0.27$. 
We also determined statistical significance using a right-tailed p-value, which describes the probability of randomly obtaining the number of neutrino events from simulation that is greater than the real observed value ($N_{observed}$). 
The right-tailed p-value for the neutrino overflow around selected TXS-like blazars is around $0.13-0.27$ as well, with the lowest value of 0.127 that occurs at $~150$ TeV. 
On the contrary, there is no neutrino excess for those selected sources at higher energy bins. 

The excess count of neutrinos around the CFS control group is gradually increasing with the reconstructed energy of the neutrino events. 
This increment indicates that there might be a correlation between blazars with highly simultaneous infrared and \gr\ flares and extremely high-energy neutrinos. 
The significance of this possible trend is not high, with a right-tailed p-value of $\sim 0.2$.   
No trial correction was performed for these pre-trail p-values as the correlations are not significant. 
We note that the excess number of neutrinos fluctuates around zero for the NFS sample and for sources with millimeter and \gr\ flaring activities similar to TXS\,0506+056. 
This implies that the number of neutrinos observed around those blazars is consistent with the number of neutrinos that are randomly nearby only by chance. 
Even though the highest overflow occurs at $\sim 500$~TeV for the NFS sample, it is regarded as random fluctuation given the extremely large trial errors. 

\begin{figure} [h!]
\begin{center}
\includegraphics[width=1.\linewidth]{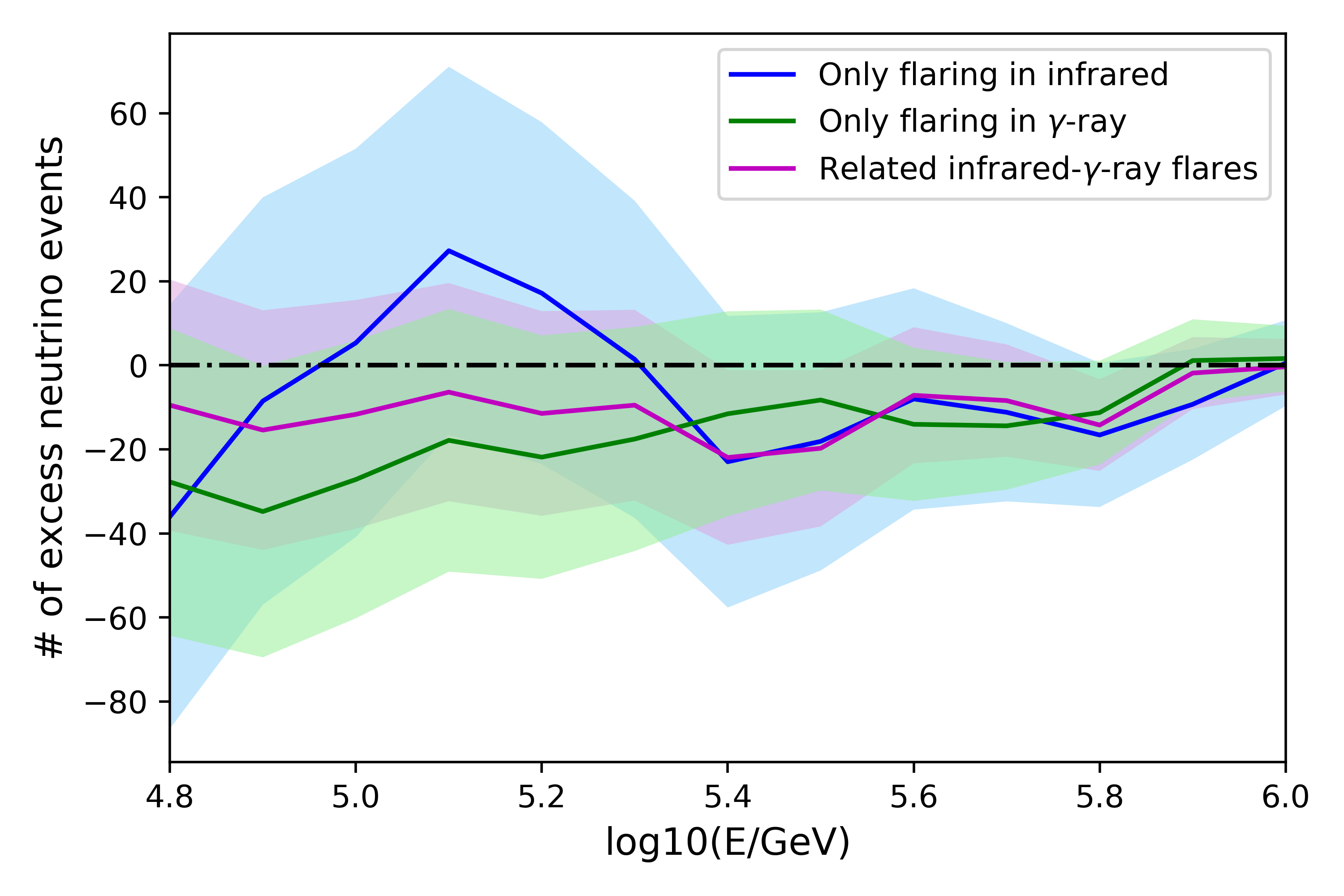}
\end{center}
\caption{Excess number of neutrinos around the position of three groups of blazars w.r.t. the reconstructed energy. Filled regions show the $1\sigma$ statistical errors from random trials of simulation. }
\label{figsimtest}
\end{figure}

According to our selection criteria, those selected sources with TXS-like flaring behaviors do have related infrared and \gr\ flares with time lag $< 500$ days, even though they are not highly simultaneous. 
Thus, we joined those 32 selected sources with the CFS sample to see if the neutrino excess around those blazars with ``Related" flares would be more significant. 
We did further tests for blazars with only infrared or \gr\ flares, but without both flares in multi-frequency light curves. 
Figure~\ref{figsimtest} shows the same analysis as described in the previous paragraph, but with three different groups of blazars: ``Related" flares, only flaring in infrared, and only flaring in \gr. 
Apparently, there is no excess of neutrino events around these three groups of blazars.

\section{Discussion} \label{discuss}
\subsection{Result 1: temporal correlation between infrared flares and IceCube alerts}
According to our analyses, there might exist a weak trend that the high-energy neutrino alerts might be temporarily correlated with infrared flares (section~\ref{alertflare}). 
From a population perspective, the neutrino emitters that left ``clues" in infrared might be weak sources, even though we have considered the emission from the whole sample of blazars within the containment regions of the alerts, there is no statistically significant neutrino signal. 
A deeper and more complete infrared monitoring catalog might lead to a much more significant result.  
If significant associations in infrared are found with higher timing resolution multi-epoch data in the future, it would be complementary to the possible radio correlation found by \citet{Plavin2021}, providing a clearer perception of the association between lower-energy wavebands and neutrinos. 
 
On the other hand, clear and significant correlation between \gr\ flaring stages and the neutrino arrival times is absent from our analyses, which is consistent with the results of \citet{Franckowiak2020}, \citet{Righi2019a}, and references therein. 
The ``direct" connection between \gr\ and neutrino becomes ambiguous if the pionic photons cascade down to much lower energies than GeV. 
Even though the protons and electrons are co-accelerated in or interact with the same photon field, we might not be able to detect related high-energy photon flux and neutrino flux. 
Not to mention that the \gr\ emissions could originate from a different region than the neutrino production locus \citep{Plavin2021}. 

\subsection{Result 2: spatial correlation between simultaneous infrared and \gr\ sources and extreme-high-energy neutrino track events} \label{result2}
A possible correlated trend between blazars with highly simultaneous infrared-\gr\ flares (CFS sample, Figure~\ref{cfslc}) and extreme-high-energy E $\gtrsim1$~PeV track-like neutrino events are shown in section~\ref{icecubeana}. 
Considering the steep spectral indexes of atmospheric neutrinos, compared to astrophysical neutrinos, the higher the energy of a neutrino event indicates a higher probability of that event being of astrophysical origin. 
Interestingly, our results also suggest that the sources with significance level $> 95\%$ in IceCube point-source searches tend to have simultaneous infrared and \gr\ flaring period (section~\ref{nusource}). 
According to all these results, blazars with \gr\ flaring phases correlated with the infrared ones are more likely to be astrophysical neutrino sources. 

The results imply that a typical single-zone Synchrotron self-Compton (SSC) model might be able to explain the emission for those CFS blazars and suggest that neutrinos might be emitted from the same region as infrared and \gr\ photons in the inner jet. 
This trend is consistent with prediction in \citet{Padovani2015} and \citet{Murase2017}, and references therein. 
They suggest that blazars might dominate the astrophysical neutrino flux at around $~$PeV, or even at higher energies of $10-100$~PeV, if the particles (primary electrons and protons) are co-accelerated in the jet. 
If so, we would expect a correlation among neutrinos, infrared flares, and \gr\ flares, with the assumption of an efficient cascade of TeV photons down to the GeV energy range, which is detectable by {\it Fermi}-LAT. 
The optical depth of the electromagnetic cascade depends on the photon field that drives the photo-hadronic process. 
In this case, the photon field is produced by the inverse Compton scattering of the jets' low-frequency synchrotron emission.

\subsection{Result 3: spatial correlation between TXS-like sources and neutrino tracks at $\sim 60-150$ TeV} \label{result3}
In section~\ref{icecubeana}, we have shown a small overflow of $\sim 10$ track-like neutrino events with energy between $\sim 60$ to $150$~TeV around blazars with infrared and \gr\ flaring behaviors like TXS\,0506+056 (Figure~\ref{txs0506}). 
Table~\ref{numtable} additionally suggests that the group of TXS-like blazars contains more {\it potential neutrino sources} than the other groups. 
These indicate that those TXS-like blazars might also contribute to the observed diffuse neutrino flux, especially at the energy of a few tens to hundreds of TeV. 

From the DCF test (section~\ref{dcfcheck}), it is known that blazars with infrared flares significantly lag to the \gr\ flares of $\sim 300$ days like TXS\,0506+056 are not common. 
The significant lag implies that their low-frequency synchrotron emission might not come from the same site as where the inverse Compton high-frequency photons are produced. 
In other words, the \gr\ emission might originate from a region closer to the central supermassive black hole, compared to the synchrotron radiation. 
The external Compton (EC) radiation may dominate the high-energy peaks in the SED of those blazars (instead of SSC radiation), and the external photon field may be highly relevant for the particles in their inner jet. 
One of the explanations for this significant lag is that when turbulence in the jet cause shock waves that propagate along with the jet, high-frequency emissions occur upstream of the jet before the low-energy synchrotron photons become transparent \citep{Boula2018,Max-Moerbeck2014}. 
Alternatively, the temporal evolution of emitting particles would also lead to the lag. 
\citet{Sahakyan2021} explained the lag of optical/UV flares regarding X-ray/\gr\ for a transient blazar, 4FGL J1544.3-0649, with an SSC model considering the acceleration of electrons and cooling of synchrotron photons. 
In their scenario, the injection of freshly accelerated electrons leads to a bright state in the high-energy band. 
We could not exclude the possibility that accelerated protons are also injected into the emission zone. 
Either explanation or the impact of the external photon fields provide efficient conditions for the production of neutrinos in those blazars. 
Especially, many studies have shown the difficulty to explain both neutrino and electromagnetic emissions from TXS\,0506+056 with a single-zone model (see \citet{Cerruti2020} for a review on the hadronic emissions from TXS\,0506+056). 

An optimistic scenario for the production of neutrinos is the existence of an external photon field. 
The remarkable point is that TXS\,0506+056 is found to be a ``masquerading BL Lac object", intrinsically an FSRQ, with a hidden broad-line region (BLR) and a radiatively efficient and geometrically thin disc accretion flow \citep{Padovani2019}. 
This hidden BLR might act as an external field for TXS\,0506+056 to increase the efficiency of the photo-hadronic process. 
Indeed, \citet{Padovani2021} suggests that the fraction of masquerading BL Lacs in their sample of 47 {\it Fermi} IBLs and HBLs in the vicinity of IceCube high-energy track-like neutrinos \citep{Giommi2020} is $>24\%$ and possibly as high as $80\%$. 
Another potential neutrino blazars, MG3 J225517+2409, reported to be (weakly) associated with ANTARES and IceCube neutrinos \citep{ANTARES2019}, is also a masquerading BL Lac \citep{Padovani2021}.
For other ``intrinsic BL Lac objects", the existence of a sheath of the jet in spine-sheath model \citep{Tavecchio2015} or the complex and relatively broad-band photon spectrum produced from radiatively inefficient accretion discs \citep{Righi2019a} might be able to act as a target field for neutrino production. 
Furthermore, the SEDs of a complete sample of 104 radio-bright blazars (all with 37 GHz flux density higher than 1 Jy) are found to be better described by an EC model with a dominant infrared external photon field that can originate from dust torus emission or molecular clouds in sphine-sheath geometry \citep{Arsioli-Chang2018}. 
Those radio-bright blazars are all with VLBI 8 GHz flux density $\geq 150$ mJy and thus are also in the samples in \citet{Plavin2021}, which shows possible correlation with the neutrinos. 
We note that \citet{Ros2020} found signs of a spine-sheath structure in the jet of TXS\,0506+056. 

If those TXS-like blazars are supposed to be related to IceCube neutrinos, the natural question that arises then is: where do neutrinos come from?
Under the assumption that high-energy and low-energy photons might not originate from the same site, we can envision multiple scenarios where neutrinos might be emitted from ({\romannumeral 1}) the same place as high-energy radiation ({\romannumeral 2}) the same place as low-energy synchrotron emission ({\romannumeral 3}) other places neither related to low nor high-frequency emissions. 
For the first possibility, \citet{Xue2019} proposed a scenario in which neutrinos might be produced from {\it p$\gamma$} process and possibly {\it pp} process in the central region with ambient gas cloud and external photon field from the BLR where the inverse Compton X-ray and \gr\ are radiated. 
On the other hand, the low-energy synchrotron radiation might originate from the outer region where external photons from BLR or accretion disc are negligible. 
The second possibility could be a scenario proposed by \citet{Plavin2021} in which neutrinos are emitted from the parsec region in blazar jets where X-ray SSC photons interact with accelerated protons. 
GeV \gr s\ (probably dominated by external Compton radiation) are supposed to be emitted from a different region than the neutrino and synchrotron radiation. 
While \citet{Neronov2021} proposed an alternative scenario suggesting that the link between radio synchrotron and neutrinos is expected in proton-proton interaction. 
In the central region of blazars, the interaction between high-energy photons and circum-nuclear medium could produce charged pions that decay into both neutrinos and synchrotron-emitting electrons. 
To explain the third possibility, a collimated neutron beam is assumed to be produced from interaction between cosmic-ray nuclei and synchrotron photons in the jet \citep{Zhang2020,Murase2018}.
The neutron beam escaping from the blazar emission zone could further interact with the external photon field and produce additional neutrinos further away from the \gr\ and radio emission zones. 


\subsection{Upper limits of neutrino flux from TXS-like blazars}

\begin{figure} [h!]
\begin{center}
\includegraphics[width=1.\linewidth]{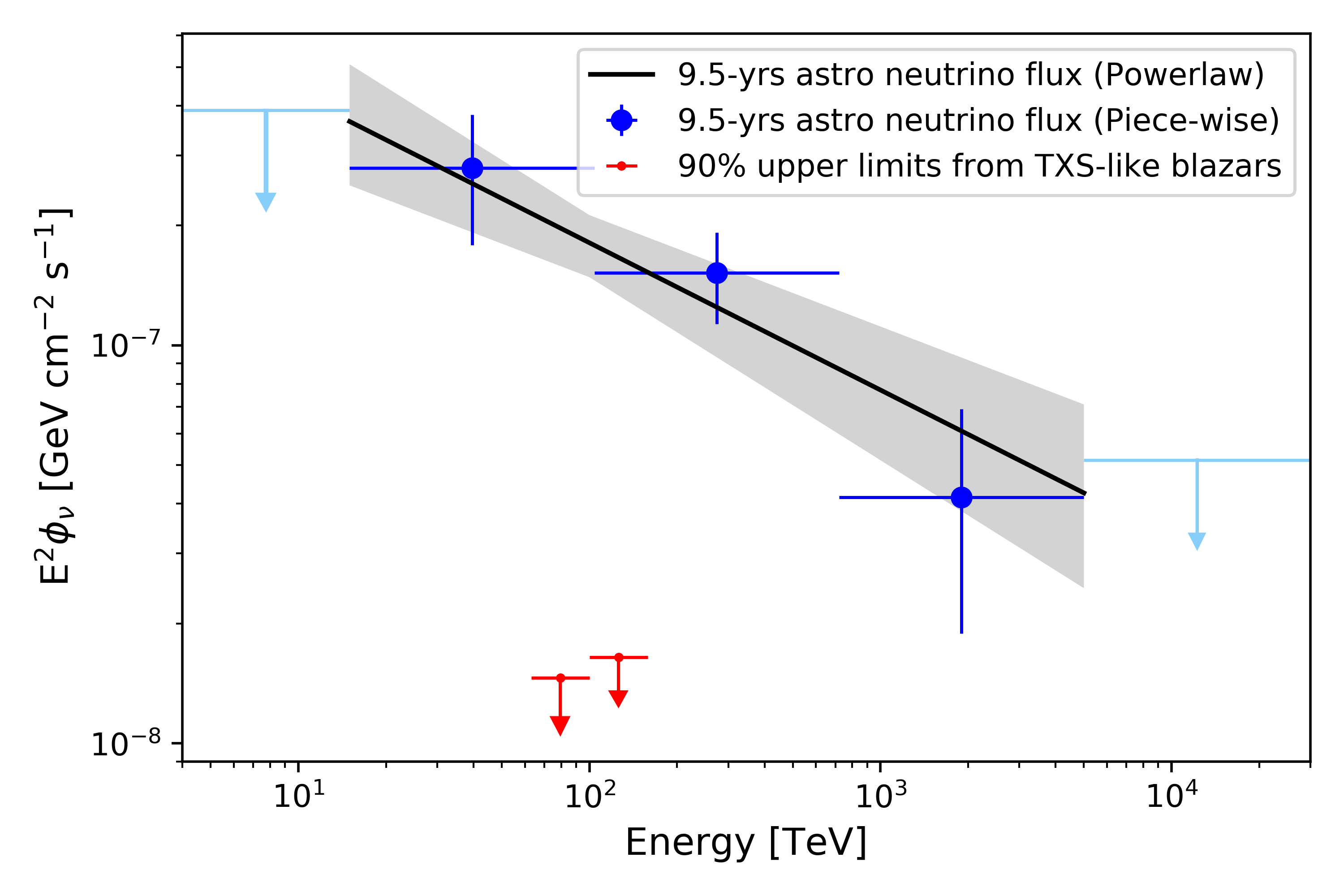}
\end{center}
\caption{Comparison between the neutrino flux upper limits from our selected TXS-like blazars and IceCube 9.5-year astrophysical neutrino flux.}
\label{astro}
\end{figure}

We could estimate the potential neutrino flux from our selected TXS-like blazars given the slight overflow of $\sim 60-150$ TeV track-like neutrinos around them. 
Since the excess is not significant (Figure~\ref{figsim}), the flux we estimated here is an upper limit. 
The expected number of astrophysical neutrinos detected during $\Delta T$ at declination $\delta$ is given by 
\begin{equation}
    N_{\nu}=\Delta T \times \int^{E_{\nu,max}}_{E_{\nu,min}} A_{eff} \phi_{\nu}(E_{\nu}) {\rm d}E_{\nu}
\end{equation}
where $\phi_{\nu}(E_{\nu})$ and $A_{eff}(E_{\nu},\delta)$ are the differential neutrino flux and the effective area, respectively \citep{IceCube2020}. 
We took the $A_{eff}$ from \citet{IceCube2021a} with two energy bins ($4.8<\log_{10}$(E/GeV)$<5$ and   $5<\log_{10}$(E/GeV)$<5.2$) and averaged the effective areas over the declinations, assuming our selected sources are uniformly distributed on the sky. 
$N$ is set to 10 as we found an overflow of $\sim 10$ track-like neutrinos around those TXS-like blazars. 
The obtained upper limits of the differential neutrino flux in the two energy bins are $\phi_{\nu}(100 {\rm TeV})$ equal to $1.459\times{10}^{-18}$ and $1.646\times{10}^{-18}~{\rm GeV}^{-1}~{\rm cm}^{-2}~{s}^{-1}$. 
Assuming neutrino spectral index of $-2$, we evaluated the contribution of our selected TXS-like blazars to the 9.5-year astrophysical muon neutrino flux from \citet{IceCube2021b}. 
Figure~\ref{astro} indicates that those TXS-like blazars contribute with $\lesssim 10\%$ of the diffuse astrophysical neutrinos. 
These results do not conflict with previous stacking limits. 
A scenario where blazars with TXS-like multi-frequency activity dominates the blazar contribution to the IceCube astrophysical neutrino flux (at energy range of $60-150$~TeV) still holds. 

\subsection{Possible reasons for the absence of significant correlations}
Even though we have shown some tiny overflows of neutrino emissions from blazars with certain multi-frequency flaring phases, additional analyses are required to improve and refine our findings. 
Neither the overflow of track-like neutrinos around TXS-like or CFS blazars nor the correlation between infrared flares and IceCube alerts is statistically significant. 
The low significance could be caused by the limited ability of current neutrino observatories to detect weak neutrino signals with astrophysical origin. 
We need more sensitive detectors with improved pointing capability and larger volume to further detect those weak neutrino sources. 
It is known that the diffuse astrophysical neutrino flux detected by IceCube is the aggregate of enormous \citep[probably at least O(50),][]{Brown2015,Murase2016}) faint neutrino emitters. 
Additionally, \citet{Palladino2019} suggested that unresolved BL Lacs with large baryonic loading might be the only sources that dominate the IceCube astrophysical neutrinos with stacking limits accounted for, assuming the neutrino flux is powered by low-luminosity blazars. 
In their hypothesis, resolved high-luminosity BL Lacs or FSRQs can only contribute to a limiting fraction of the observed astrophysical neutrino flux. 

Alternatively, we could not exclude the possibility that the majority of blazars are inefficient neutrino emitters at the sub-TeV to sub-PeV range. 
Blazars are supposed to be more relevant to account for neutrinos with higher energy $\gtrsim 10-100 {\rm PeV}$ (as discussed in Section~\ref{result2}), which constitute only an extremely small fraction ($\lesssim 0.5\%$) of the 10-year track-like neutrinos. 
On the contrary, majority of the IceCube diffuse flux must have come from other sources. 
Tidal Disruption Events (TDE) is one of the possible candidates \citep{Reusch2022}. 
While \citet{Stein2019} constrained the contribution of the TDE to be $\leq 27\%$, \citet{Bartos2021} suggested that probably more than $50\%$ of IceCube astrophysical flux might come from either AGN or TDE at $90\%$ confidence level. 
Furthermore, at least a fraction of $10 \%$ of the astrophysical neutrinos might come from sources other than AGNs and TDEs with $80 \%$ probability \citep{Bartos2021}. 
Those neutrinos might originate from unresolved objects or truly diffuse processes, such as intergalactic shocks and dark matter annihilation, which might also dominate the diffuse \gr\ background below 100 GeV.  

The incompleteness of our multi-frequency and multi-epoch data might also bias our results, especially the large blank from 55600 to 56600 in NEOWISE and AllWISEMEP data. 
This gap in all the light curves of our blazars might lead to a number of sources with TXS-like multi-frequency behaviors not being properly selected because some of the infrared flaring stages fell exactly into the ``blank period", causing a lower significance level with insufficient sources. 
Apart from the gap and large time bin of the WISE data, the ACC and Swift XRT data used in this paper sometimes are taken from target of opportunity observations and are not equally binned. 
Unfortunately, there are no other complete and equally-binned light curves available. 
In the future, with a better sensitivity of next-generation neutrino observatories and a more complete multi-frequency coverage for the light curves, we could further confirm or refute the hypothesis that those blazars are indeed efficient neutrino emitters. 

\section{Conclusions} \label{conclude}
We have performed a series of analyses to search for potential neutrino emissions from multi-frequency catalogs (ACC, WISE, {\it Swift} XRT, and {\it Fermi} 4FGL-DR2) and various blazar samples (3HSP, 5BZCAT, WIBRaLS, and 4LAC), investigating possible correlations with IceCube alerts and 10-year track-like events. 
The associations between IceCube neutrinos and astrophysical sources are thoroughly discussed in the paper by examining the multi-frequency flaring stages of sources within the containment regions of IceCube alerts and blazars with differents types of multi-frequency activity. 
A time-dependent analysis to investigate the coincidence between the bright stages of multi-frequency sources and the arrival times of the neutrino alerts suggests a possible correlation trend between the infrared flares and the IceCube alerts. 
The cross-matching between blazars with various infrared-\gr\ flaring behaviors and 10-year track-like neutrino events shows a small overflow of neutrinos around the blazars with flaring phases highly similar to TXS\,0506+056 (with a significant lag in infrared) or highly simultaneous (CFS sample). 
In a nutshell, we have shown that considering the infrared and \gr\ flaring behaviors, the CFS sample and the TXS-like sources might be the most likely neutrino-source candidates among blazars. 
Our results are consistent with the prediction that blazars might dominate the astrophysical neutrino flux at PeV or higher energies, and consistent with the current limits on the blazars' contribution to the IceCube astrophysical neutrino flux. 
Moreover, the possible neutrino emitting sites from the TXS-like blazars are discussed in detail, accounting for several models from the literature. 


An unstable jet or changing accretion rate would lead to the formation of a jet blob, which expands during its propagating outwards along blazars' jet \citep{Chen2021}. 
The radio outburst might result from this inflating blob region, caused by long-term expansion when the synchrotron radiations transitioning from optically thick to thin. 
This expansion effect does not have to lead to an outburst at higher frequency (e.g., \gr), while the charged particles accelerated inside this large-scale blob may account for high energy neutrino emissions. 
The (plausible) statistical correlation between the radio flaring phases and the arrival of neutrinos \citep{Plavin2020} might be a clue to this scenario (and maybe also include our infrared result, see Figure~\ref{timedelaywibrl}). 
Furthermore, the small perturbations in this blob would bring about flares at higher frequencies with a shorter and non-simultaneous time scale, considering the radio outburst related to the inflating blob. 
This is not in conflict with our results (Figure~\ref{figsim}). 
A model considering a larger scale of accelerated sites, inspired by ({\romannumeral 1}) the statistic correlation found in radio (and/or infrared) and ({\romannumeral 2}) the multi-frequency activity of plausible neutrino blazars,  would be of importance to understand the neutrino emitting mechanism of those blazars. 

Although the putative correlations investigated along this work are not significant, our results suggest that additional studies $-$ with more complete multi-frequency light curves and more sensitive neutrino detectors $-$ are worth to consider. 
In the future, with the next generation of neutrino observatories \citep[like TRIDENT, IceCube-Gen2, P-ONE, KM3NeT, and Baikal-GVD:][]{Ye2022,IceCube-gen22021,Agostini2020,km3net,baikal2018}, we expect to better investigate the neutrino signatures from those sources, hopefully unveiling neutrino/hadronic processes taking place in blazars. 

\begin{acknowledgments}
We thank Ke Fang, Segev BenZvi, Weikang Lin, Neng-Hui Liao, and Ting-Gui Wang for helpful comments and discussions. YLC, DLX, and WLL acknowledge the National Natural Science Foundation of China (NSFC) grant (No.~12175137) on ``Exploring the Extreme Universe with Neutrinos and multi-messengers" and acknowledge the Double First Class start-up fund provided by Shanghai Jiao Tong University. YLC and BA thank the support from China Postdoctoral Science Foundation Funded Project (BR4260008). LC thanks the support of NSFC grant (No.~12173066). We acknowledge the \textit{Centro de Computa\c{c}\~{a}o John David Rogers} (CCJDR) at IFGW Unicamp, Campinas-Brazil, for granting access to the Feynman and Planck Clusters and the IcraNet - IT for the granted access to Joshua Computer Cluster (IT), on which the parallel computation of hundreds of \gr\ light curves were processed with the Fermi Science Tools. 


\end{acknowledgments}

\vspace{5mm}
\facilities{ALMA, WISE, {\it Swift} (XRT), {\it Fermi} (LAT)}






\begin{table*}[h!]
\begin{center}
    \begin{tabular}{cccc}
    \hline
    \hline
    \multicolumn{4}{c}{\textbf{Searching for neutrino emissions (Section~\ref{flares} and \ref{blazarlc})}} \\
    \textbf{Analyses} & \textbf{Source lists} & \textbf{Source selection criteria} & \textbf{Motivation}\\
    \hline
     Temporal correlations & WIBRaLS$^a$ & ${\rm F}_{100~{\rm GHz}} \geq 100~{\rm mJy}$ & Focus on potential flaring sources \\
     with IceCube alerts & {\it Fermi} 4FGL-DR2$^a$ & Variability $\geq 18.48$ & to probe the time correlations \\
     & {\it Swift} XRT$^a$ & \# of detection $\geq 5$ & \\
     & ACC$^a$ & \# of detection $\geq 5$ & \\
    \hline
     Spatial correlations with& FIBS TXS-like sources$^c$ & According to multi-frequency & Explore the correlations taking \\
    IceCube 10-year track events & FIBS CFS sample$^c$ & activity of blazars (section~\ref{txsblazar}) & into account the multi-frequency\\
     &  FIBS NFS sample$^c$ & &behaviors of sources\\
     &  ACC-{\it Fermi} TXS-like sources$^c$ & & \\
    \hline
    \hline
    \multicolumn{4}{c}{\textbf{Multi-frequency properties of blazars (Section~\ref{dcfcheck} and \ref{nusource})} } \\
    \textbf{Analyses} & \textbf{Source lists} & \textbf{Source selection criteria} & \textbf{Motivation} \\
    \hline
    Correlations between & FIBS sources & All FIBS sources & Understand the typical correlations \\
    multi-frequency flares & &  &  of blazars' multi-frequency flares \\
    \hline
     Simultaneity of & FIBS sources & Flaring in infrared and \gr & Compare the simultaneous flaring \\
     multi-frequency flares & FIBS sources & Within IceCube alert region & stages between blazars related to \\
     & FIBS sources & IceCube warm spots$^d$ & IceCube and the whole population \\
    \hline
    \hline
    \end{tabular}
\caption{Analyses, source lists as well as selection criteria and motivation throughout the paper. Detailed descriptions for multi-frequency data and soure lists are written in section~\ref{mwcatalog} and \ref{sample}.} \label{anatable}
\end{center}
\footnotesize{$^a$ cut with $b > |10^{\circ}|$, $\delta < |40^{\circ}|$} \\
\footnotesize{$^b$ {\it Fermi}-Infrared Blazar Sample.} \\
\footnotesize{$^c$ There are four groups of blazars tested with Icecube 10-year track events. One experimental group of selected TXS-like sources (Figure~\ref{txs0506}) from FIBS, two control groups of CFS (Figure~\ref{cfslc}) and NFS samples, and a comparison group of selected TXS-like sources from ACC-{\it Fermi} sub-sample. See section~\ref{txsblazar} for selection details.} \\
\footnotesize{$^d$ Sources with p-value $\leq 0.05$ in IceCube previous point-source analyses. } \\
\end{table*}

\clearpage
\appendix

\section{Tables of selected sources and CFS sources} \label{tabletxscfs}

\begin{table} [h!]
\begin{center}
    \begin{tabular}{cc|cc}
    \hline
     Name & 4FGL & Name  & 4FGL \\
    \hline
  J0112+3208          &  4FGLJ0112.8+3208 & J1246$-$2547           & 4FGLJ1246.7$-$2548 \\
  J0143$-$3200          &  4FGLJ0143.5$-$3156 & J1248+5128           & 4FGLJ1248.7+5127 \\
  J0509+0541          &  4FGLJ0509.4+0542 & 5BZQJ1256$-$0547       & 4FGLJ1256.1$-$0547 \\
  J0646$-$3903          &  4FGLJ0646.7$-$3913 & J1258$-$1800           & 4FGLJ1258.6$-$1759 \\
  3HSPJ064850.5v694522&  4FGLJ0648.4$-$6941 & 3HSPJ125848.0$-$044745 & 4FGLJ1258.7$-$0452 \\
  J0701$-$4634          &  4FGLJ0701.5$-$4634 & 3HSPJ131146.0+395317 & 4FGLJ1311.8+3954 \\
  J0729+6129          &  4FGLJ0728.5+6128 & J1318$-$1235           & 4FGLJ1318.7$-$1234 \\ 
  J0733$-$5445          &  4FGLJ0733.5$-$5445 & J1354$-$1041           & 4FGLJ1354.8$-$1041 \\
  J0808$-$0751          &  4FGLJ0808.2$-$0751 & 3HSPJ150644.5+081400 & 4FGLJ1506.6+0813 \\
  J0816+5739          &  4FGLJ0816.3+5739 & J1546+1817           & 4FGLJ1546.5+1816 \\
  J0916+3854          &  4FGLJ0916.7+3856 & J1616+4632           & 4FGLJ1616.6+4630 \\
  J0921+2335          &  4FGLJ0921.7+2336 & J1716+6836           & 4FGLJ1716.1+6836 \\
  J0930+3503          &  4FGLJ0930.7+3502 & 5BZQJ1753+2848       & 4FGLJ1753.7+2847 \\
  J0944+6135          &  4FGLJ0943.7+6137 & J1806+6949           & 4FGLJ1806.8+6949 \\
  5BZQJ1145$-$6954      &  4FGLJ1145.7$-$6949 & J1844+1614           & 4FGLJ1845.0+1613 \\
  5BZBJ1149+6243      &  4FGLJ1149.2+6246 & J2142$-$0437           & 4FGLJ2142.7$-$0437 \\
    \hline
    \end{tabular}
\caption{Selected sources with infrared and \gr\ light curves similar to that of TXS\,0506+056.}
\label{txstable}
\end{center}
\end{table}

\begin{table} [h!]
\begin{center}
    \begin{tabular}{cc|cc}
    \hline
     Name  & 4FGL & Name & 4FGL \\
    \hline
  3HSPJ003552.6+595004  &    4FGLJ0035.9+5950 & J0958+4725           &     4FGLJ0958.0+4728 \\
  J0050$-$5727            &    4FGLJ0050.0$-$5736 & J1037$-$2823           &     4FGLJ1037.7$-$2822 \\
  3HSPJ005116.6$-$624204  &    4FGLJ0051.2$-$6242 & J1337$-$1257          &     4FGLJ1337.6$-$1257 \\
  J0056-2117          &    4FGLJ0056.4$-$2118 & J1351+0031              &     4FGLJ1351.0+0029 \\
  J0113-3551           &    4FGLJ0113.1$-$3553 & J1422+3223           &     4FGLJ1422.3+3223 \\
  J0113+4948           &    4FGLJ0113.4+4948 & J1424+0434            &     4FGLJ1424.2+0433 \\
  3HSPJ011546.1+251953  &    4FGLJ0115.8+2519 & J1436+2321            &     4FGLJ1436.9+2321 \\
  J0210$-$5101           &    4FGLJ0210.7$-$5101 & J1443+2501            &     4FGLJ1443.9+2501 \\
  J0231$-$4746         &    4FGLJ0231.2$-$4745 & J1457$-$4248            &     4FGLJ1457.3$-$4246 \\
  J0231+1322            &    4FGLJ0231.8+1322 & J1459+7140            &     4FGLJ1459.0+7140 \\
  J0239+0416             &    4FGLJ0239.7+0415 & J154824.38+145702.8   &     4FGLJ1548.3+1456 \\
  J0243+7120            &    4FGLJ0243.4+7119 & J1551$-$1755          &     4FGLJ1550.8$-$1750 \\
  J0253$-$5441            &    4FGLJ0253.2$-$5441 & J1617$-$7717          &     4FGLJ1617.9$-$7718 \\
  3HSPJ033859.6$-$284619  &    4FGLJ0338.9$-$2848 & 3HSPJ172818.6+501310  &     4FGLJ1728.3+5013 \\  
  5BZQJ0354$-$1616        &    4FGLJ0354.7$-$1617 & J1743$-$0350             &     4FGLJ1744.2$-$0353 \\
  J0359+5057             &    4FGLJ0359.6+5057 & 3HSPJ181335.0+314417   &     4FGLJ1813.5+3144 \\
  J0449+6332             &    4FGLJ0449.2+6329 & J1825$-$5230           &     4FGLJ1825.1$-$5231 \\
  J0607+6720             &    4FGLJ0608.0+6721 & 5BZQJ1833$-$2103        &     4FGLJ1833.6$-$2103 \\
  5BZUJ0620$-$2515       &    4FGLJ0620.5$-$2512 & 5BZBJ1925$-$1018       &     4FGLJ1925.1$-$1019 \\
  5BZUJ0622+3326        &    4FGLJ0622.9+3326 & J2000$-$1748           &     4FGLJ2000.9$-$1748 \\
  J0644$-$6712          &    4FGLJ0644.4$-$6712 & J2022$-$4513           &   4FGLJ2022.3$-$4513\\ 
  J071304.54+573810.2   &    4FGLJ0713.0+5738 & 3HSPJ213151.5$-$251558  &     4FGLJ2131.7$-$2515 \\
  5BZQJ0733+0456       &    4FGLJ0733.8+0455 & J2134$-$0153           &     4FGLJ2134.2$-$0154 \\
  J0739+0137           &    4FGLJ0739.2+0137 & 5BZQJ2148+0657       &     4FGLJ2148.6+0652 \\
  J0750+7909           &    4FGLJ0751.0+7908 & J2212+0646           &     4FGLJ2212.8+0647 \\
  5BZQJ0751+3313       &    4FGLJ0752.2+3313 & J2237$-$3921            &     4FGLJ2237.0$-$3921 \\
  3HSPJ075936.1+132117 &    4FGLJ0759.6+1321 & 3HSPJ224017.1$-$524113  &     4FGLJ2240.3$-$5241 \\
  3HSPJ085409.9+440830 &    4FGLJ0854.3+4408 & 5BZQJ2249+2107        &     4FGLJ2248.9+2106 \\
  5BZQJ0912+4126       &    4FGLJ0912.2+4127 & J2311+3425           &     4FGLJ2311.0+3425 \\
  J0923+4125          &    4FGLJ0923.5+4125 & 5BZQJ2323$-$0617        &     4FGLJ2323.6$-$0617 \\
  J0930$-$8534         &    4FGLJ0931.2$-$8533 & J2336$-$4115            &     4FGLJ2336.6$-$4115 \\
    \hline 
    \end{tabular}
\caption{Sources with highly simultaneous infrared and \gr\ flaring light curves.}
\label{txstable}
\end{center}
\end{table}

\clearpage
\section{Multi-frequency light curves for selected sources.} \label{mwlcselect}

\begin{figure}[h!]
\includegraphics[width=0.493\linewidth]{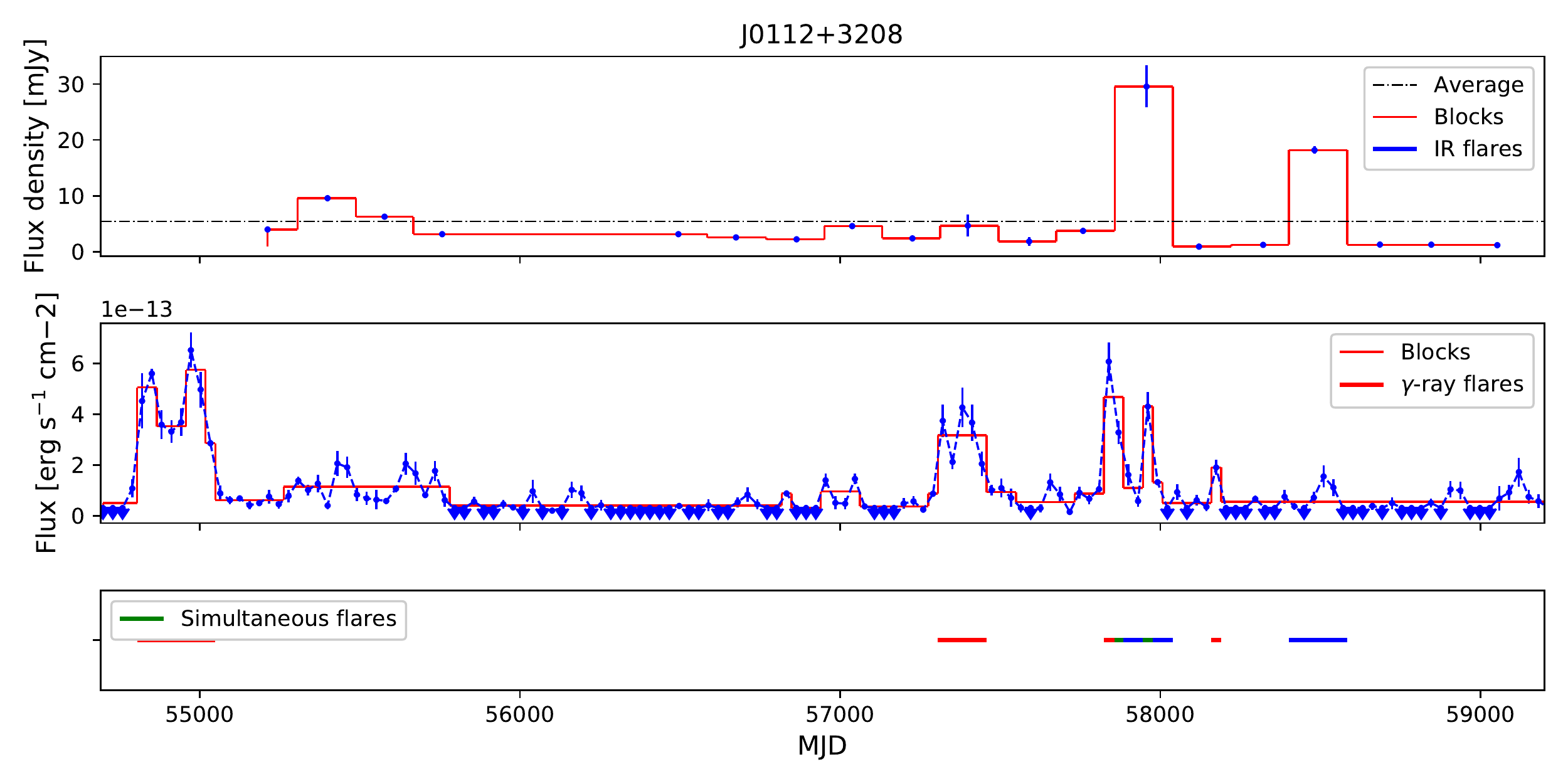}
\includegraphics[width=0.493\linewidth]{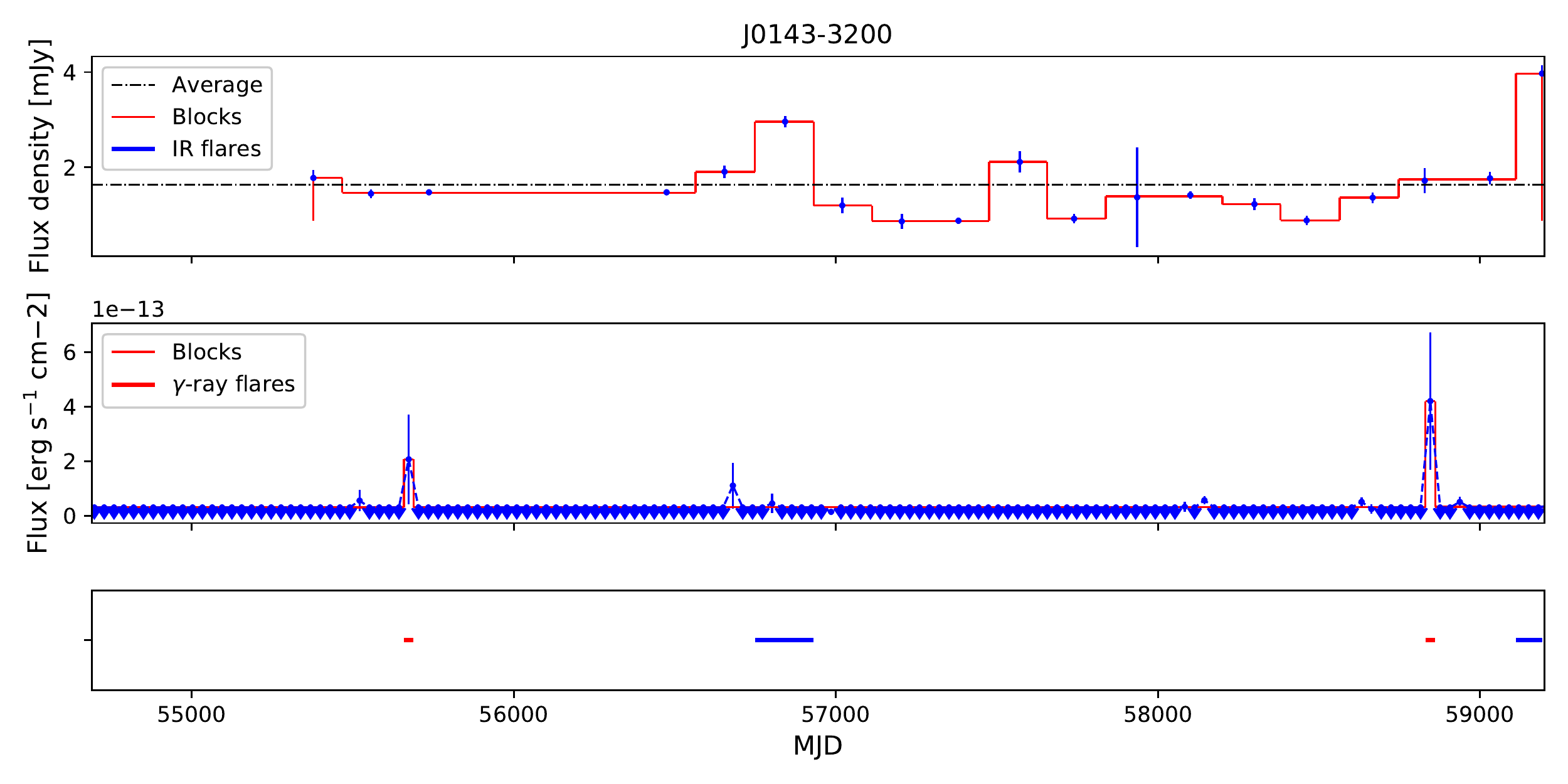}

\includegraphics[width=0.493\linewidth]{LC601.eps}
\includegraphics[width=0.493\linewidth]{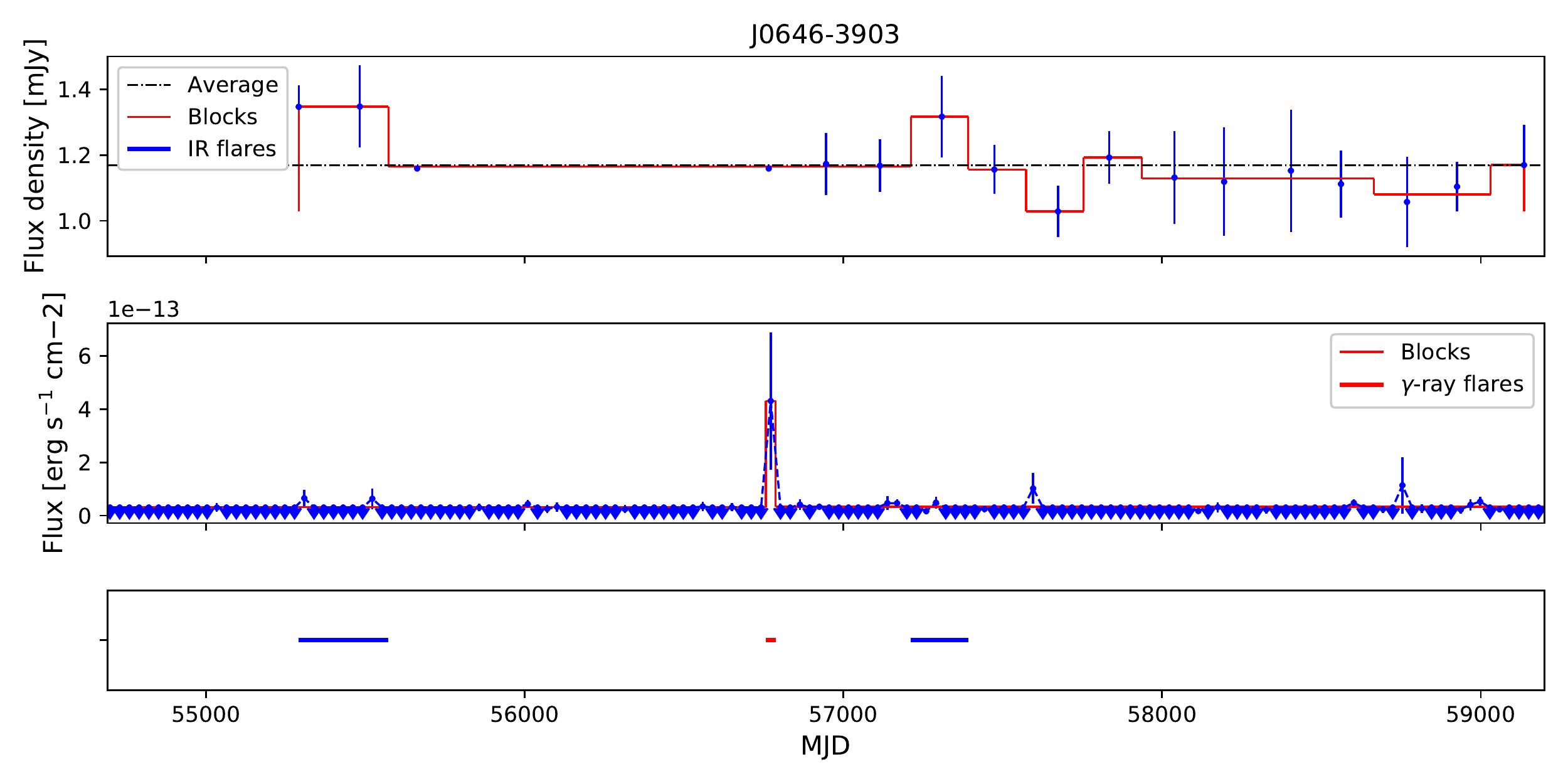}

\includegraphics[width=0.493\linewidth]{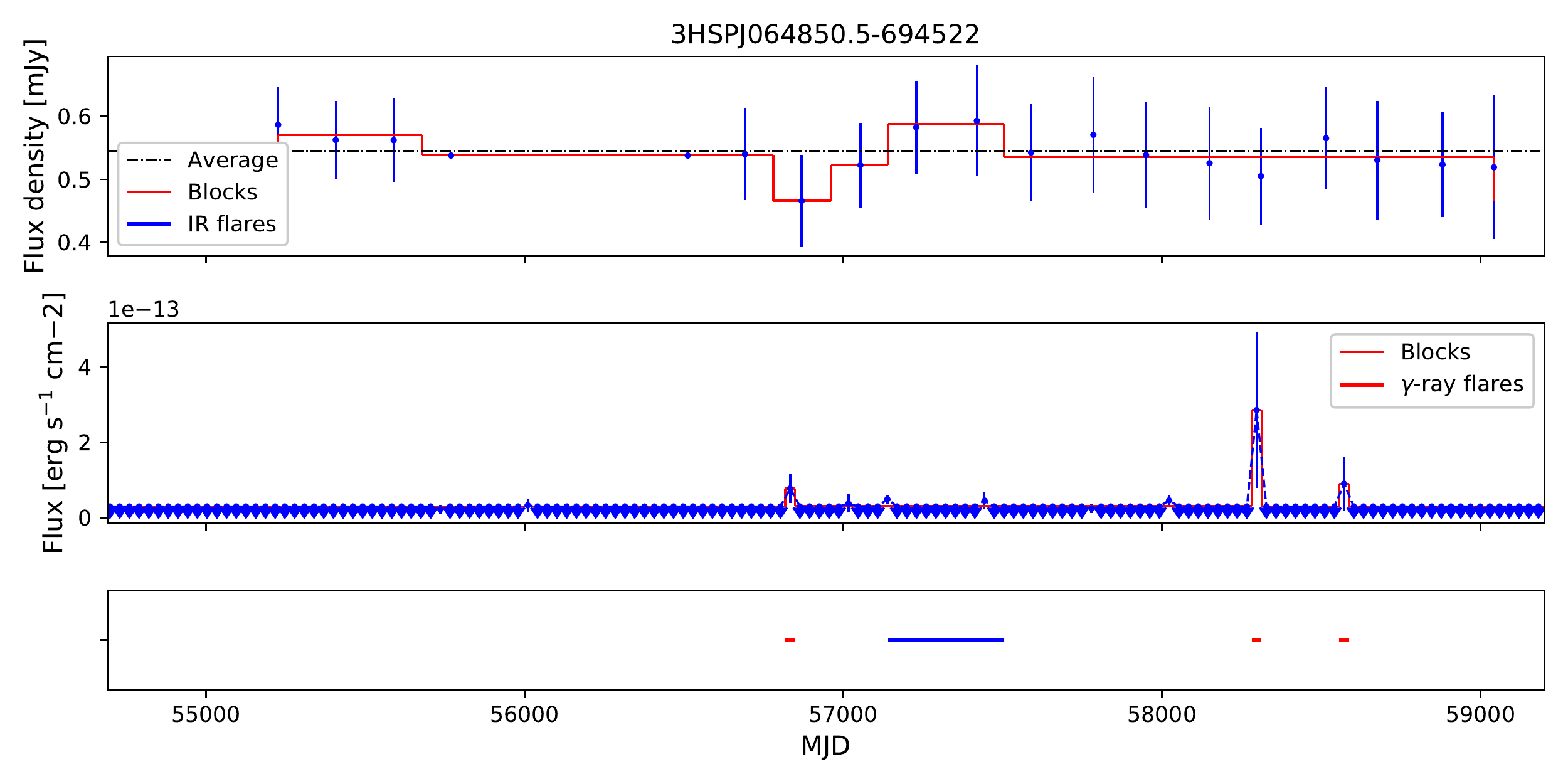}
\includegraphics[width=0.493\linewidth]{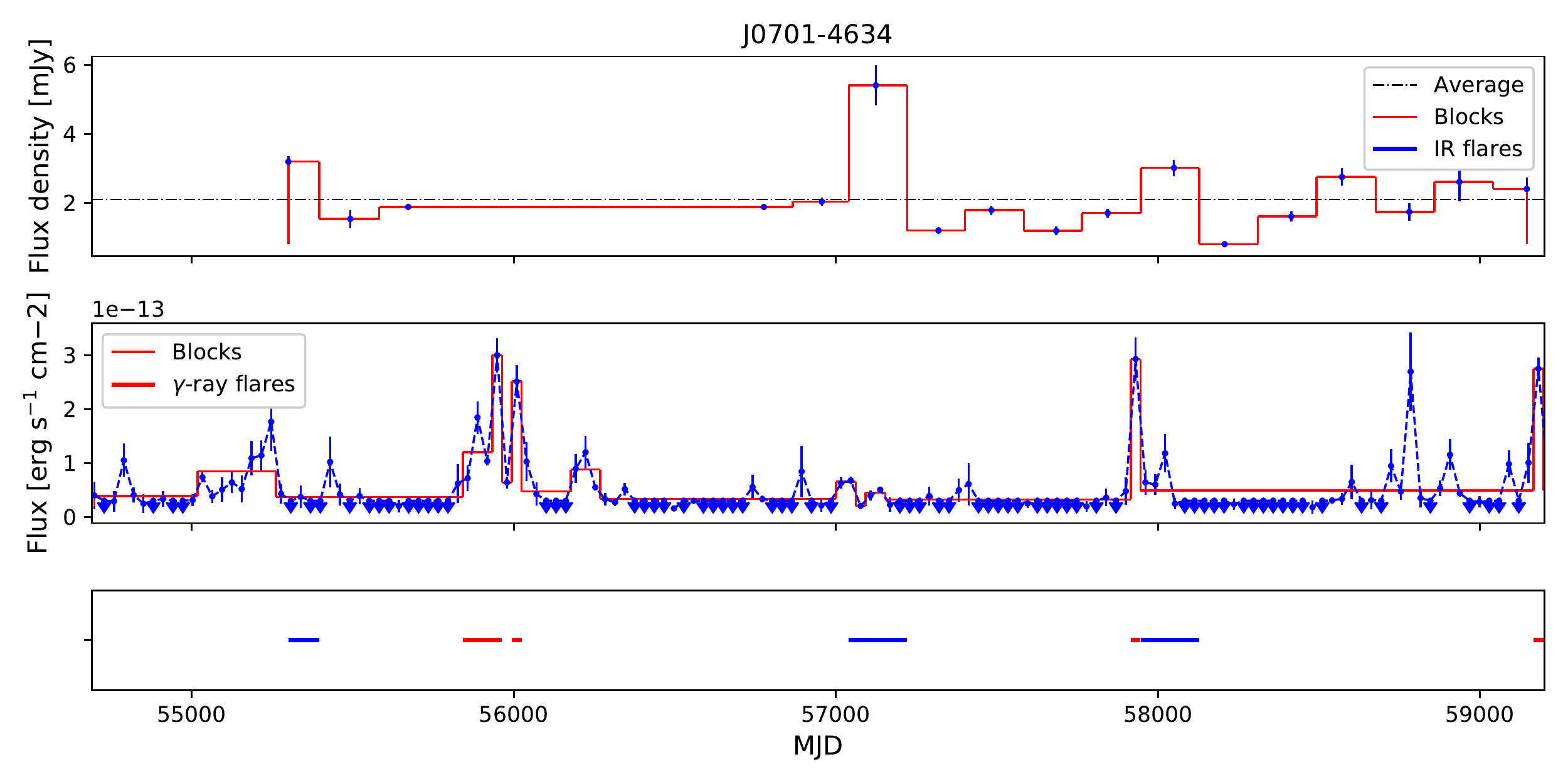}

\includegraphics[width=0.493\linewidth]{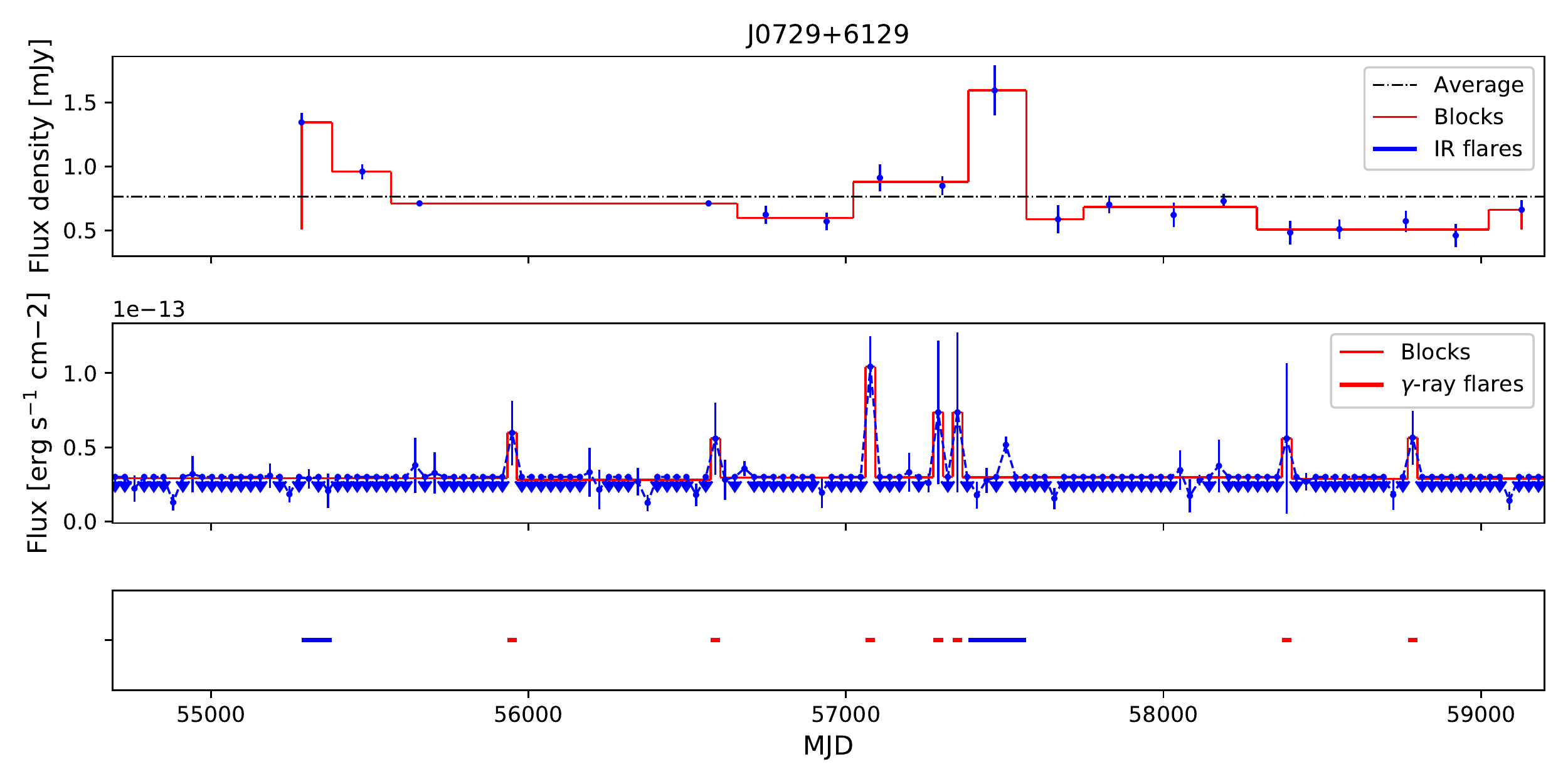}
\includegraphics[width=0.493\linewidth]{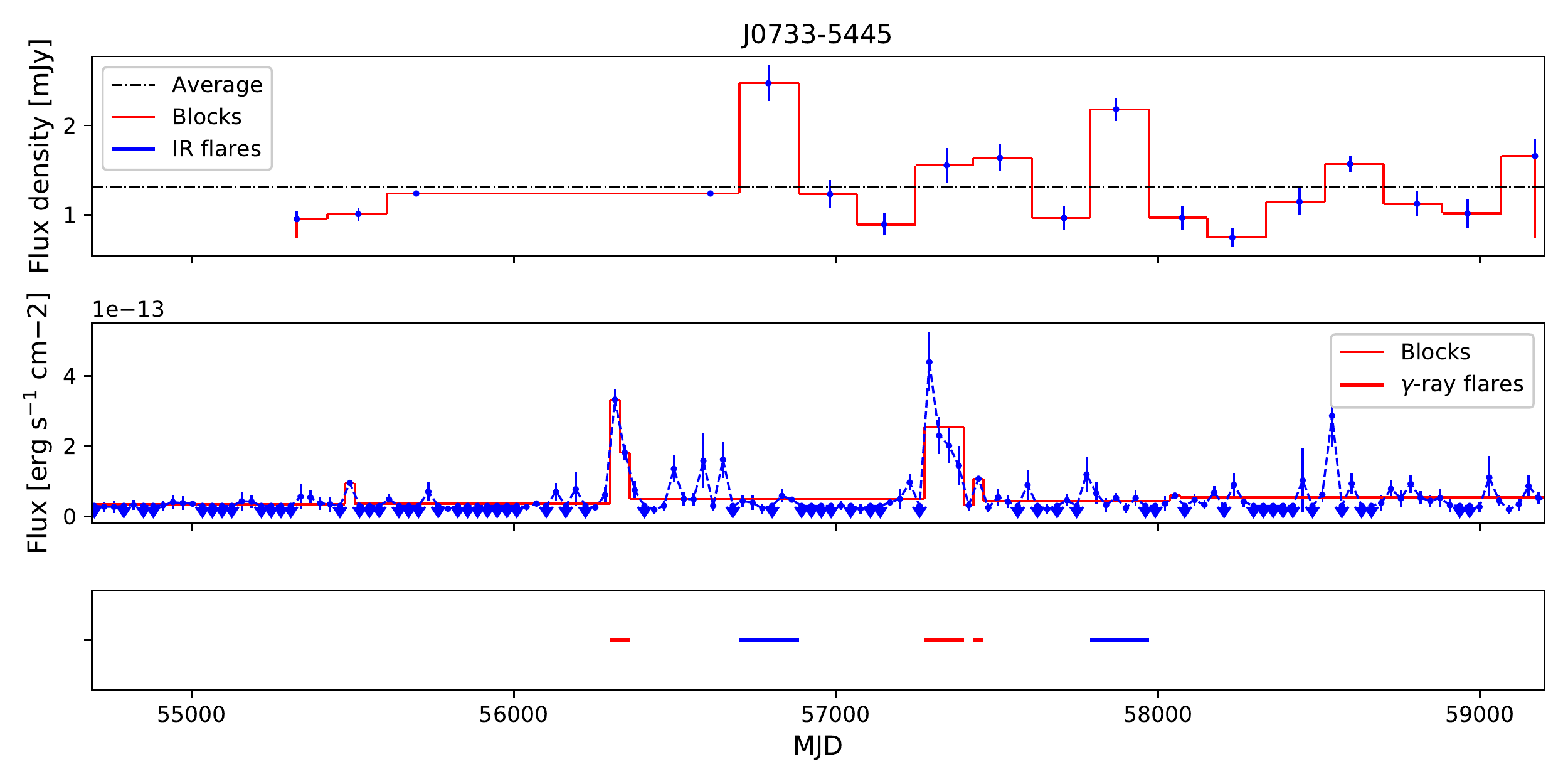}
\caption{Light curves of selected TXS-like sources. The upper panel is the infrared light curve from WISE multi-epoch data, and the middle panel is the {\it Fermi} \gr\ light curve analyzed with Fermi Science Tool. The lower panel represents the flaring periods in infrared (blue thick lines) and \gr\ (red thick lines) as well as the simultaneous flaring stages (green lines). Black lines represent the arrival time of the neutrino alerts.}
\label{lightcurve}
\end{figure}

\begin{figure}[h!]
\includegraphics[width=0.493\linewidth]{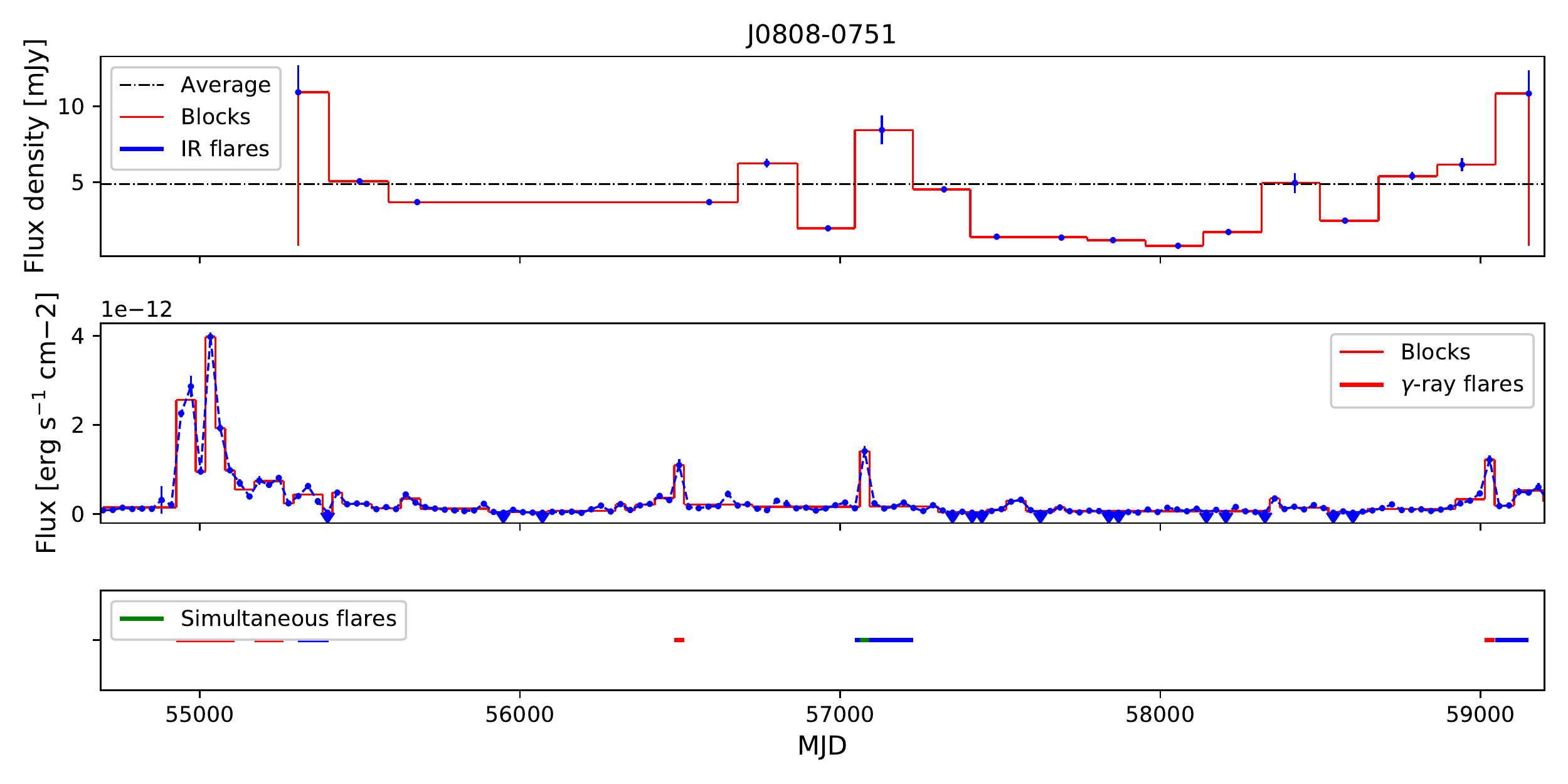}
\includegraphics[width=0.493\linewidth]{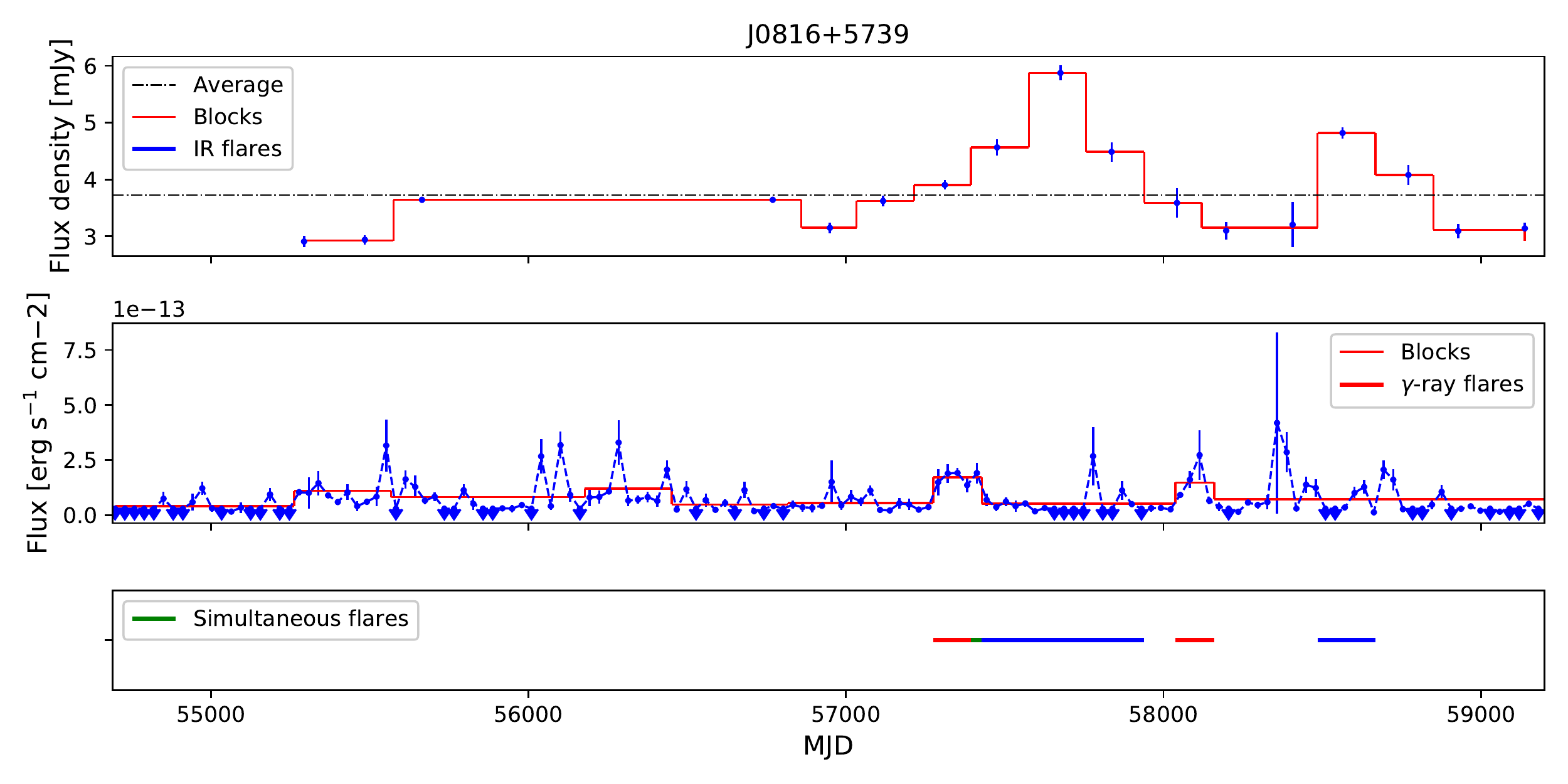}

\includegraphics[width=0.493\linewidth]{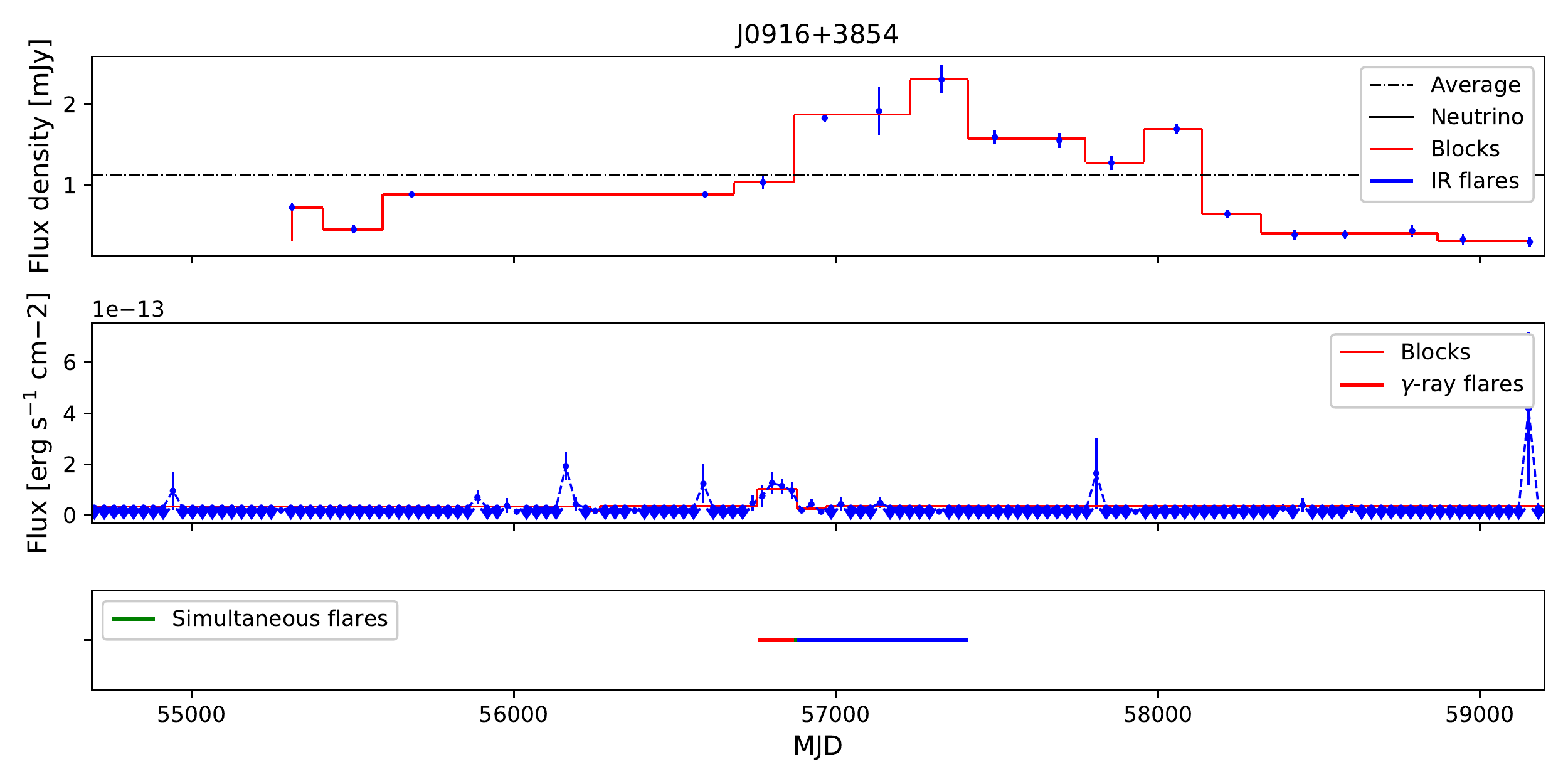}
\includegraphics[width=0.493\linewidth]{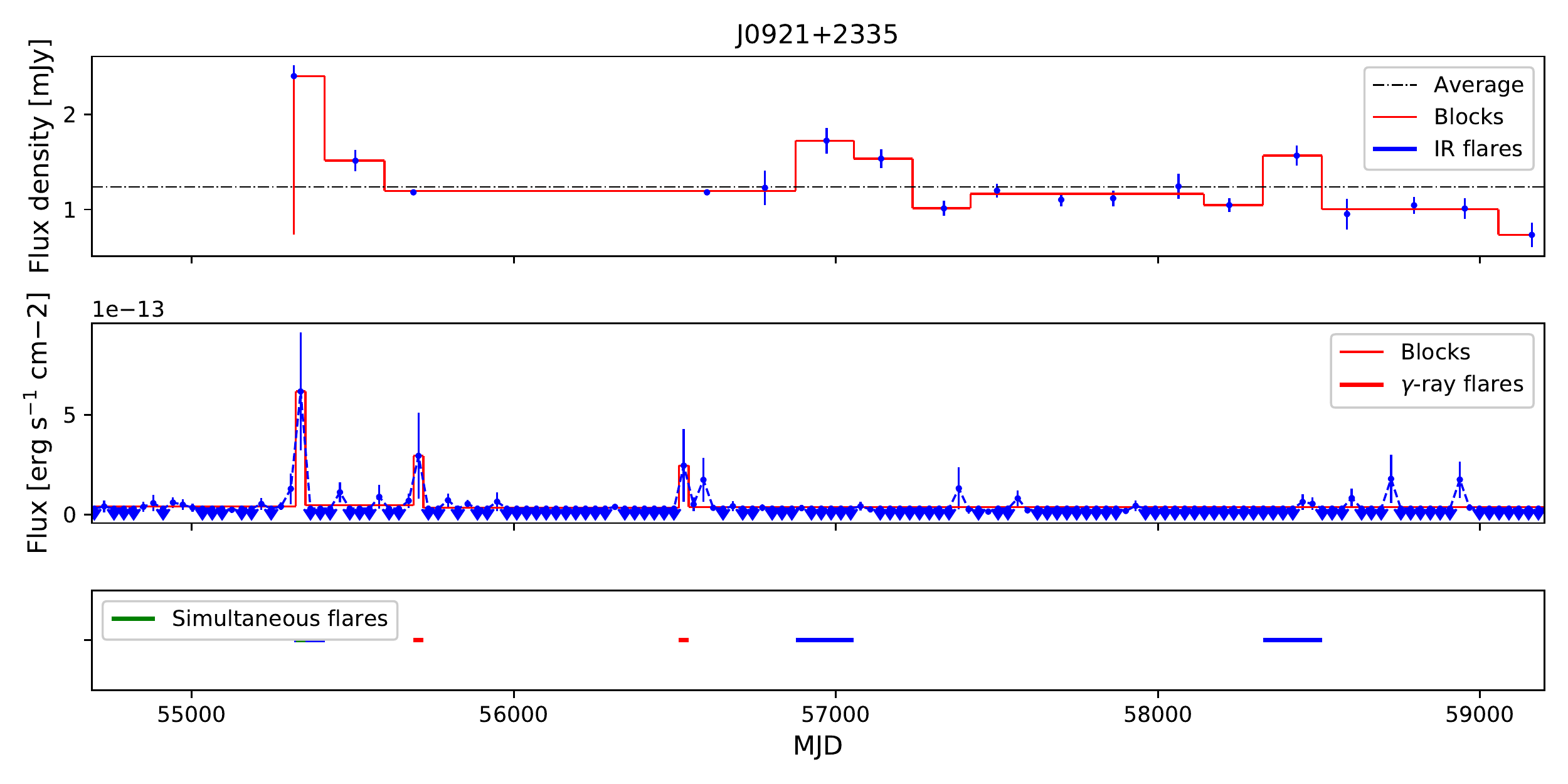}

\includegraphics[width=0.493\linewidth]{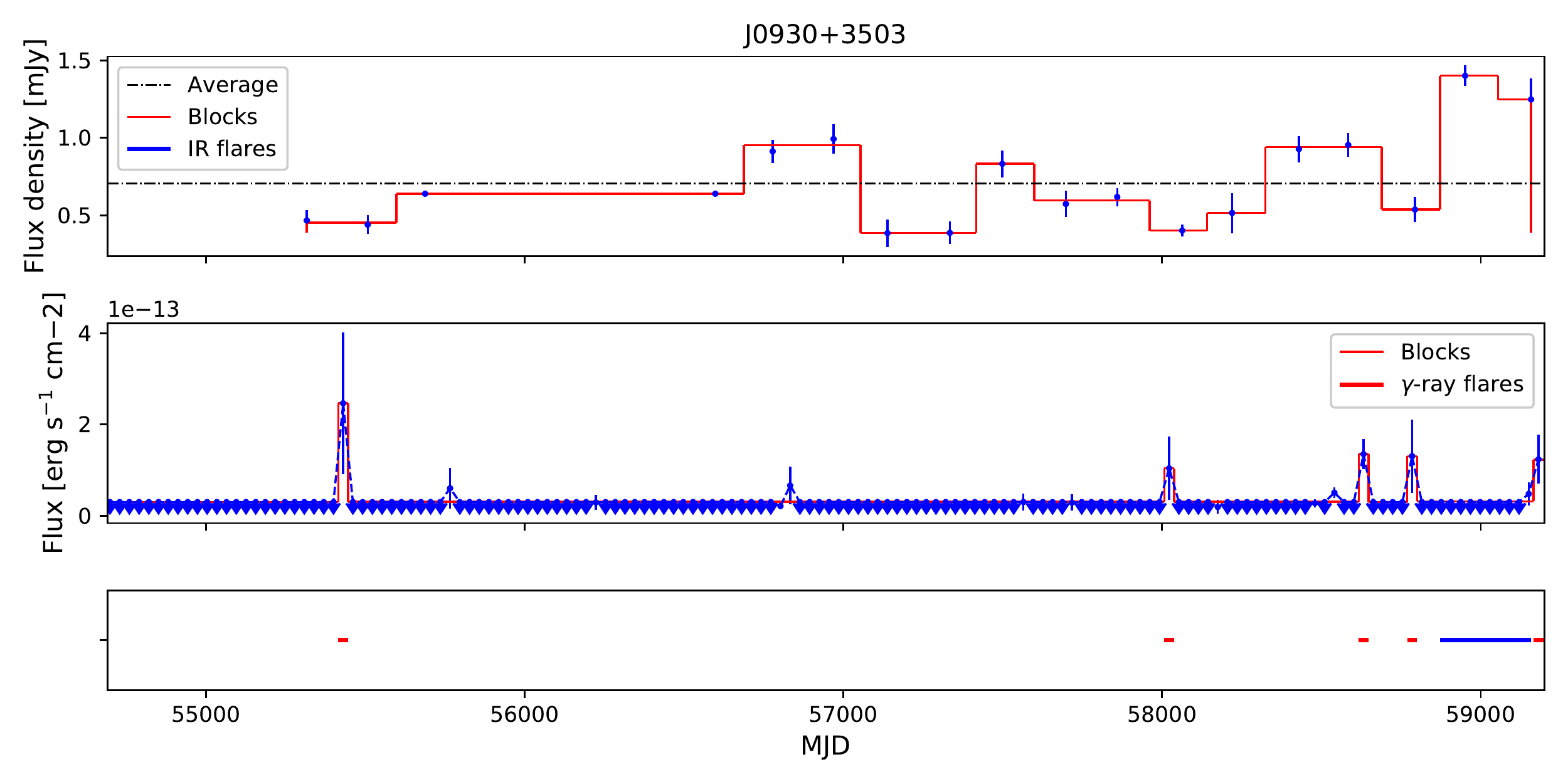}
\includegraphics[width=0.493\linewidth]{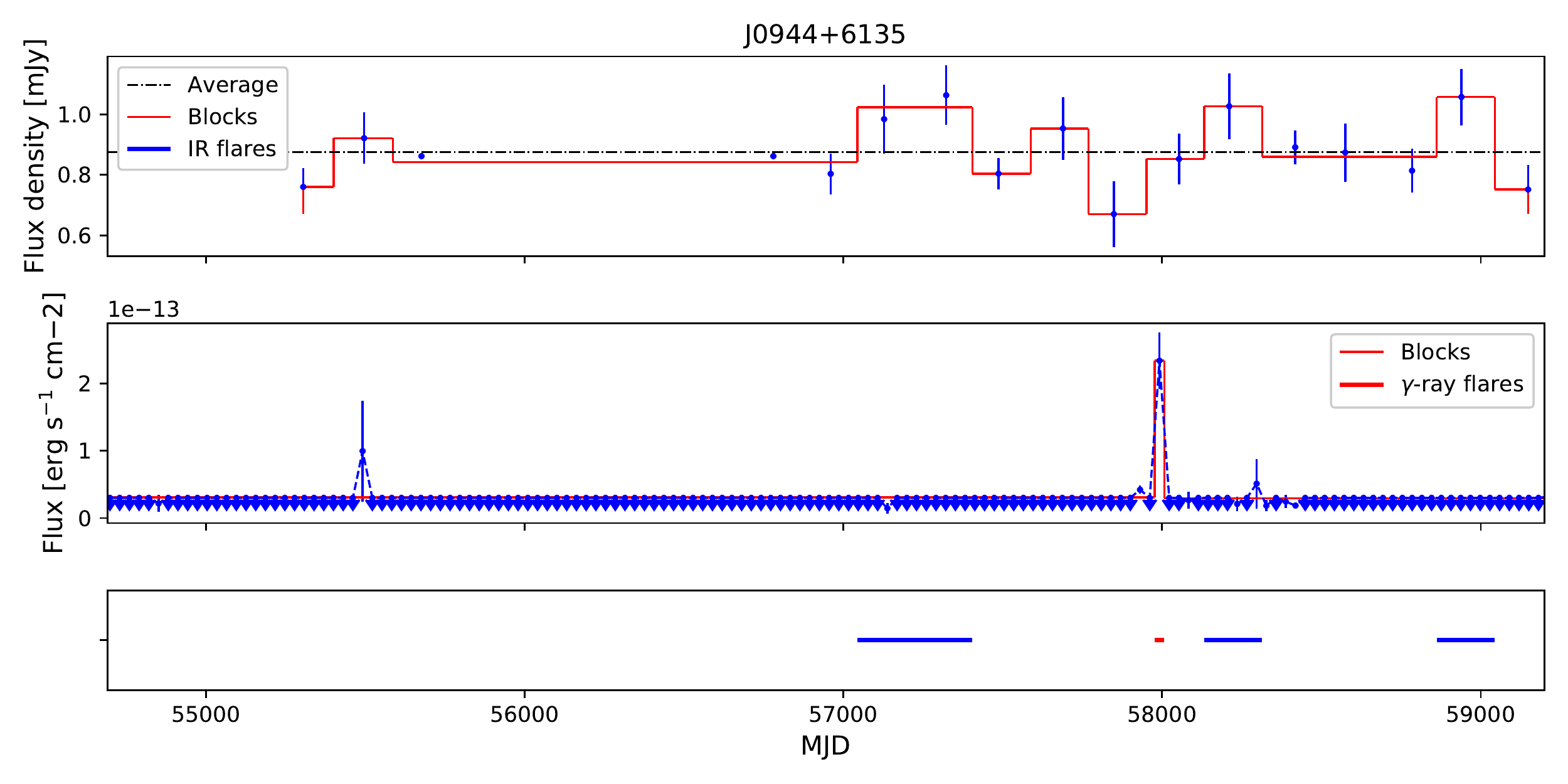}

\includegraphics[width=0.493\linewidth]{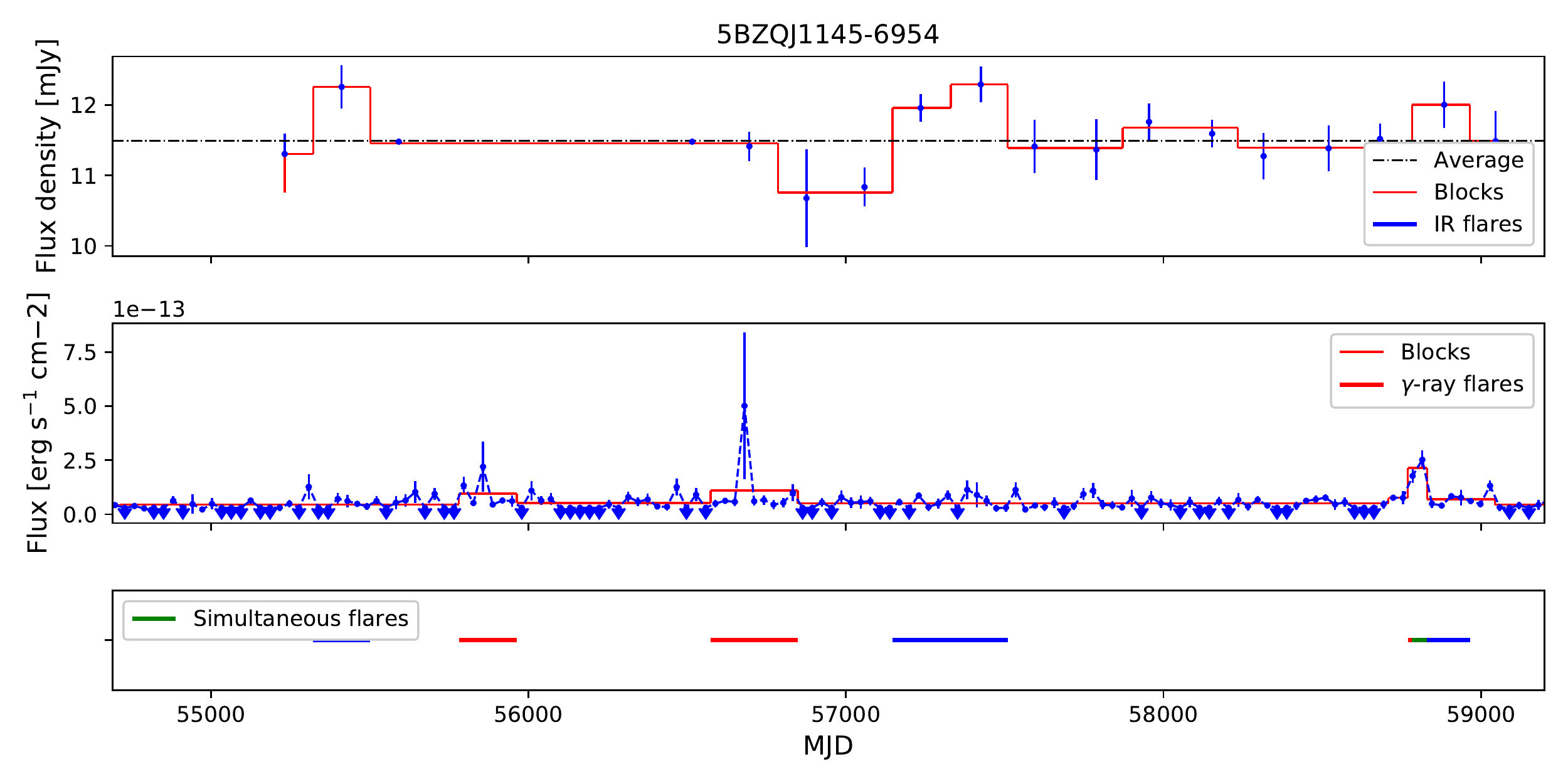}
\includegraphics[width=0.493\linewidth]{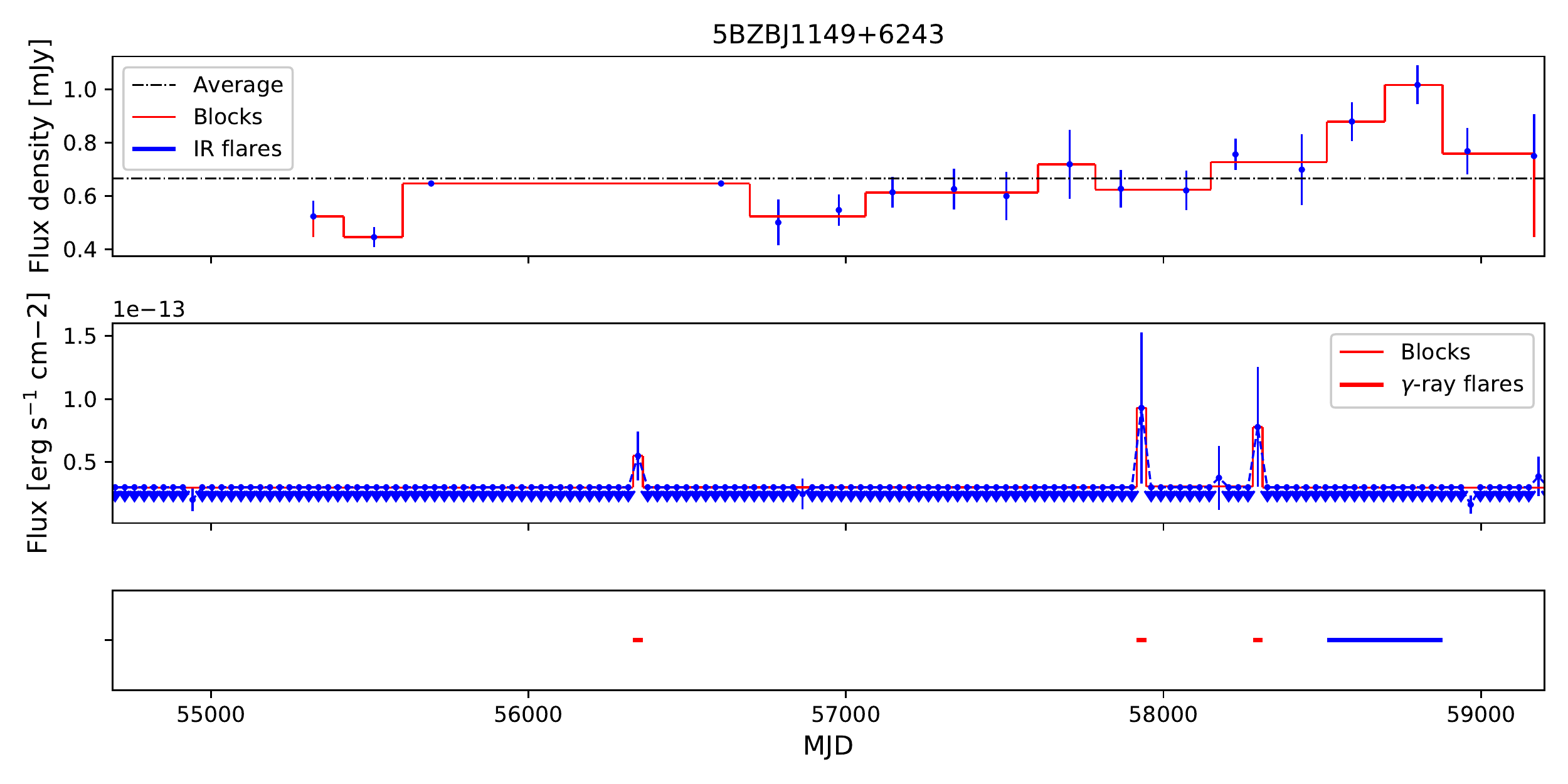} \par
Figure~\ref{lightcurve} continued. Light curves of selected TXS-like sources.  The upper panel is the infrared light curve from WISE multi-epoch data, and the middle panel is the {\it Fermi} \gr\ light curve analyzed with Fermi Science Tool. The lower panel represents the flaring periods in infrared (blue thick lines) and \gr\ (red thick lines) as well as the simultaneous flaring stages (green lines). Black lines represent the arrival time of the neutrino alerts.
\end{figure}

\begin{figure}[h!]
\includegraphics[width=0.493\linewidth]{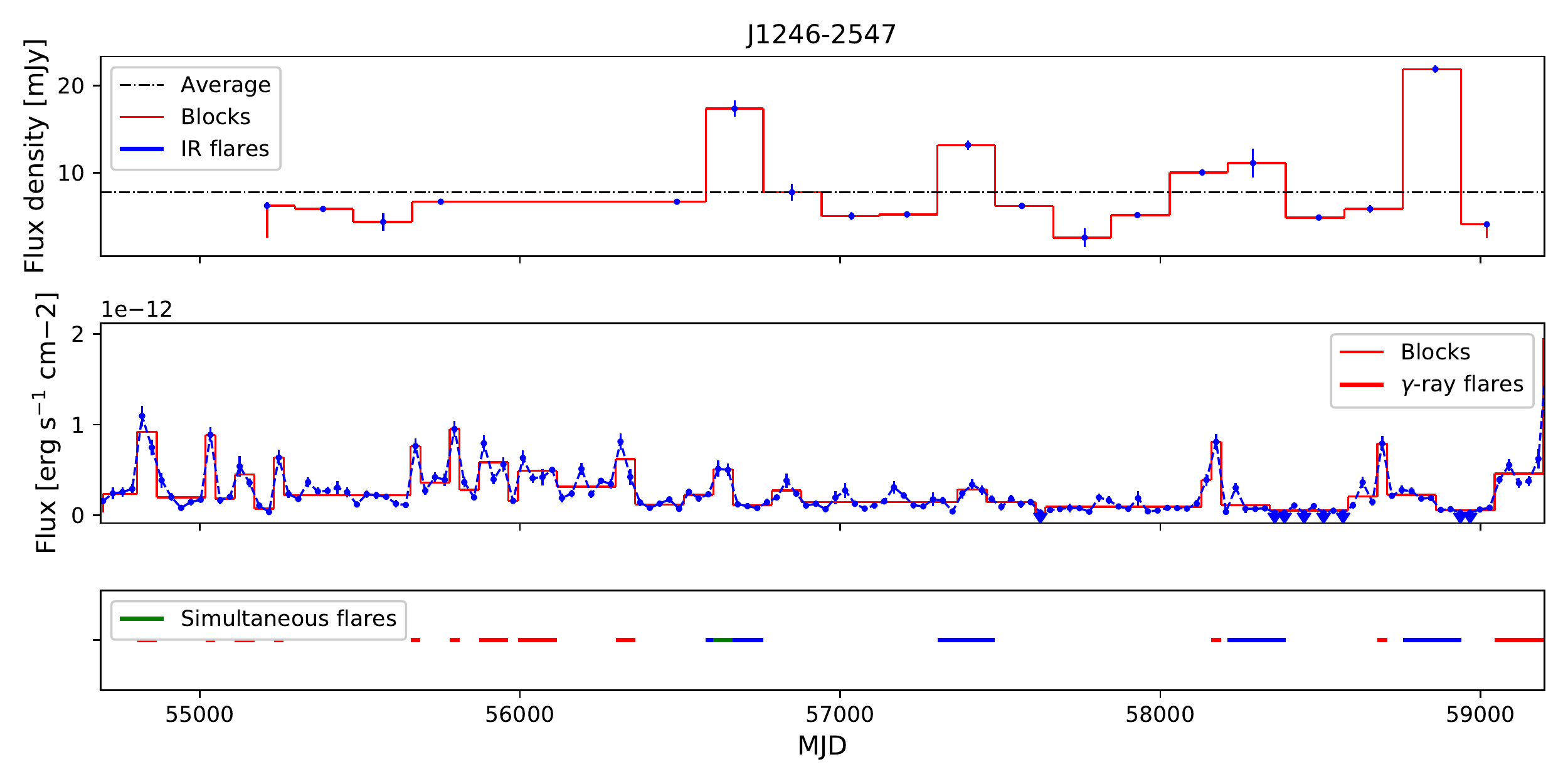}
\includegraphics[width=0.493\linewidth]{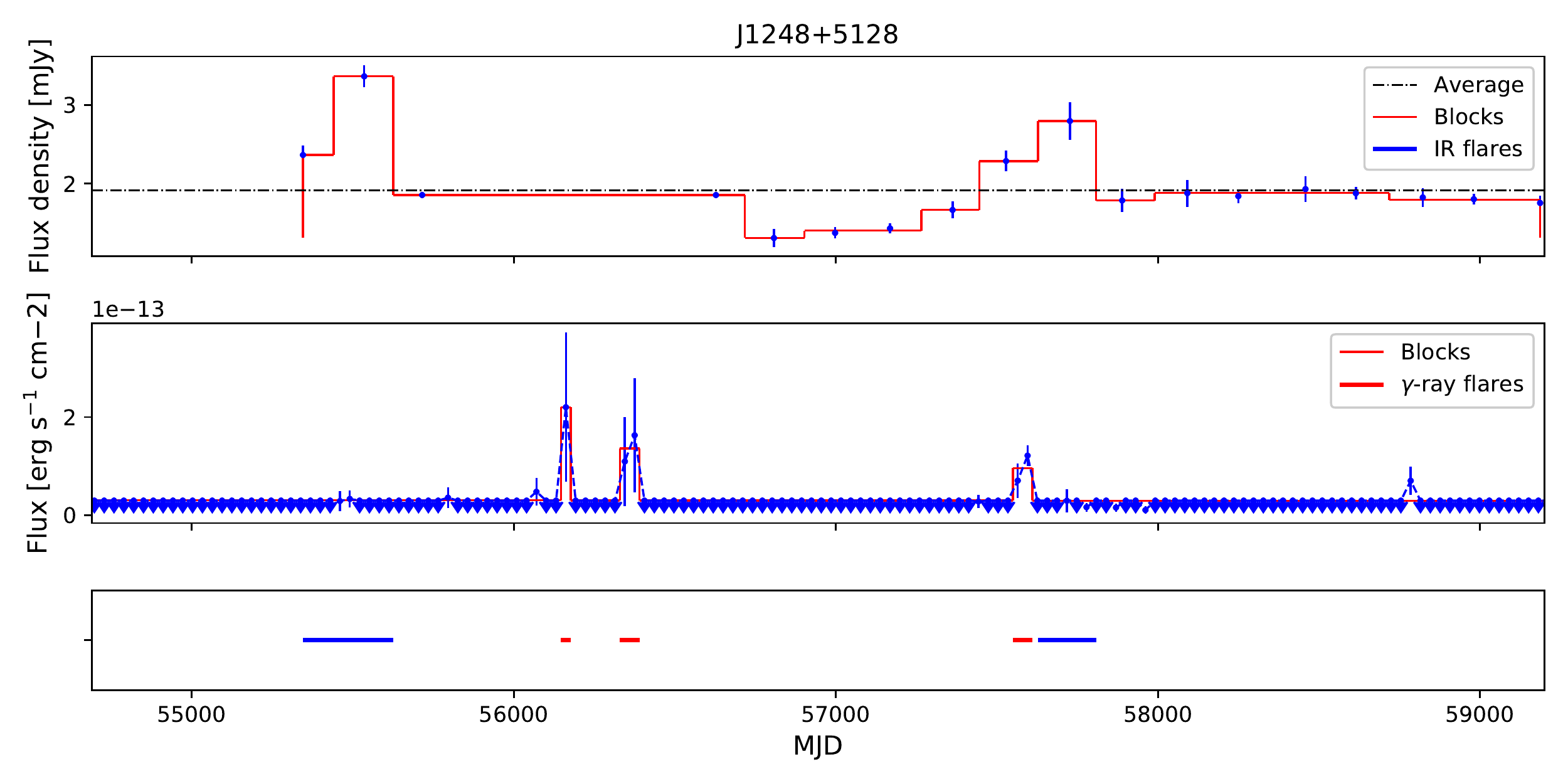}

\includegraphics[width=0.493\linewidth]{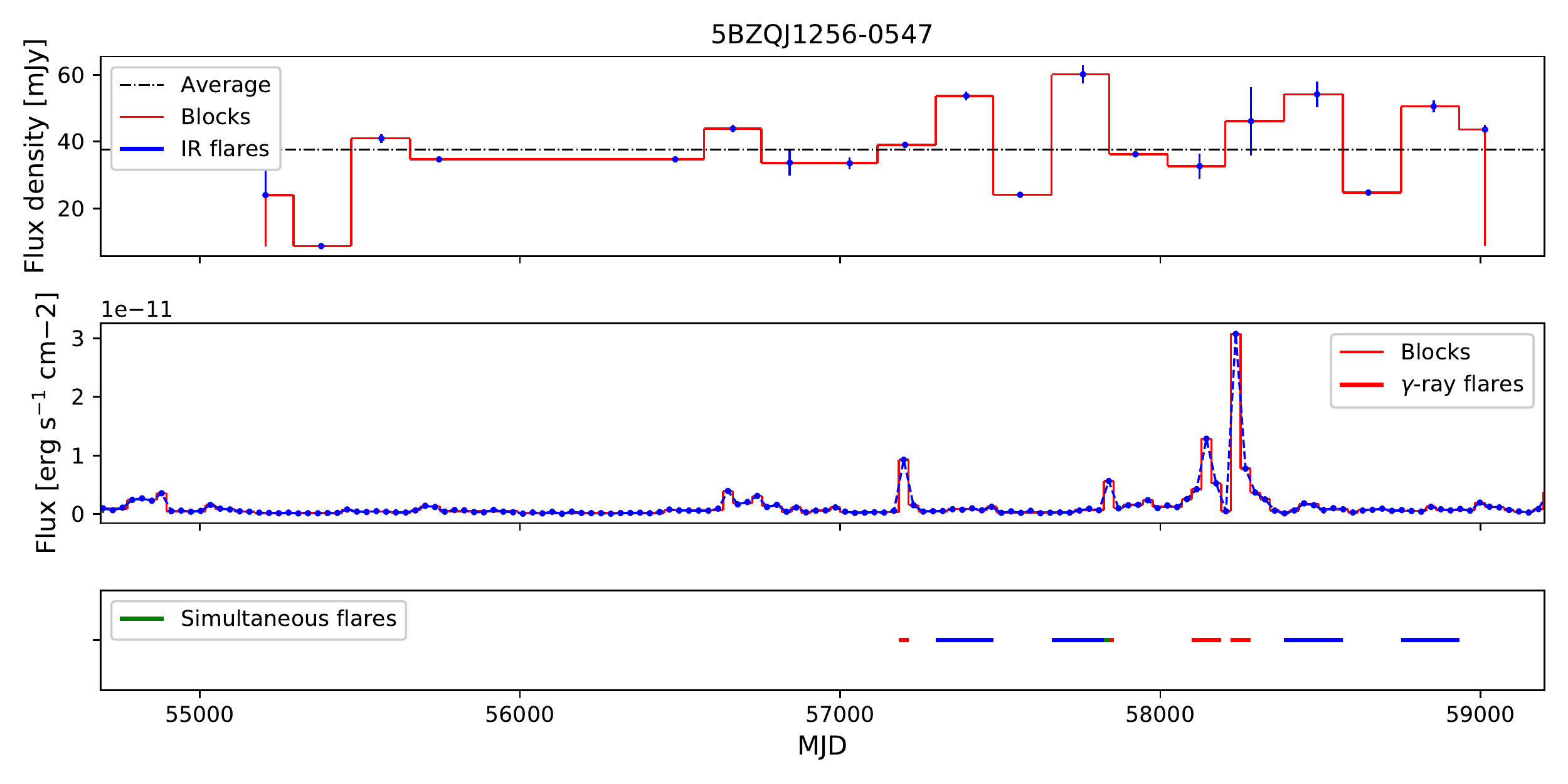}
\includegraphics[width=0.493\linewidth]{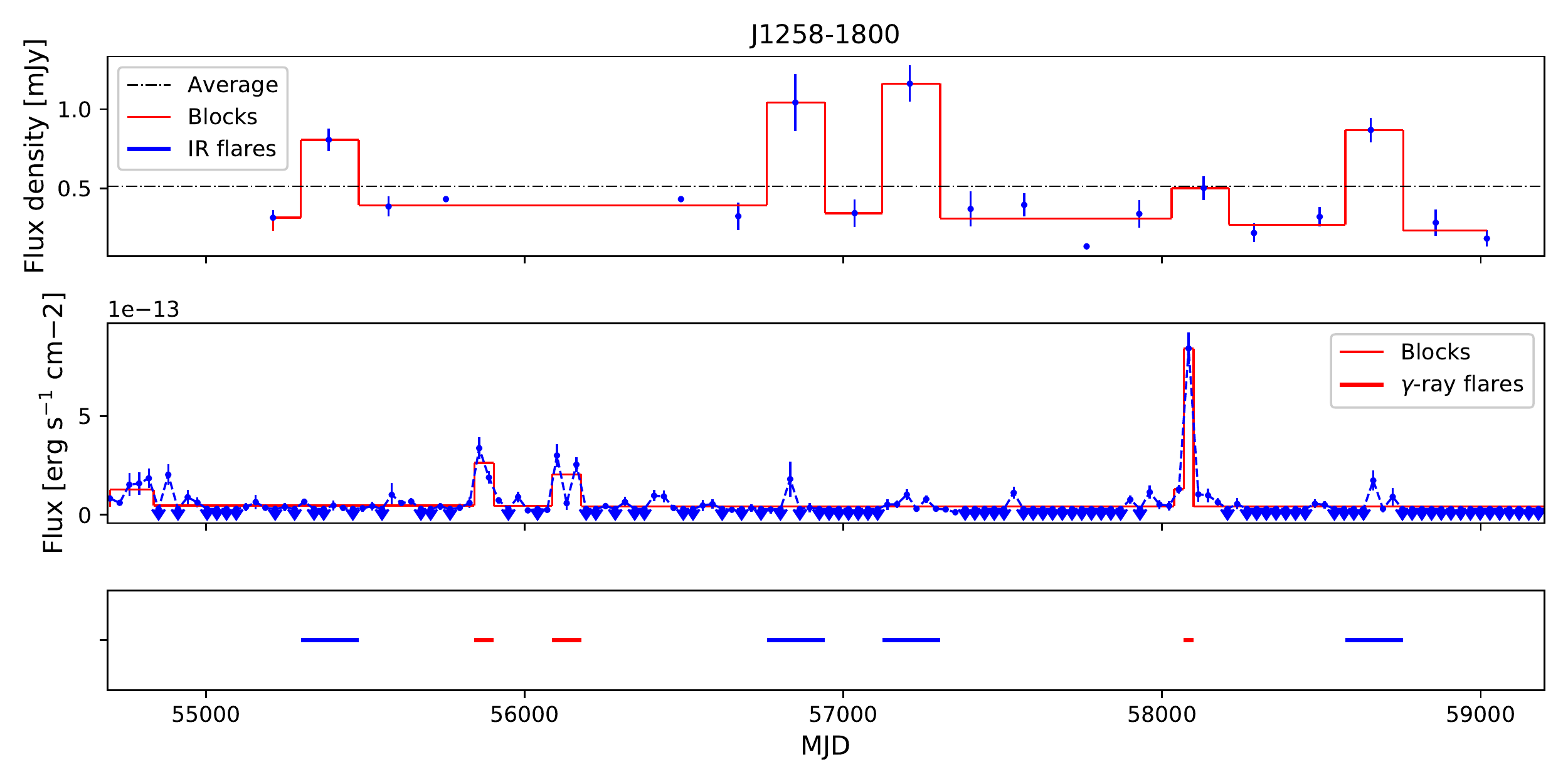}

\includegraphics[width=0.493\linewidth]{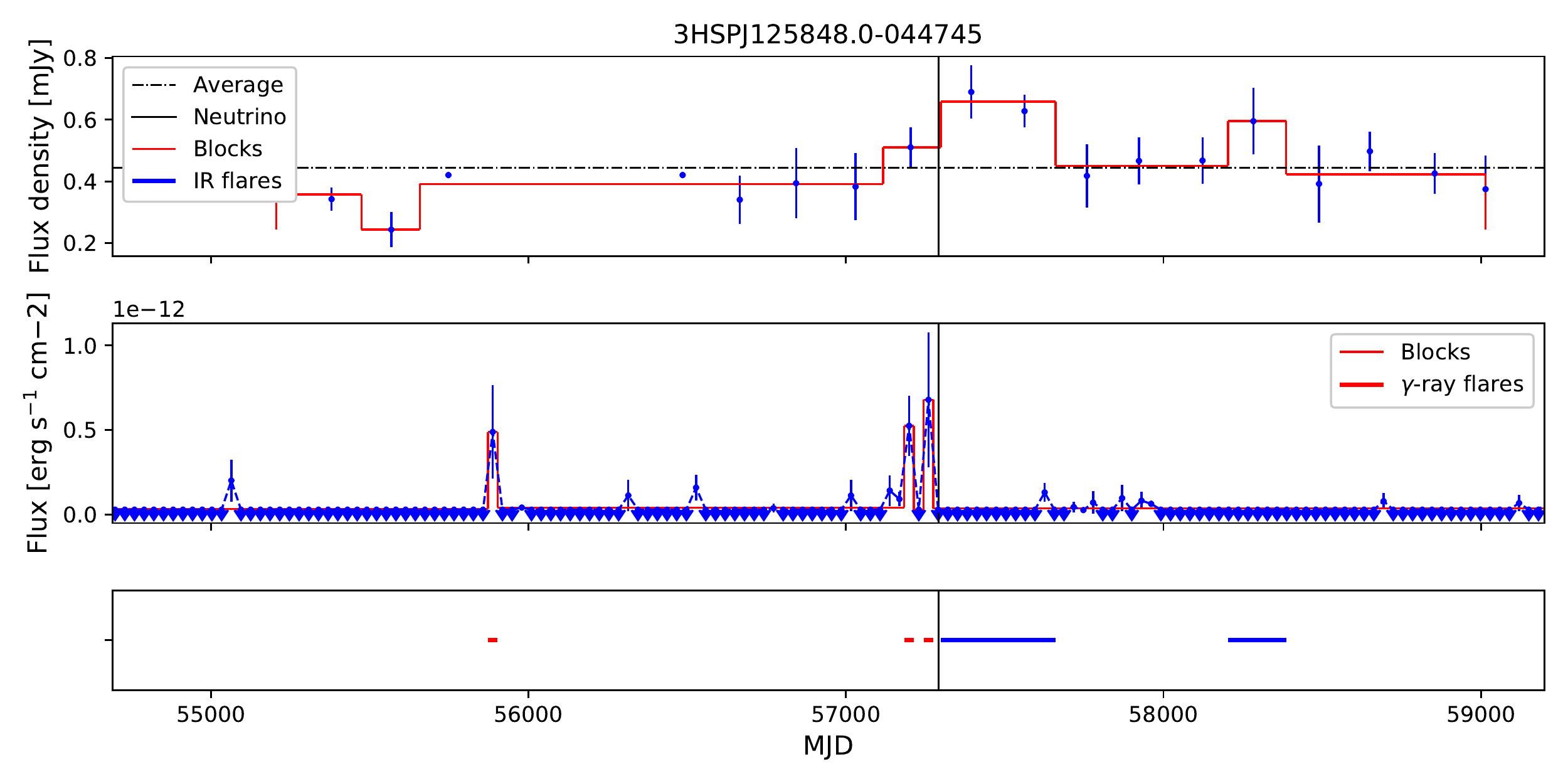}
\includegraphics[width=0.493\linewidth]{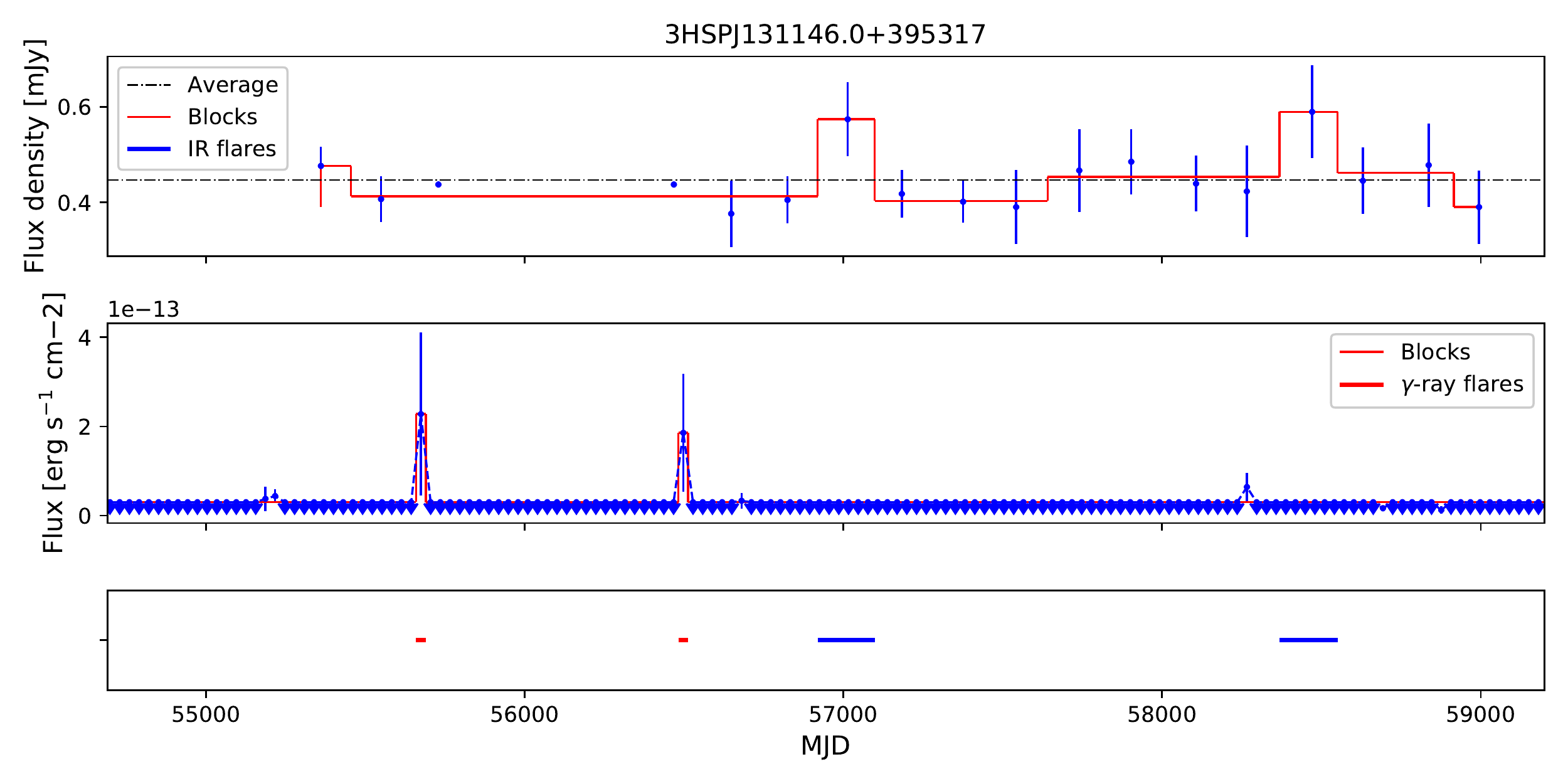}

\includegraphics[width=0.493\linewidth]{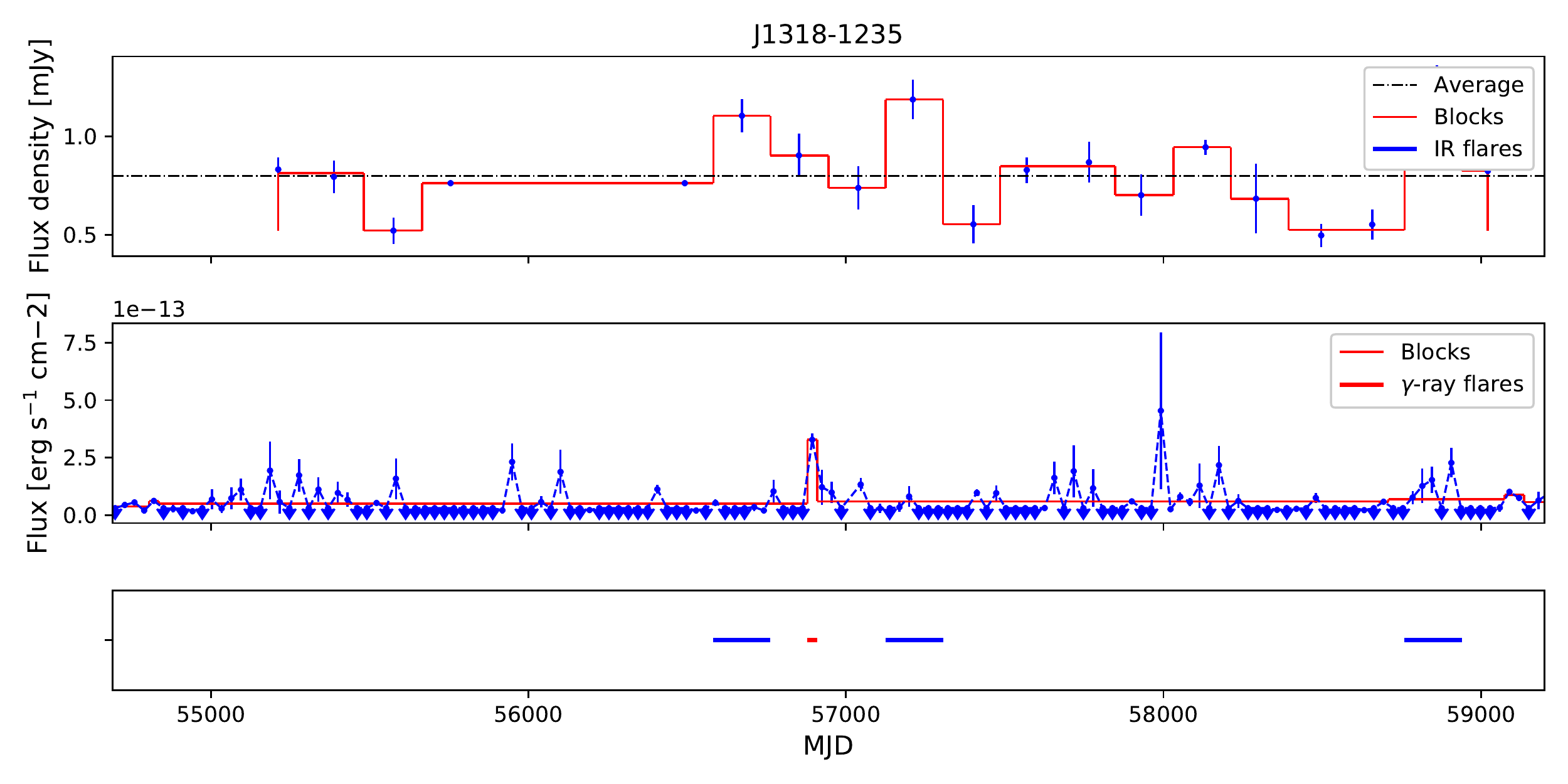}
\includegraphics[width=0.493\linewidth]{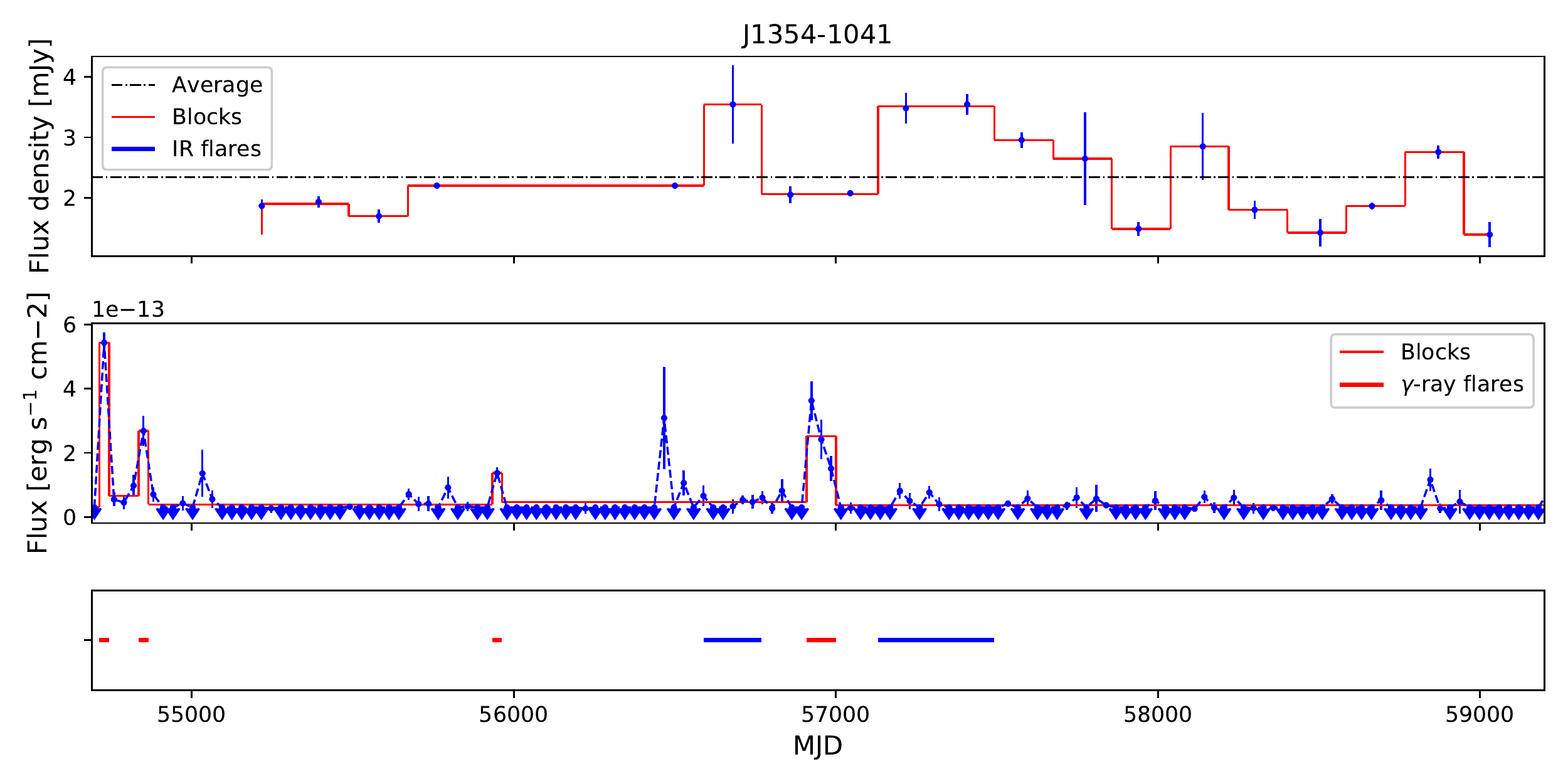} \par
Figure~\ref{lightcurve} continued. Light curves of selected TXS-like sources.  The upper panel is the infrared light curve from WISE multi-epoch data, and the middle panel is the {\it Fermi} \gr\ light curve analyzed with Fermi Science Tool. The lower panel represents the flaring periods in infrared (blue thick lines) and \gr\ (red thick lines) as well as the simultaneous flaring stages (green lines). Black lines represent the arrival time of the neutrino alerts.
\end{figure}

\begin{figure}[h!]
\includegraphics[width=0.493\linewidth]{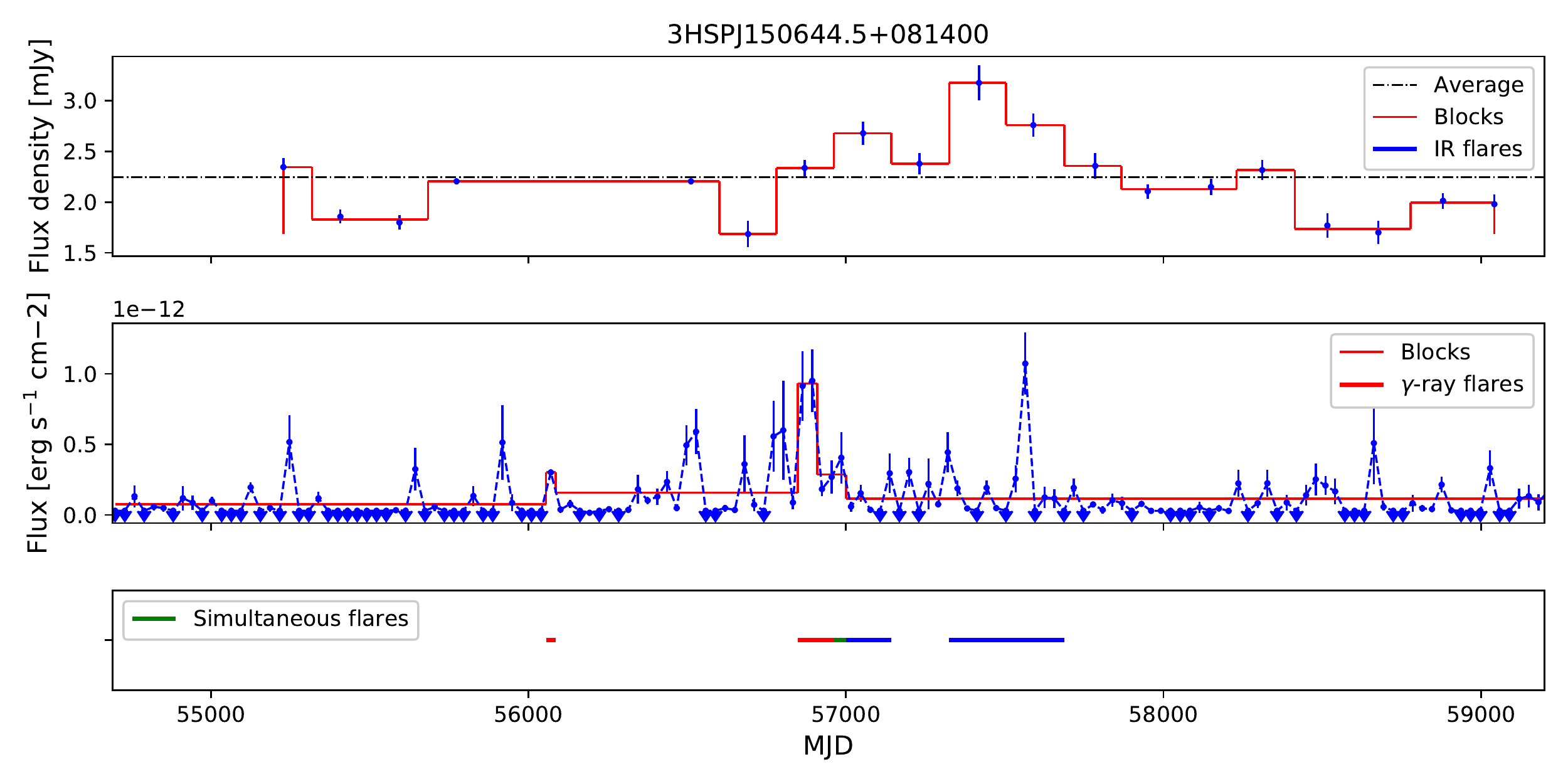}
\includegraphics[width=0.493\linewidth]{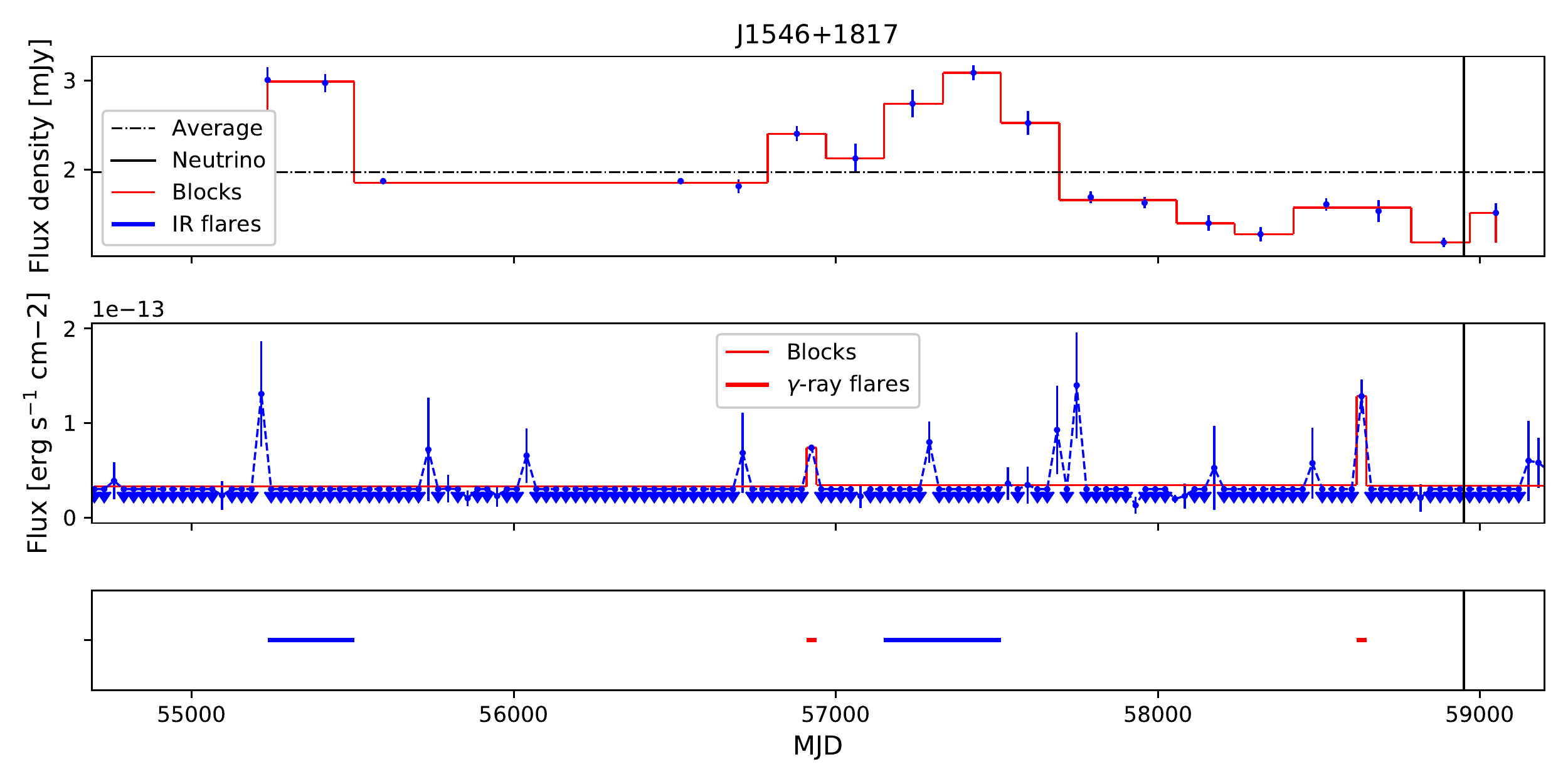}

\includegraphics[width=0.493\linewidth]{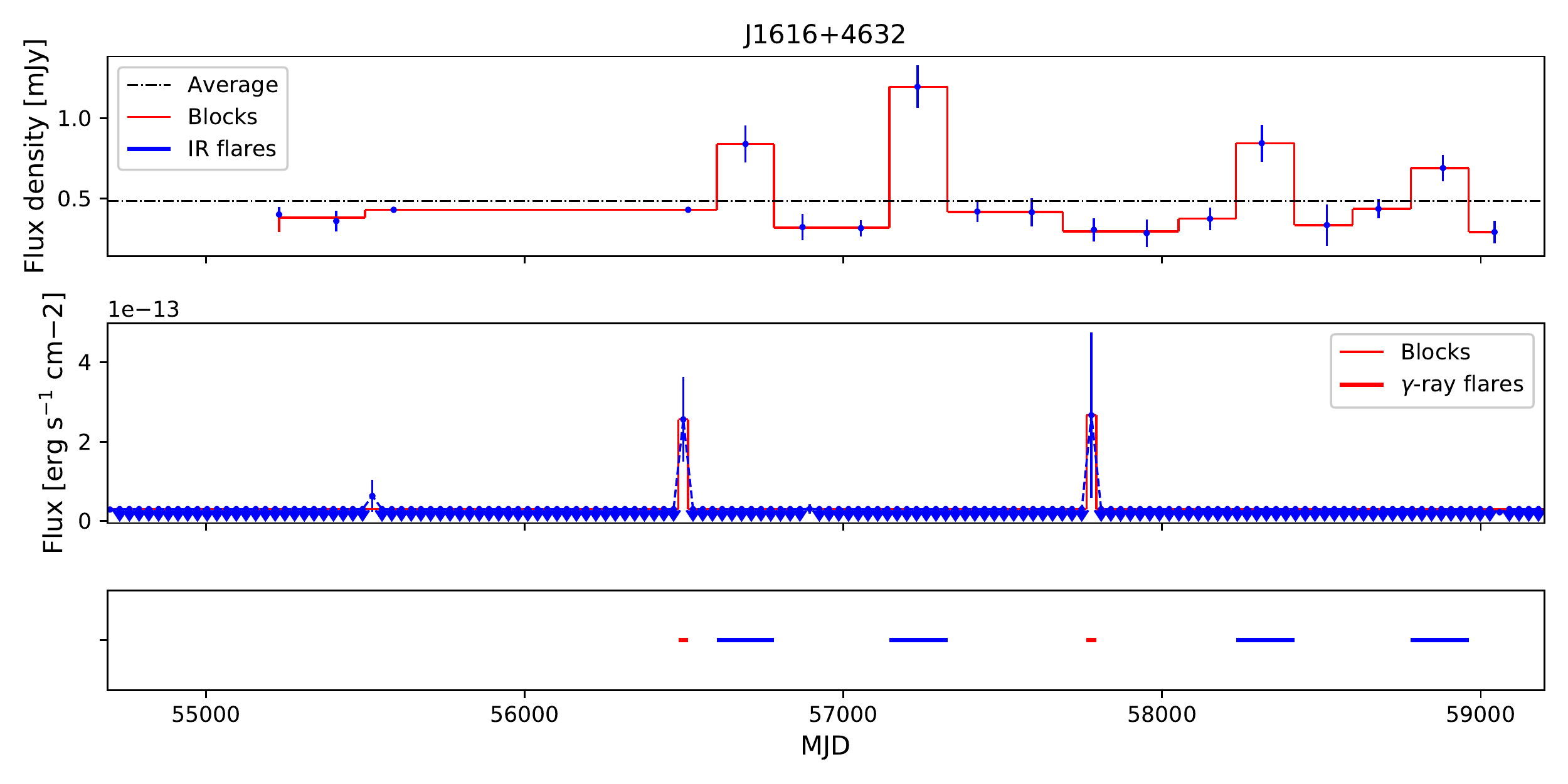}
\includegraphics[width=0.493\linewidth]{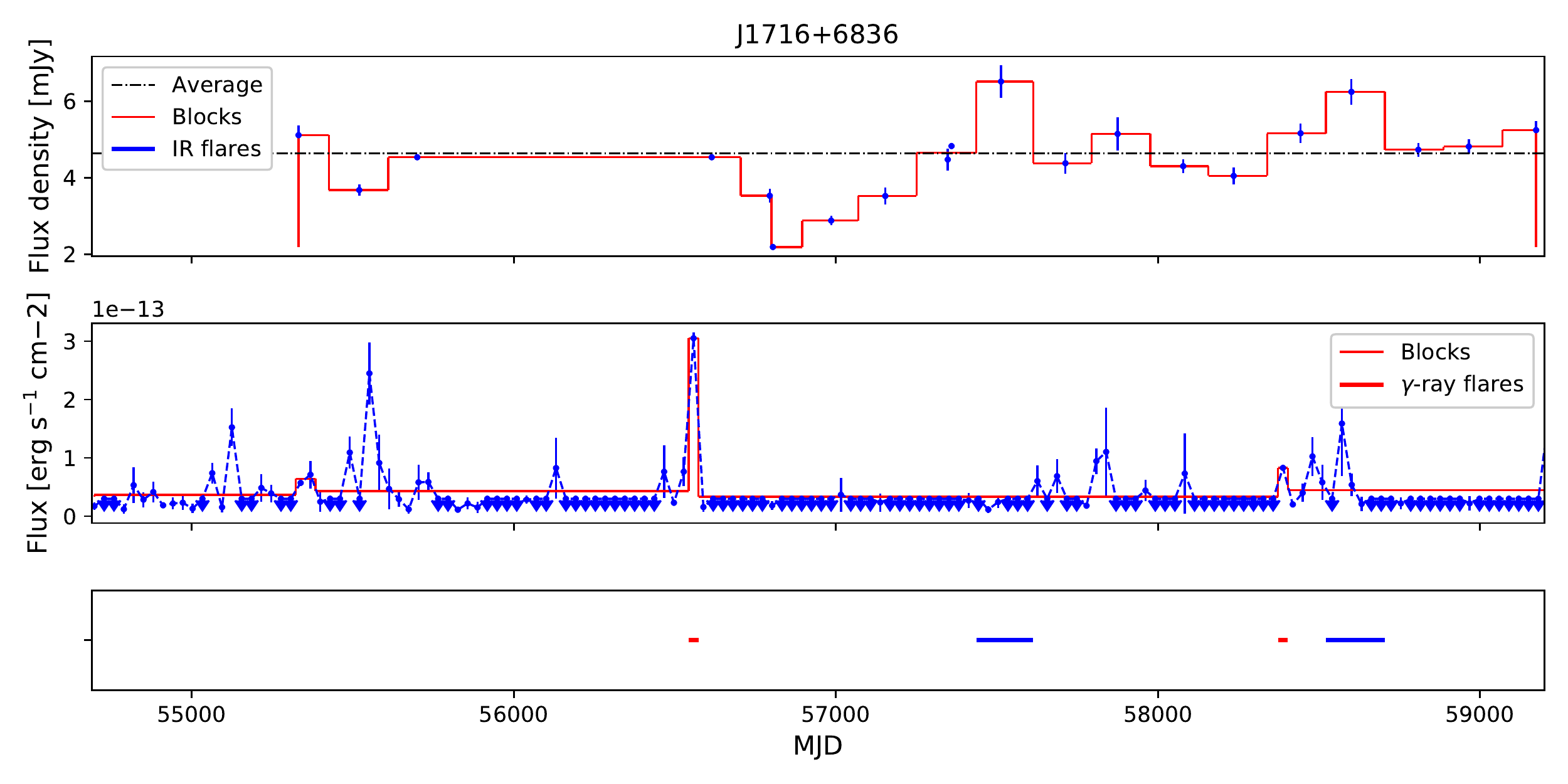}

\includegraphics[width=0.493\linewidth]{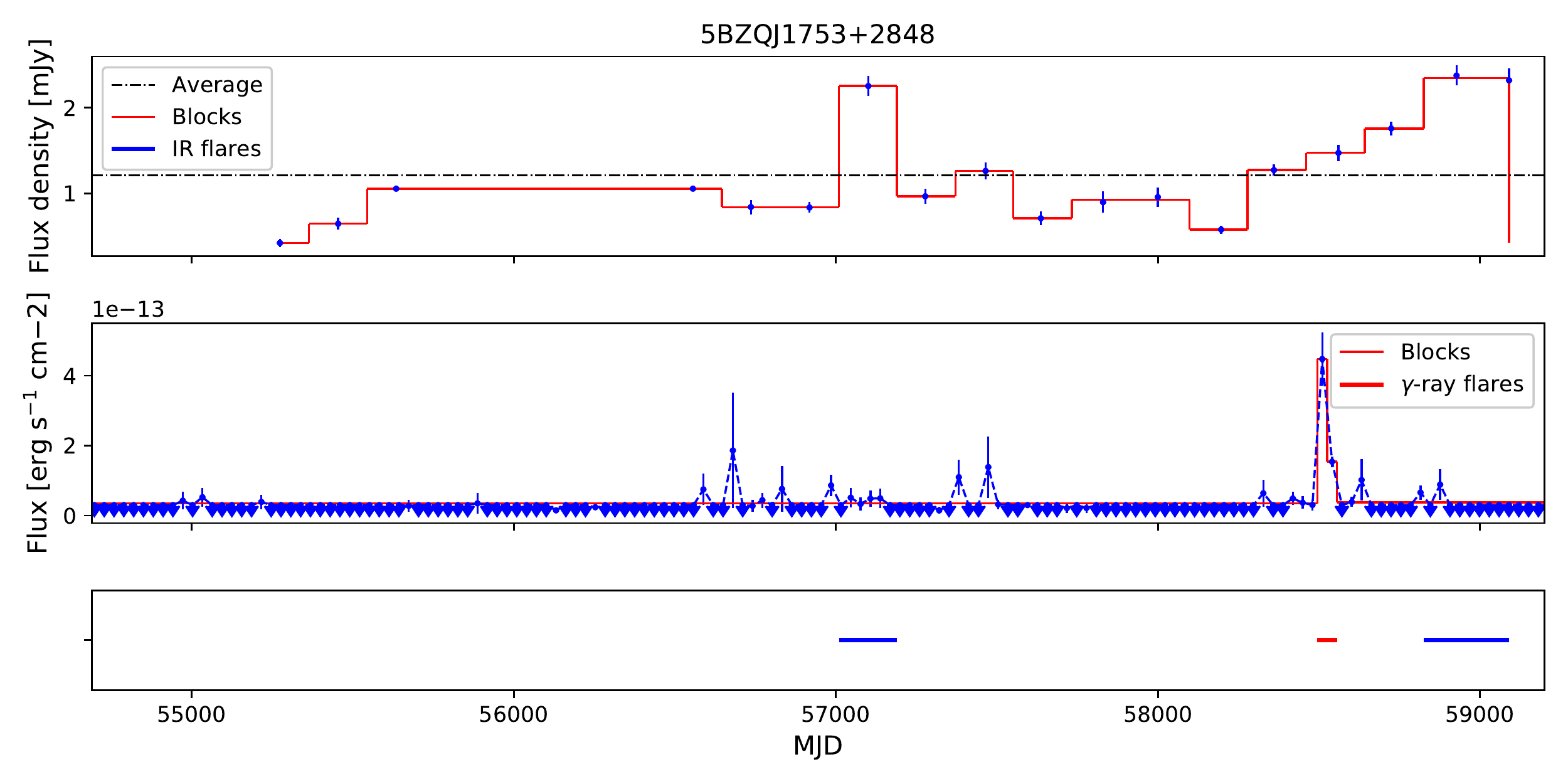}
\includegraphics[width=0.493\linewidth]{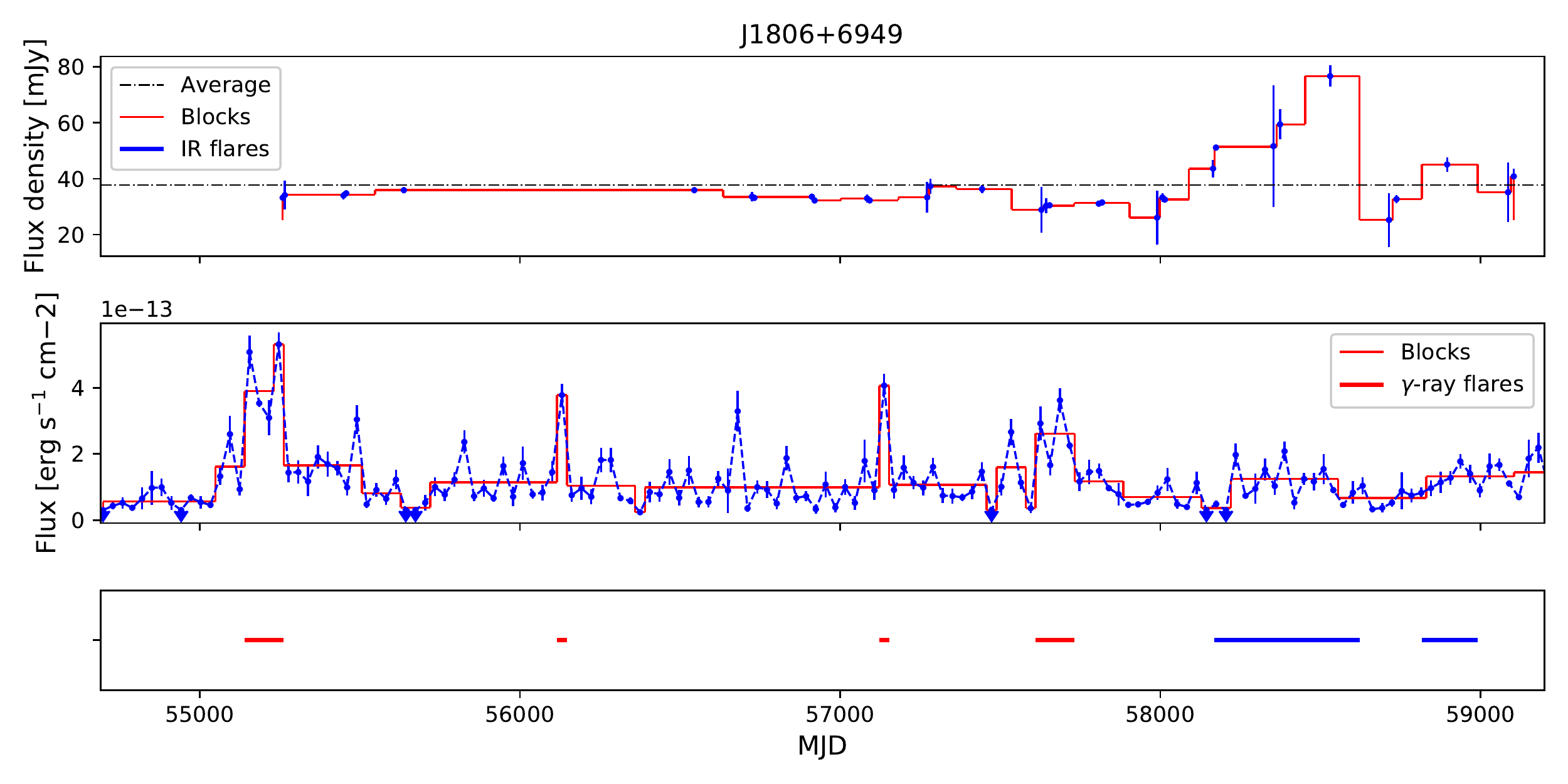}

\includegraphics[width=0.493\linewidth]{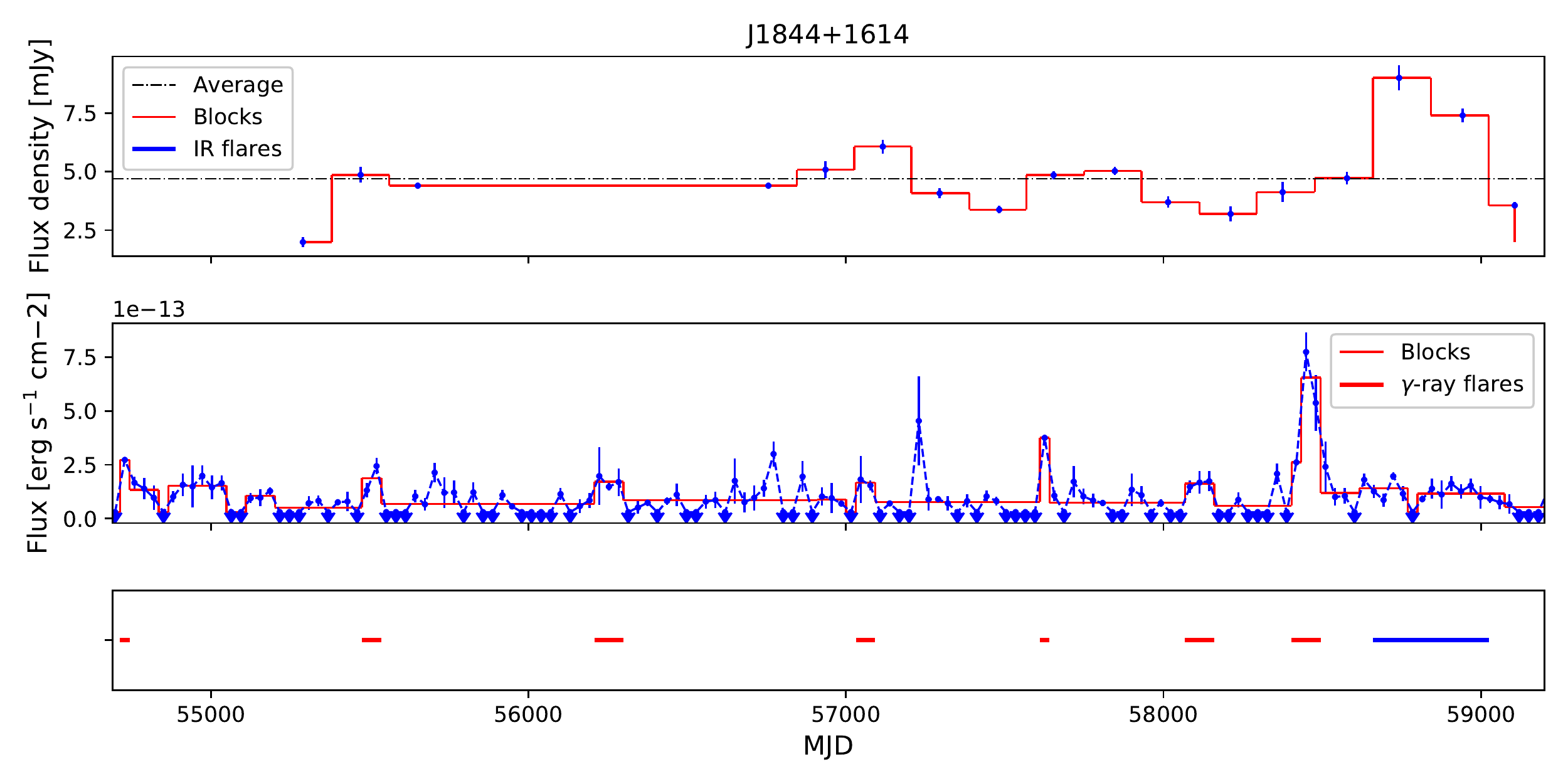}
\includegraphics[width=0.493\linewidth]{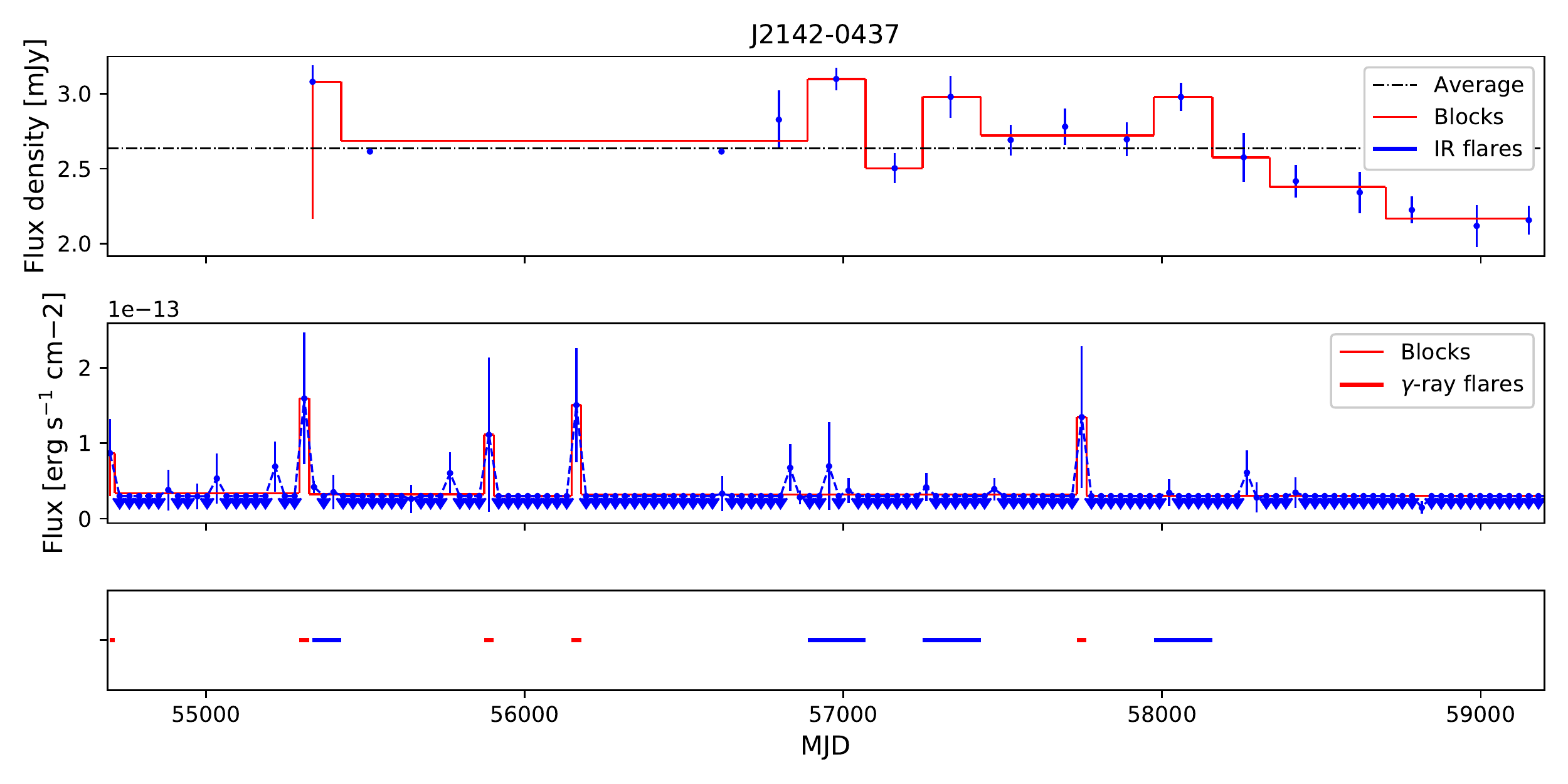} \par
Figure~\ref{lightcurve} continued. Light curves of selected TXS-like sources.  The upper panel is the infrared light curve from WISE multi-epoch data, and the middle panel is the {\it Fermi} \gr\ light curve analyzed with Fermi Science Tool. The lower panel represents the flaring periods in infrared (blue thick lines) and \gr\ (red thick lines) as well as the simultaneous flaring stages (green lines). Black lines represent the arrival time of the neutrino alerts.
\end{figure}

\clearpage
\section{Table of neutrino counts around the 4 groups of blazars} \label{tableresult}

\begin{table} [h!]
\begin{center}
    \begin{tabular}{cccccc}
    \hline
     Log10(E/GeV) &  & Selected TXS-like & CFS & NFS & ACC selected \\
    \hline
    \hline
    4.8 & $N_{measured}$ & 243 & 595 & 3084 & 158 \\
        & $N_{expected}$ & 232 & 610 & 3071 & 156 \\
        & $1\sigma$ & 15 & 25 & 60 & 13 \\
    \hline
    4.9 & $N_{measured}$ & 223 & 551 & 2862 & 146 \\
        & $N_{expected}$ & 214 & 570 & 2859 & 145 \\
        & $1\sigma$ & 14 & 24 & 58 & 12 \\
    \hline
    5.0 & $N_{measured}$ & 209 & 505 & 2629 & 135 \\
        & $N_{expected}$ & 194 & 526 & 2627 & 133 \\
        & $1\sigma$ & 14 & 23 & 55 & 11 \\
    \hline
    5.1 & $N_{measured}$ & 188 & 463 & 2370 & 122 \\
        & $N_{expected}$ & 173 & 480 & 2379 & 119 \\
        & $1\sigma$ & 13 & 22 & 52 & 11 \\
    \hline
    5.2 & $N_{measured}$ & 159 & 416 & 2106 & 114 \\
        & $N_{expected}$ & 153 & 430 & 2116 & 103 \\
        & $1\sigma$ & 12 & 21 & 48 & 10 \\
    \hline
    5.3 & $N_{measured}$ & 135 & 366 & 1855 & 94 \\
        & $N_{expected}$ & 132 & 376 & 1855 & 88 \\
        & $1\sigma$ & 12 & 20 & 45 & 9 \\
   \hline
    5.4 & $N_{measured}$ & 108 & 304 & 1560 & 75 \\
        & $N_{expected}$ & 113 & 319 & 1568 & 75 \\
        & $1\sigma$ & 11 & 18 & 41 & 9 \\
   \hline
    5.5 & $N_{measured}$ & 80 & 252 & 1261 & 62 \\
        & $N_{expected}$ & 92 & 258 & 1260 & 59 \\
        & $1\sigma$ & 10 & 16 & 36 & 8 \\
   \hline
    5.6 & $N_{measured}$ & 59 & 207 & 936 & 38 \\
        & $N_{expected}$ & 70 & 202 & 942 & 41 \\
        & $1\sigma$ & 8 & 14 & 30 & 6 \\
   \hline
    5.7 & $N_{measured}$ & 46 & 131 & 642 & 26 \\
        & $N_{expected}$ & 48 & 136 & 629 & 28 \\
        & $1\sigma$ & 7 & 12 & 25 & 5 \\
   \hline
    5.8 & $N_{measured}$ & 25 & 82 & 407 & 18 \\
        & $N_{expected}$ & 31 & 89 & 406 & 18 \\
        & $1\sigma$ & 6 & 9 & 19 & 4 \\
   \hline
    5.9 & $N_{measured}$ & 12 & 62 & 261 & 15 \\
        & $N_{expected}$ & 20 & 56 & 257 & 11 \\
        & $1\sigma$ & 5 & 7 & 15 & 3 \\
   \hline
    6.0 & $N_{measured}$ & 7 & 39 & 171 & 7 \\
        & $N_{expected}$ & 12 & 34 & 161 & 7 \\
        & $1\sigma$ & 3 & 6 & 12 & 3 \\
    \hline
    \end{tabular}
\caption{Number of neutrino events around blazars in our selected lists and control groups. $N_{measured}$ represents the real number of observed neutrinos, $N_{expected}$ represents the average number of neutrino counts from the simulation with blazars RA randomly scrambled 10000 times, and $1\sigma$ is the statistical error at $68\%$ significance level of the simulation. } \label{simresult}
\end{center}
\end{table}



\clearpage
\bibliographystyle{aasjournal}
\bibliography{ICalert}{}

\end{document}